\documentclass[usenatbib,letterpaper]{mn2e}
\usepackage[totalwidth=480pt,totalheight=680pt]{geometry}
\usepackage{psfig}
\usepackage{aas_macros}
\usepackage{amssymb}

\usepackage{color}
\definecolor{dr}{rgb}{0.6,0,0}
\definecolor{db}{rgb}{0,0,0.6}

\usepackage[hyperfootnotes={false},
dvips,
pdftitle={Are Debris Disks Self-Stirred?},
pdfauthor={Grant M. Kennedy and Mark C. Wyatt},
pdfstartview={FitH},
bookmarksopen={true} ]{hyperref}
\hypersetup{colorlinks={true}, urlcolor={db}, citecolor={dr}}

\newcommand{\dbl}{$D_{\rm bl}$}
\newcommand{\um}{$\mu$m}
\newcommand{\rmid}{$r_{\rm mid}$}
\newcommand{\rmidmin}{$r_{\rm mid,min}$}
\newcommand{\rin}{$r_{\rm in}$}
\newcommand{\rinf}{$r_{24-70}$}
\newcommand{\chir}{r}
\newcommand{\rout}{$r_{\rm out}$}
\newcommand{\rstir}{$r_{\rm stir}$}
\newcommand{\tstir}{$t_{\rm stir}$}
\newcommand{\tc}{$t_{\rm c}$}
\newcommand{\taueff}{$\tau_{\rm eff}$}
\newcommand{\qd}{$Q_{\rm D}^\star$}
\newcommand{\dc}{$D_{\rm c}$}
\newcommand{\fmax}{$f_{\rm max}$}
\newcommand{\xdel}{$x_{\rm delay}$}

\begin{document}

\title[Are debris disks self-stirred?]
  {Are debris disks self-stirred?}

\author[G. M. Kennedy \& M. C. Wyatt]
  {G. M. Kennedy\thanks{Email: gkennedy@ast.cam.ac.uk} and M. C. Wyatt\\
  Institute of Astronomy, University of Cambridge, Madingley Road,
  Cambridge CB3 0HA, UK}

\maketitle

\begin{abstract} This paper aims to consider the evidence that debris disks are
  self-stirred by the formation of Pluto-size objects. A semi-analytical model for the
  dust produced during self-stirring is developed and applied to the statistics for A
  stars. We show that there is no significant statistical difference between fractional
  excesses of A-stars $\lesssim$50\,Myr old, and therefore focus on reproducing the broad
  trends, the ``rise and fall'' of the fraction of stars with excesses that the
  pre-stirred model of \citet{2007ApJ...663..365W} does not predict. Using a population
  model, we find that the statistics and trends can be reproduced with a self-stirring
  model of planetesimal belts with radius distribution $N(r) \propto r^{-0.8}$ between
  15--120\,AU, with width $dr = r/2$. Disks must have this 15\,AU minimum radius in order
  to show a peak in disk fraction, rather than a monotonic decline. However, the marginal
  significance of the peak in the observations means that models with smaller minimum
  radii also formally fit the data. Populations of extended disks with fixed inner and/or
  outer radii fail to fit the statistics, due mainly to the slow 70\,\um\ evolution as
  stirring moves further out in the disk. This conclusion, that debris disks are narrow
  belts rather than extended disks, is independent of the significance of 24\,\um\ trends
  for young A-stars. Although the rise and fall is naturally explained by self-stirring,
  we show that the statistics can also be reproduced with a model in which disks are
  stirred by secular perturbations from a nearby eccentric planet. Detailed imaging,
  which can reveal warps, sharp edges, and offsets in individual systems is the best way
  to characterise the stirring mechanism. From a more detailed look at $\beta$ Pictoris
  Moving Group and TW Hydrae Association A-stars we find that the disk around $\beta$
  Pictoris is likely the result of secular stirring by the proposed planet at
  $\sim$10\,AU; the structure of the HR 4796A disk also points to sculpting by a
  planet. The two other stars with disks, HR 7012 and $\eta$ Tel, possess transient hot
  dust, though the outer $\eta$ Tel disk is consistent with a self-stirred origin. We
  suggest that planet formation provides a natural explanation for the belt-like nature
  of debris disks, with inner regions cleared by planets that may also stir the disk, and
  the outer edges set by where planetesimals can form.
\end{abstract}


\section{Introduction}\label{sec:intro}

Debris disks are the disks of dust found around nearby main sequence stars through their
thermal emission \citep[for a recent review see][]{2008ARA&A..46..339W}. The dust itself
is short-lived compared to the lifetime of the star so is believed to be continually
replenished through collisions between km-sized planetesimals
\citep{2002MNRAS.334..589W,2007MNRAS.380.1642Q}, much in the same way that dust in the
zodiacal cloud is replenished through collisions in the asteroid belt and Kuiper belt
\citep{dermott01,2003AJ....125.2255M}.  While there remain difficulties in growing dust
from its initially sub-$\mu$m size into $>$km-sized planetesimals
\citep{2008ARA&A..46...21B}, the widely invoked coagulation and core accretion models for
planet formation rely on relative velocities in the protoplanetary disk being low enough
that collisions result in net accretion, rather than destruction. As it seems that the
opposite is the case in debris disks, these disks must have been stirred at some point so
that collisions occur at $\gtrsim$1--10\,m/s, which typically corresponds to
eccentricities and inclinations for the disk material of $e \gtrsim 10^{-3}$ to $10^{-2}$
\citep[e.g.][]{2008ApJS..179..451K}.

To explain debris disks it is not normally necessary to understand how (or when) they
were stirred. It is sufficient to invoke a ``pre-stirred'' debris disk---one that was
stirred when the star was born. For example, it is possible to explain the statistics of
dust found around A-stars of ages $>$10\,Myr by assuming that all stars are born with a
pre-stirred planetesimal belt that evolves due to steady-state collisional erosion
\citep{2007ApJ...663..365W}. The diversity seen at different ages then reflects their
different initial masses and radii. A similar conclusion was reached for debris disks
around sun-like stars \citep{2008ApJ...673.1123L}.

However, recent results on the presence of debris disks around young A-stars are
challenging this view by showing that the fractional excess at 24\,$\mu$m from hot dust
increases from 3\,Myr (the time at which most protoplanetary disks have dissipated) to a
peak at $\sim$10--30\,Myr, followed by a slow decline as expected by steady-state
evolution \citep{2006ApJ...652..472H,2008ApJ...672..558C,2008ApJ...688..597C}. This peak
has been interpreted as evidence for self-stirring, where debris are created when the
largest objects become massive enough to stir planetesimals to fragmentation
velocities.\footnote{Self-stirring is also called delayed stirring in the literature, but
  following \citet{2008ARA&A..46..339W} we consider that what we are calling
  self-stirring is a sub-set of possible delayed-stirring models.}

Self-stirring models follow the evolution of an extended planetesimal belt from the
protoplanetary disk stage, allowing both the size distribution and the velocity
distribution to evolve in a self-consistent manner. These models find that stirring of
the planetesimal belts occurs when planets reach Pluto-size, which depends strongly on
their distance from the star, as well as on the mass surface density of solid material
there: $t_{\rm Pluto} \propto P / \Sigma$, where $P$ is the orbital period and $\Sigma$
is the surface density of planetesimals \citep[e.g.][]{1987Icar...69..249L}. For a
typical disk model $\Sigma = \Sigma_0 \, r^{-1.5}$, so $t_{\rm Pluto}$ depends strongly
on radius $r$:
\begin{equation}
  t_{\rm{Pluto}} \propto r^3 / \, \left( \Sigma_0 \, \sqrt{M_\star} \right) \, .
  \label{eq:tpluto}
\end{equation}
Thus, compared to pre-stirred disks, self-stirring means that farther disk regions are
stirred at later times because it takes longer for Pluto-size objects to form there. This
evolution means that individual disks can show a peak in emission at $\sim$10--30\,Myr,
set by the time to form Pluto-size objects at the inner disk edge
\citep{2006ApJ...652..472H,2008ApJ...672..558C,2008ApJS..179..451K,2008ARA&A..46..339W}. That
is, the observed evolution occurs if debris disks tend to have $\sim$10\,AU inner
holes. The overall dust content at $<$10\,Myr is low because such times are needed for
Plutos to form at the inner edge of the disk. This lack of dust implies that accretion
with minimal fragmentation is ongoing in young ($<$10\,Myr) debris disks and the outer
regions of older debris disks.

Though the \citet{2008ApJS..179..451K} self-stirring models are qualitatively consistent
with the $\sim$10--30\,Myr peak in fractional excesses, there has been no quantitative
test of whether these models can reproduce the observed A-star statistics, or how the
observations constrain disk parameters. In this paper, our aim is to address this issue
using the \citet{2007ApJ...658..569W} debris disk evolution model, modified to include
self-stirring \citep[as outlined in][]{2008ARA&A..46..339W}.

In particular, there are several issues to address. Because the peak in excesses occurs
when Plutos form at the inner disk edge, the timing of this peak is highly dependent on
the radius of the cleared region and on mass surface density (Eq. \ref{eq:tpluto}). Yet
we know that protoplanetary disks have a range of masses and surface densities
\citep[e.g.][]{2000prpl.conf..559N,2005ApJ...631.1134A,2007ApJ...671.1800A}, and it seems
unlikely that the inner clearing would be at the same radius for all disks. Therefore, we
wish to find whether self-stirring models can reproduce the A-star statistics, and if so,
whether they put a strong constraint on disk inner radii as Equation (\ref{eq:tpluto})
suggests.

Another issue is the extent of debris disks. \citet{2006ApJ...637L..57K} note that disks
resolved in scattered light appear to be either relatively narrow belts 20--30\,AU wide,
or extended disks wider than 50\,AU (however, a disk that appears extended may result
from blowout of small grains created in a relatively narrow planetesimal belt). In
fitting the statistics for ($\gtrsim$10\,Myr) A-stars \citet{2007ApJ...663..365W}
considered disks to be the former; belts centred at radial distance $r$ with an assumed
width of $dr = r/2$. However, self-stirring allows the interesting possibility that disks
similar in extent to observed protoplanetary disks---from near the star to many hundreds
of AU---may evolve to look like narrow belts because only regions of recent Pluto
formation may be luminous enough to be detected.  Therefore, we also wish to evaluate
whether debris disks tend to be ``narrow belts'' or ``extended disks'' (we use these
terms consistently throughout to refer to these types of disks).

A final issue is whether debris disks could be stirred by something other than the
formation of Pluto-sized objects. Though self-stirring has been the only proposed
mechanism, secular perturbations from planets (another subset of delayed stirring) seems
to be an equally viable way to stir debris disks \citep{2009MNRAS.399.1403M}. This
hypothesis is partly motivated by stars with debris disks known or predicted to harbour
planets, such as Fomalhaut and $\beta$ Pic
\citep{1997MNRAS.292..896M,2006MNRAS.372L..14Q,2008Sci...322.1345K,2009A&A...493L..21L}. The
question is therefore whether we can identify the more important stirring mechanism, both
at a population level and for individual objects.

The layout of this paper is as follows. In \S \ref{sec:model} we outline the
\citet{2007ApJ...658..569W,2007ApJ...663..365W} model, and the modifications made to
include self-stirring. We empirically fit some model parameters to reproduce the
self-stirring models of \citet{2008ApJS..179..451K} in \S \ref{sec:single}. As
self-stirring results in not only excess evolution, but also spatial evolution as Plutos
form farther from the central star, we show how disk surface density profiles evolve and
vary with model parameters. In \S \ref{sec:astar} a population model is compared with the
statistics for A-stars. We briefly consider a planet-stirred population model, take a
more detailed look at resolved $\beta$ Pictoris Moving Group and TW Hydrae A-stars in the
context of delayed stirring, and discuss other influences on debris disk structure in \S
\ref{sec:discussion}. We summarise our main conclusions in \S \ref{sec:conclusions}.

\section{Analytical self-stirring model}\label{sec:model}

This section describes the analytical model. It is an extension of the model described in
\citet{2007ApJ...658..569W,2007ApJ...663..365W} to include delayed stirring by splitting
disks into a series of concentric annuli. The implementation uses the equations in \S 2
of \citet{2007ApJ...658..569W,2007ApJ...663..365W} and sums over 100 logarithmically
spaced annuli to create a disk with radial extent. The self-stirring prescription is
semi-empirical, based on the more detailed models of \citet{2008ApJS..179..451K} (see \S
\ref{sec:comp}). Here, we briefly summarise the model and refer the reader to \S 2 of
\citet{2007ApJ...658..569W,2007ApJ...663..365W} for details omitted here.

We use a Minimum Mass Solar Nebula \citep[MMSN,][]{1977Ap&SS..51..153W} surface density
profile to specify the disk mass
\begin{equation}\label{eq:mmsn}
  \Sigma = \eta \, M_\star \, \Sigma_0 \, r^{-\delta} \, ,
\end{equation}
where $\eta$ is a scaling parameter reflecting a range of disk masses, $M_\star$ is in
Solar units, radial distance $r$ is in AU, and the surface density at 1\,AU is
$\Sigma_0=30$\,g\,cm$^{-2}$ (1.1\,$M_\oplus/{\rm AU}^2$ in our units). The power-law
index $\delta$ is typically 1--1.5; we use 1.5 as our canonical value. This surface
density provides roughly the amount of solid material contained in the outer Solar
System. The surface density scales linearly with stellar mass, consistent with mm
observations of protoplanetary disks.  However, the range of stellar masses in our model
is much smaller than the expected range of $\eta$, so the factor $M_\star$ is relatively
unimportant \citep{2000prpl.conf..559N,2005ApJ...631.1134A}. For extended disks the inner
and outer radii are independent and specified by \rin\ and \rout. Narrow belts are
specified by their radii $r_{\rm mid}$ and width $dr$ (so for belts $r_{\rm in,out} =
r_{\rm mid} \pm dr/2$).

Where \citet{2007ApJ...663..365W} considered a range of total masses $M_{\rm tot}$, we
consider a log-normal distribution about $\eta_{\rm mid}$.\footnote{Previously each disk
  was assigned a total mass distributed about $M_{\rm mid}$ independent of disk
  location. This led to typical disks at $r = 120$\,AU having about 6 times lower
  $\Sigma_0$ than disks at 3\,AU (for $\delta = 3/2$). However, this difference is
  smaller than the dispersion in surface density and disk mass so is not particularly
  important.}  There is as yet no evidence of a positive correlation between disk mass
and radius (which might argue for the distribution of $\eta$ to be narrower than that for
$M_{\rm tot}$), and thus we use the 1\,$\sigma$ width of 1.14\,dex observed for disk
masses in Taurus \citep{2005ApJ...631.1134A}.

The planetesimal disk is assumed to be in collisional equilibrium with a size
distribution in each annulus defined by $N(D) = K D^{2-3q}$, where $K$ is a constant, $q
= 11/6$ for an infinite collisional cascade \citep{1969JGR....74.2531D}, and $D$ is
planetesimal diameter. That distribution is assumed to hold from the largest planetesimal
in the disk, of diameter \dc\ (in km), down to the size below which particles are blown
out by radiation pressure as soon as they are created, \dbl\ (in \um). If $q$ is in the
range 5/3 to 2 then most of the mass is in the largest planetesimals while the
cross-sectional area $\sigma_{\rm tot}$ is dominated by the smallest particles:
\begin{equation}\label{eq:stot}
  \sigma_{\rm{tot}} = 3.5 \times 10^{-17} K(3q-5)^{-1} (10^{-9}D_{\rm{bl}})^{5-3q}
\end{equation}
in AU$^2$.

We assume that particles act like blackbodies, so the fractional luminosity of grains $f
= L_{\rm IR}/L_\star = \sigma_{\rm tot}/4 \pi r^2$ for the annulus at $r$. The grain
blowout size is
\begin{equation}\label{eq:dbl}
  D_{\rm bl} = 0.8 \, (L_\star/M_\star) \, (2700/\rho) \, , 
\end{equation}
where \dbl\ is in $\mu$m, $L_\star$ and $M_\star$ are in Solar units, and density
$\rho$ is in kg\,m$^{-3}$. The dust temperature can be worked out from:
\begin{equation}\label{eq:tbb}
  T = 278.3 \, L_\star^{0.25} \, r^{-0.5} \, .
\end{equation}
We also assume that the central star acts like a blackbody.

Because we use a fixed size distribution with $q = 11/6$ in our model, the long-term
evolution of the disk is determined by the removal of mass from the top end of the
cascade. Planetesimals have a disruption threshold $Q_{\rm{D}}^\star$, and eccentricity
$e$. The collisional lifetime of the largest planetesimals of size \dc\ at a radius $r$
in an MMSN disk is
\begin{equation}\label{eq:tc}
  t_{\rm c} = 2.2 \times 10^{-10} r^{23/6} \, D_{\rm c} \, {Q_{\rm D}^\star}^{5/6} \, 
  e^{-5/3} \, M_\star^{-7/3} \, \Sigma_0^{-1} \, \eta^{-1} \,
\end{equation}
in Myr, obtained by substituting $\Sigma = M_{\rm tot} / 2 \pi r dr$ into
\citet{2007ApJ...663..365W} Equation (9). We have simplified the relation here by
assuming that $X_{\rm c} = D_{\rm cc}/D_{\rm c} \ll 1$, where $D_{\rm{cc}}$ is the
smallest planetesimal that has enough energy to catastrophically destroy a planetesimal
of size $D_{\rm{c}}$. This condition applies for the $e \gtrsim 0.01$ eccentricities
found for self-stirring models \citep{2007ApJ...658..569W}.

Assuming that collisions are the only process affecting the evolution of the surface
density, then
\begin{equation}
  \Sigma(r,t) = \left\{ \begin{array}{ll} 
      \Sigma(r,0) & t < t_{\rm stir} \\
      \Sigma(r,0)/[1+\left( t-t_{\rm stir}(r) \right)/t_{\rm{c}}(r,0)]  & t > t_{\rm stir}
    \end{array}
  \right.
\end{equation}
where $\Sigma(0)$ is the initial surface density profile, $t_{\rm stir}$ is the delay
until stirring \citep[assumed to be 0 in][]{2007ApJ...663..365W}, and $t_{\rm{c}}(r,0)$
is the collisional lifetime at $r$ at that initial epoch. Where $t < t_{\rm stir}$, the
initial surface density profile applies.

Because more massive disks process their mass faster ($t_{\rm{c}} \propto 1/\eta$), the
surface density at late times ($t \gg t_{\rm c}$ and $t > t_{\rm stir}$) at a given
radius is independent of initial surface density. Written in terms of the surface density
of cross-sectional area in the disk (or simply ``surface density'' because $\tau_{\rm
  eff} \propto \Sigma$, $\tau_{\rm eff} = \sigma_{\rm tot} / 2 \pi r dr = 2 f r / dr$).
\begin{eqnarray}\label{eq:tmax}
  \tau_{\rm{eff,max}} & = & 1.2 \times 10^{-9} \, r^{7/3} \, D_{\rm{c}}^{0.5} \,
  {Q_{\rm{D}}^\star}^{5/6} \nonumber \\
  & & \times \, \, e^{-5/3} \, M_\star^{-5/6} \, L_\star^{-0.5} \left(t -
    t_{\rm{stir}}\right)^{-1}
\end{eqnarray}
This prescription means that for a disk of known size at any given age, there is a
maximum infrared luminosity, \fmax, that can remain due to collisional processing.  While
\fmax\ can be applied to unresolved stars based on the $24 - 70$\,\um\ colour and some
assumption about disk extent \citep{2007ApJ...658..569W}, $\tau_{\rm eff,max}$
necessarily requires resolved observations for comparison. We return to the application
of $\tau_{\rm eff,max}$ to self-stirred disks in \S \ref{sec:trans}.

The final component of the model is the implementation of self-stirring, where successive
annuli can be stirred at later times with increasing $r$ (such as
Eq. \ref{eq:tpluto}). Other mechanisms for delayed-stirring, such as secular
perturbations from an eccentric planet \citep[e.g.][]{2009MNRAS.399.1403M}, can be
implemented with different \tstir\ (see \S \ref{sec:selfvplanet}). The introduction of
self-stirring also requires us to specify the level of disk emission before a disk is
stirred. We assume that the emission is proportional to the surface density, but reduced
by a factor \xdel\ with respect to that expected from Equation (\ref{eq:stot}) because
the collisional cascade has not yet begun. We estimate \xdel\ through comparison with the
\citet{2008ApJS..179..451K} models below.

\section{Debris disk evolution for individual stars}\label{sec:single}

In this section we consider model evolution for individual disks, first making
comparisons with the \citet{2008ApJS..179..451K} models, and then illustrating the
surface density evolution. We then consider the implications for transience in the
context of the \citet{2007ApJ...658..569W} model and the effects of disk extent on excess
evolution.

\subsection{Comparison with \citeauthor{2008ApJS..179..451K} models}\label{sec:comp}

\begin{figure*}
  \begin{center}
    \vspace{-0.2in}
   \begin{tabular}{cc}
      \hspace{-0.35in} \psfig{figure=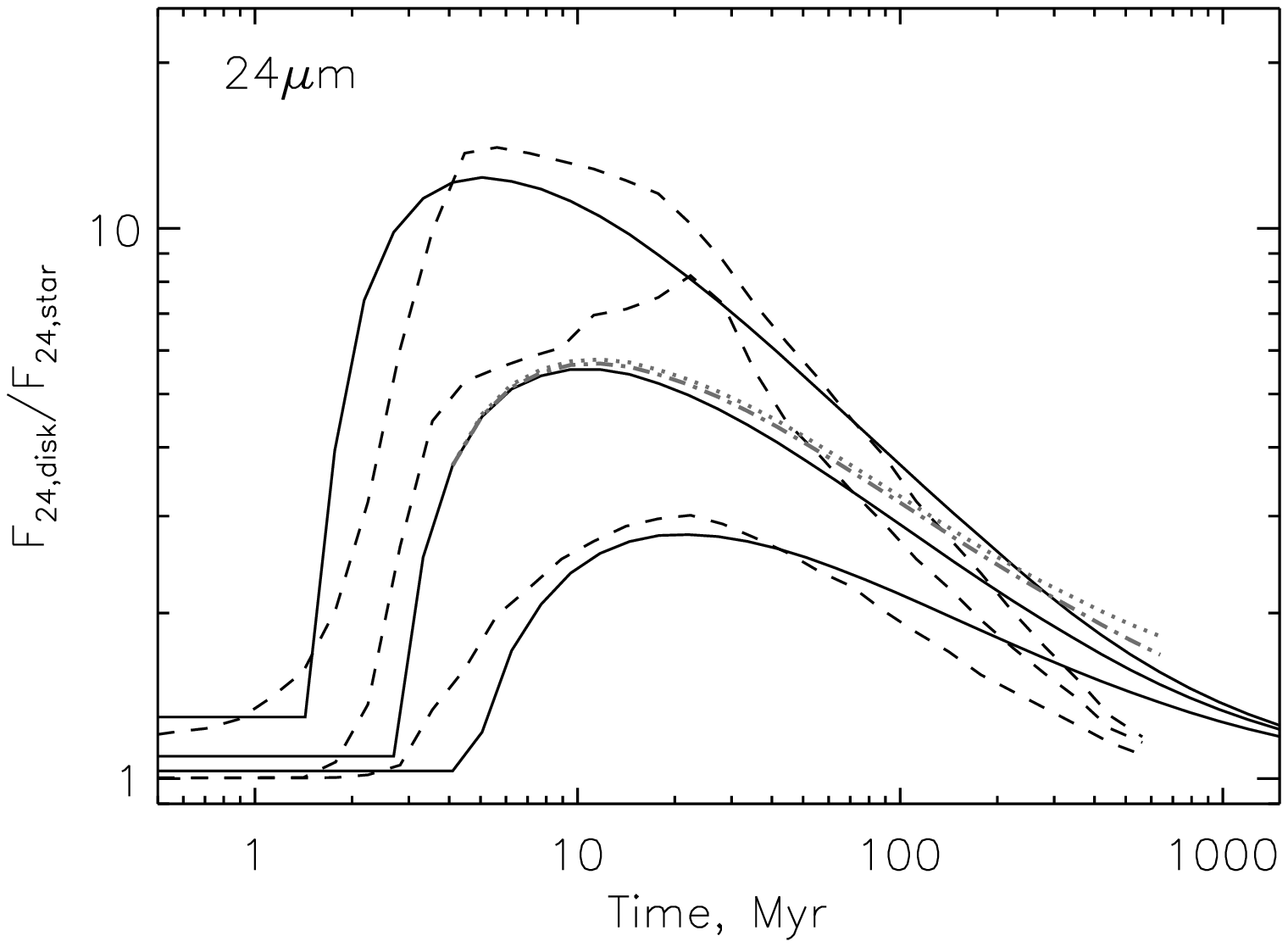,width=0.52\textwidth}&
      \hspace{-0.35in} \psfig{figure=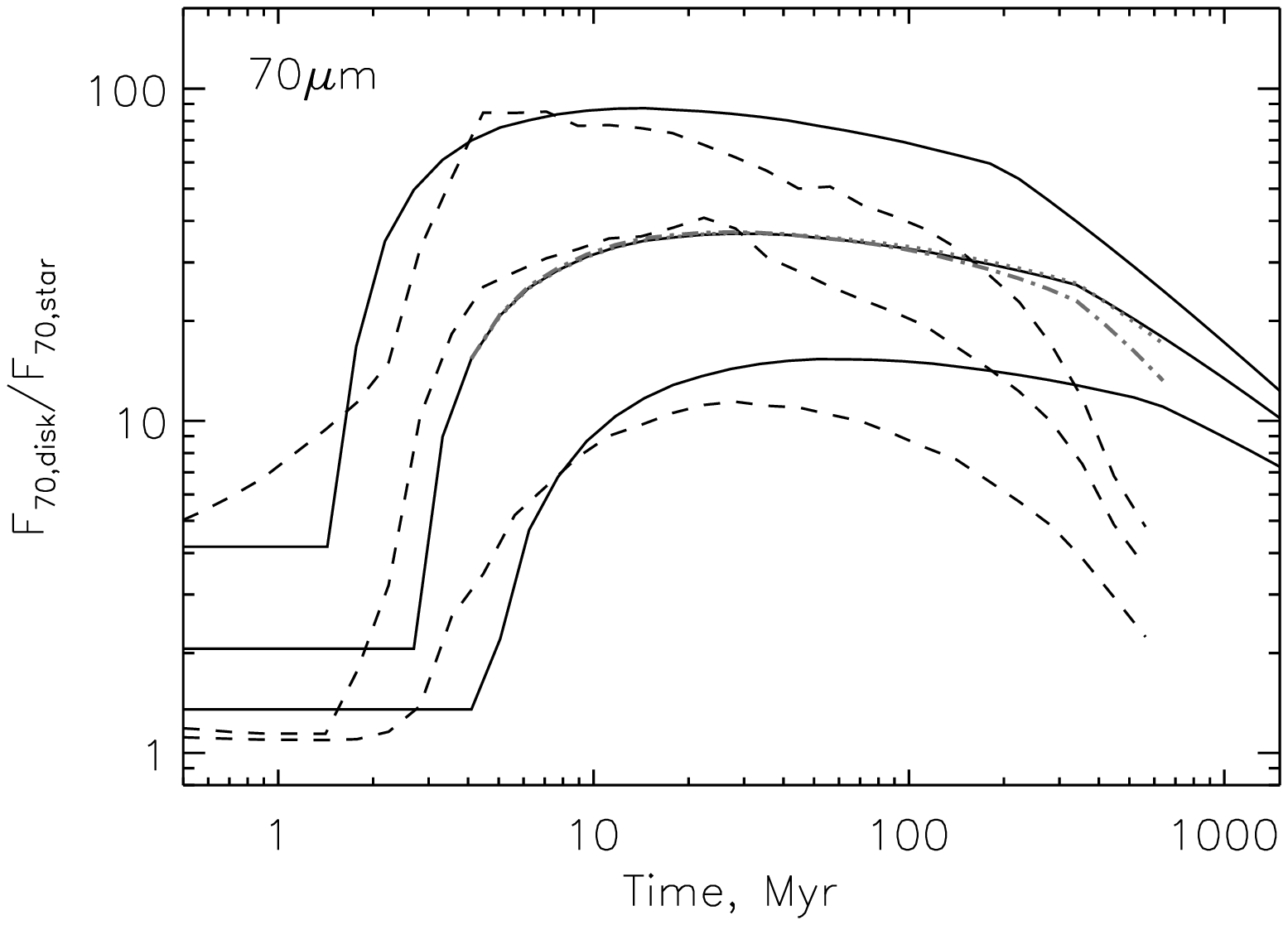,width=0.52\textwidth}\\
    \end{tabular}
    \caption{Evolution of 24\,$\mu$m (\emph{left}) and 70\,$\mu$m (\emph{right}) excesses
      for $\eta = 1/3$, 1, and 3 for a disk from 30 to 150\,AU from our model
      (\emph{solid lines}) compared to models from \citet{2008ApJS..179..451K}
      (\emph{dashed lines}), for an A2 (2.5\,$M_\odot$) star. The main difference between
      the models at 70\,\um\ is caused by continued accretion and subsequent stirring in
      the \citet{2008ApJS..179..451K} models (see text). Also shown is the effect of the
      increasing main-sequence luminosity on an $\eta = 1$ disk with $D_{\rm bl} =
      1$\,\um\ (\emph{grey dotted line}) and then also allowing \dbl\ to vary (\emph{grey
        dot-dashed line}). The overall effect of stellar evolution due to changing \dbl\
      and hotter grains (i.e. dot-dashed line) is a small increase in 24\,\um\ excesses
      and a similar decrease in 70\,\um\ excesses.}\label{fig:24comp}
  \end{center}
\end{figure*}

We compare our model with the \citeauthor{2008ApJS..179..451K} simulations of planet
formation and debris disk evolution around A-stars to derive an empirical self-stirring
time and \xdel\ that reproduces their excess evolution. Though
\citet{2008ApJS..179..451K} derive relations for the time to reach the peak excess (their
Eq. 55), and the radial time dependence for forming 1000\,km objects (their Eq. 41), they
do not derive a relation for the peak excess at different radial locations. Because our
aim is to reproduce their excess evolution, we use the following empirical relation.
\begin{equation}\label{eq:tstir}
  t_{\rm stir} = 4 \times 10^{-4} \, r^3 \, \eta^{-1/2} \, M_\star^{-3/2}
\end{equation}
in Myr. We first derived the numerical pre-factor and the exponent of $\eta$ in this
relation with a $\chi^2$ minimisation procedure over models with $1.5<M_\star<3$ and
$1/3<\eta<3$ at 24 and 70\,\um. We fix the radial and stellar mass dependence because
these are set by the orbital period and our assumption of $\Sigma \propto
M_\star$. However, this formal method ignores some important model differences, in
particular the continued stirring that appears to cause the \citet{2008ApJS..179..451K}
models to decay faster at 70\,\um\ (discussed below). We therefore arrive at Equation
(\ref{eq:tstir}) by focusing on the 24\,\um\ peak excesses. This different emphasis
changes the values obtained in the formal fit at 24\,\um\ by
$\sim$10--20\%.\footnote{Similar variation is also found depending on how the
  \citeauthor{2008ApJS..179..451K} models are weighted in the minimisation procedure. For
  example, lower weights for smaller excesses makes the peak excesses fit better, whereas
  constant weights fit the late time evolution better but underestimates the peak
  excesses. Allowing \xdel\ to vary results in values in the range $\sim$0.01--0.02.}

Though the dependence on $\eta$ in Equation (\ref{eq:tstir}) is weaker than implied by
Equation (\ref{eq:tpluto}) ($-1/2$ vs. $-1$), the two should not necessarily agree.
\citet{2008ApJS..179..451K} find that the time to peak dust production is $\propto
\eta^{-1} \, M_\star^{-1.5}$ (as predicted by Eq. \ref{eq:tpluto}), but that the time to
peak luminosity is $\propto \eta^{-2/3} \, M_\star^{-1}$ and takes much longer (their
Eqs. 48 \& 55). In comparing these proportionalities, it is important to remember that
the peak dust production and peak luminosity are linked, but not directly. The observed
luminosity depends on the history of the dust production rate and the rate at which small
grains are being removed by radiation forces. Further, our empirical relation sets the
evolution at specific $r$, rather than describing the global evolution of specific disk
properties. Though Equation (\ref{eq:tstir}) is derived largely from the 24\,\um\
evolution, we reproduce the $f$ evolution, from which the
\citeauthor{2008ApJS..179..451K} peak luminosity is derived equally well. Our scaling of
\tstir\ with $\eta$ is probably weaker than their $-2/3$ because we set the \emph{onset}
of stirring, not the peak excess and higher mass disks reach their peak excess more
rapidly once stirring starts (Fig. \ref{fig:24comp}). We refer to the radial location
where stirring has just begun (i.e. where $t = t_{\rm stir}$) as \rstir.

Figure \ref{fig:24comp} shows a comparison of the \citeauthor{2008ApJS..179..451K} model
for an A2 (2.5\,$M_\odot$) star with an equivalent one from our model. The 24\,\um\
excess increases to a peak around 10\,Myr, and then declines \citep[see][for comparison
with
observations]{2006ApJ...652..472H,2008ApJ...672..558C,2008ARA&A..46..339W,2008ApJS..179..451K}.
In this example, the disk extends from 30--150\,AU, and has $\eta = 1/3$, 1, and 3. We
set the level of disk emission before stirring to $x_{\rm delay} = 0.01$, which gives a
reasonable match to the excesses prior to the formation of Pluto size objects. We take
other model parameters from the best fit model of \citet{2007ApJ...663..365W}; $e =
0.05$, $D_{\rm c} = 60$\,km, and $Q_{\rm D}^\star = 150$\,J\,kg $^{-1}$. Though the
largest objects in the disk are roughly Pluto size when the collisional cascade begins,
\citeauthor{2008ApJS..179..451K} find that objects larger than $\sim$1--10\,km continue
growing. Thus the effective maximum planetesimal size \dc\ for our model is of order
10\,km because the smallest dust derives from a reservoir of objects of this
size. Different choices for \dc, $e$, and \qd\ change the rate at which excesses increase
after the disk is stirred, though still give peak excesses that agree with the
\citeauthor{2008ApJS..179..451K} models within factors of a few.

To achieve the agreement shown in Figure \ref{fig:24comp} we must temporarily modify our
model to match some \citeauthor{2008ApJS..179..451K} assumptions about dust properties;
they set $D_{\rm bl} = 1$\,$\mu$m, and use a ``greybody'' emission law $F_\nu \propto
B_\nu(T) \, (1-\exp(\lambda_0/\lambda))$ with $\lambda_0 = 10$\,$\mu$m. They use
\citet{2001ApJS..136..417Y} luminosities, somewhat brighter than the
\citet{1982lbor.book.....ASK} luminosities used in our model. Finally, to ensure a good
match between models we apply a factor 2.5 decrease to the
\citeauthor{2008ApJS..179..451K} excesses. The need for this factor is unsurprising
considering the model differences and could, for example, arise from differences of a few
percent in the average size distribution index $q$ between \dc\ and \dbl. With these
differences taken into account, both models produce fairly similar results. The delay in
the rise of 24 and 70\,$\mu$m excesses is similar, as are their magnitudes over the range
A0--F2 ($\sim$1.5--3\,$M_\odot$) and $\eta = 1/3$--3.

The main difference is the decline in 70\,\um\ excesses at late times. Our model shows a
fairly slow decline until after 200--600\,Myr, when stirring reaches the outer disk edge
and the excess decreases more rapidly because mass is being lost at all radii. The
\citet{2008ApJS..179..451K} models show a faster decay. In their model, the largest
objects continue to accrete once they reach Pluto size, thus increasing the rate of decay
through removal of mass from the collisional cascade. More important is the increased
stirring due to their continued growth, and the subsequent higher collision rates
(S. Kenyon, \emph{priv. communication}). This importance can be understood from the
strong $e$ dependence in Equation (\ref{eq:tc}).

Figure \ref{fig:24comp} also shows the effect of including stellar evolution in our model
(scaled to show relative differences). For fixed \dbl, increasing $L_\star$ increases
grain temperatures and the 24\,\um\ excess is higher. If \dbl\ is allowed to vary, the
disk emission drops slightly due to smaller $\sigma_{\rm tot}$ as \dbl\ increases and the
overall effect is a small increase in 24\,\um\ emission and a similar decrease at
70\,\um.

After a few hundred Myr, the \citeauthor{2008ApJS..179..451K} 70\,\um\ excesses also show
a break towards faster decay. They attribute this decrease to stellar evolution as the
star nears the end of its main-sequence lifetime (of 650\,Myr for 2.5\,$M_\odot$). Higher
stellar luminosity increases the importance of PR drag relative to collisions and the
disk emission drops more rapidly as grains spiral toward the star. However, any mass lost
to PR drag is probably due to the assumption of a fixed \dbl, as these models are of
observable disks, which are generally not tenuous enough to suffer PR drag before grains
collide \citep{2005A&A...433.1007W}. For the A2 star in Figure \ref{fig:24comp}, \dbl\
should increase from about 10 to 20\,\um\ during the MS lifetime, so fixing $D_{\rm bl} =
1$\,\um\ allows PR drag to remove grains that would have instead been blown out of the
system. Also, a disk is unlikely to become PR dominated by realistic increases in stellar
luminosity.\footnote{For fixed $M_\star$ and $r$ the orbit decay timescale for smallest
  grains (which are most affected by PR drag) is $t_{\rm PR} \propto D_{\rm
    bl}/L_\star$. However, because \dbl\ increases with $L_\star$ (Eq. \ref{eq:dbl}), the
  PR timescale for the smallest grains is constant. The collisional lifetime of the
  smallest grains increases slowly as $t_{\rm coll} \propto \sqrt{D_{\rm bl}}$. For fixed
  $\beta = 0.5$, the relative importance of PR drag vs. collisions $\eta_0 \equiv t_{\rm
    PR}/t_{\rm coll} \propto 1/\sqrt{L_\star}$, and therefore changes little for the
  factor $\sim$few changes in main-sequence $L_\star$.} Our model shows a break at
200\,Myr for the $\eta = 3$ (top) line at 70\,\um\ when stirring reaches the outer edge
of the disk. Given that PR drag is unlikely to be the cause, the further drop in the
\citeauthor{2008ApJS..179..451K} models after $\sim$200\,Myr may in part be for the same
reason.

In summary, we have shown that our simplified model is in good agreement with the
\citet{2008ApJS..179..451K} models, with the main remaining discrepancies arising for
older stars at 70\,\um. The effect of changing main-sequence luminosities is small, and
continued accretion by the largest bodies leads to smaller 70\,\um\ excesses after the
peak is reached. We retain our prescription for $D_{\rm bl}$, the main cause of the
initial differences because it is well established that grains should respond to
radiation forces \citep[e.g.][]{1979Icar...40....1B}. We allow $x_{\rm delay}$ to vary,
though it is usually unimportant because most emission comes from stirred regions.

\subsection{Evolution of  surface brightness}\label{sec:surf}

\begin{figure*}
  \begin{center}
    \vspace{-0.2in}
   \begin{tabular}{cc}
     \hspace{-0.35in} \psfig{figure=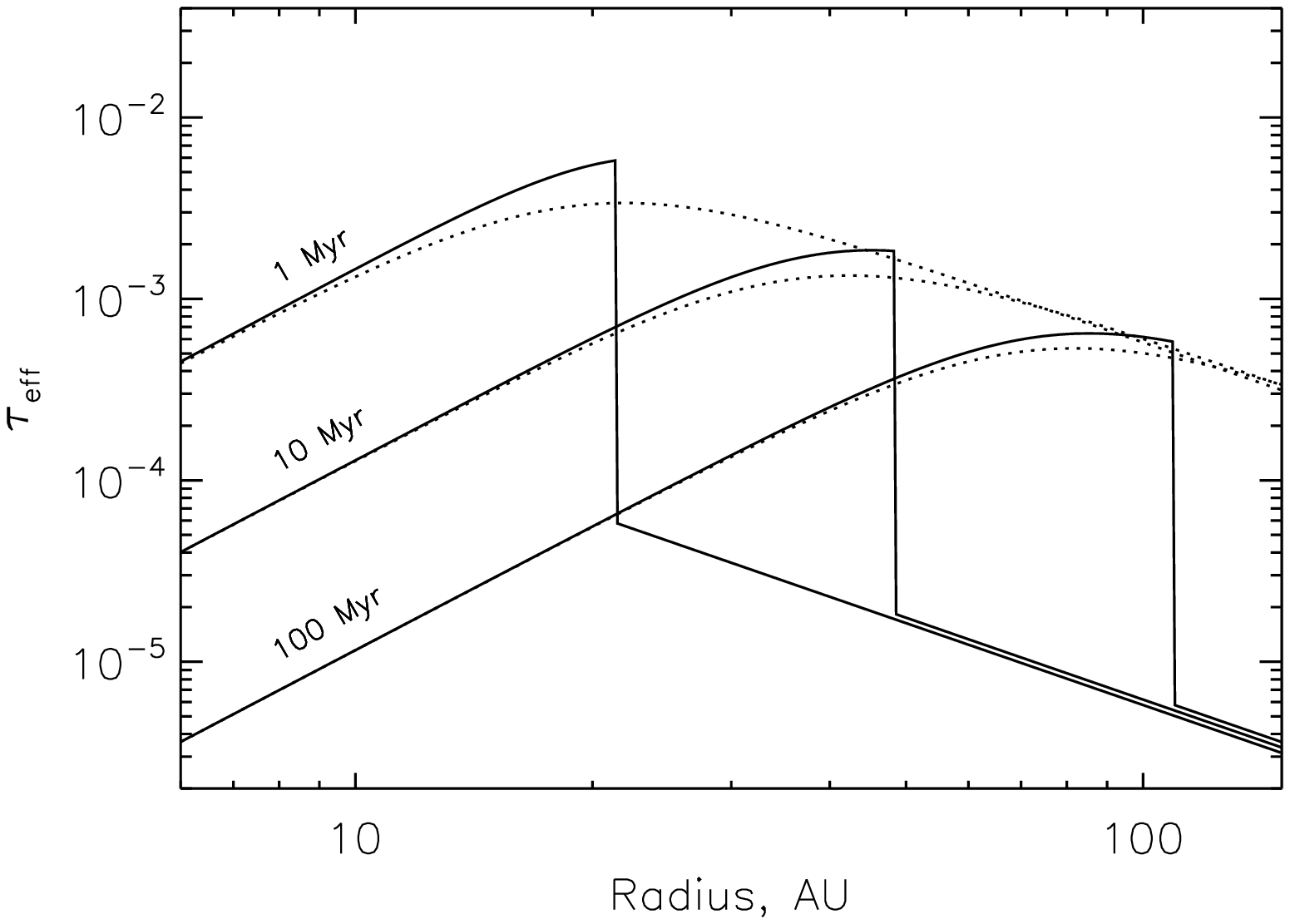,width=0.52\textwidth}&
     \hspace{-0.35in} \psfig{figure=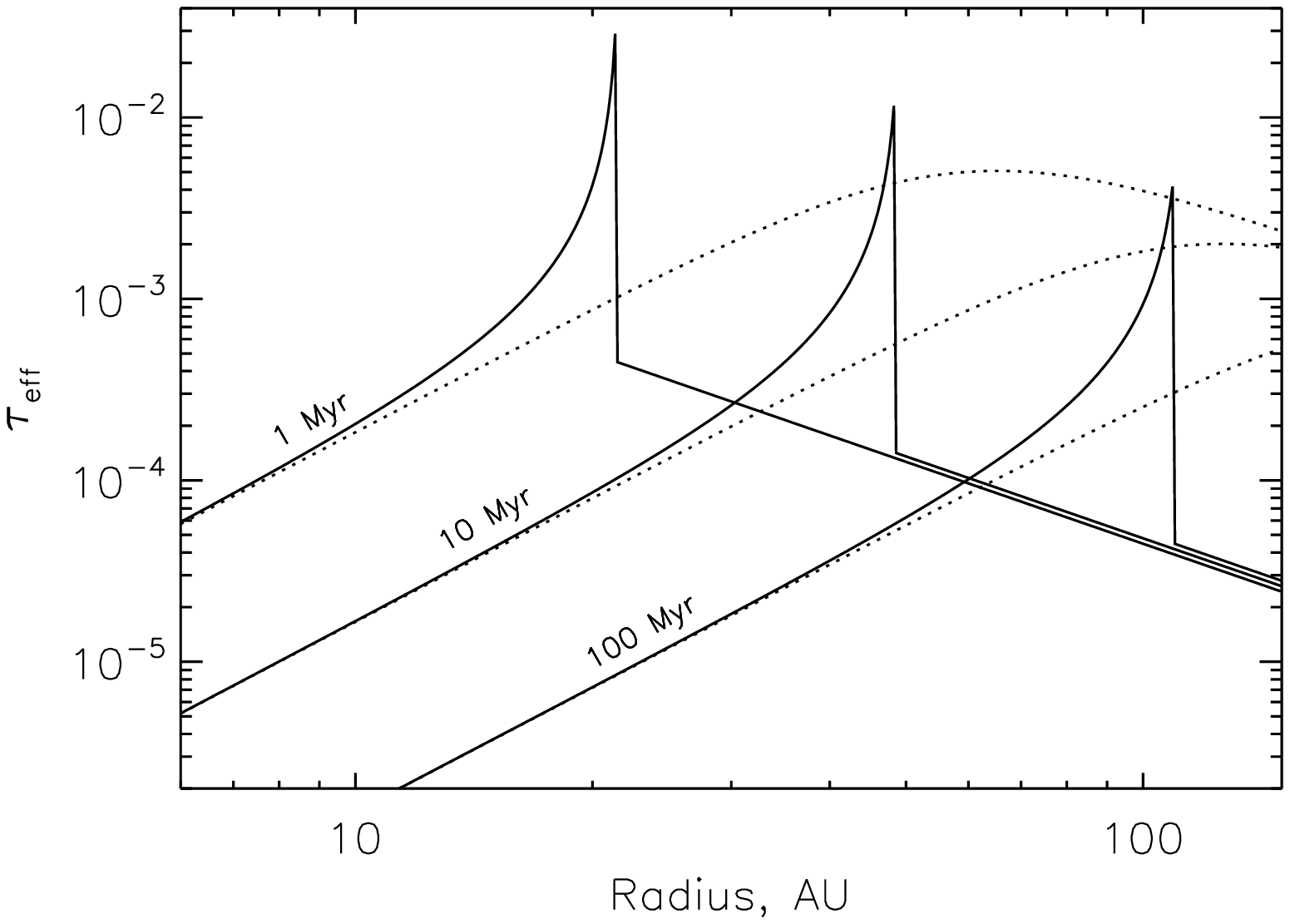,width=0.52\textwidth}\\
   \end{tabular}
   \caption{Evolution of $\tau_{\rm eff}$ for self-stirred (\emph{solid lines}) and
     pre-stirred (\emph{dotted lines}) disks at 1, 10, and 100\,Myr. In the left panel
     $t_{\rm c} \gg t_{\rm stir}$ so evolution is ``slow'' and similar to the pre-stirred
     case. In the right panel $t_{\rm c} \ll t_{\rm stir}$ so the evolution at $r_{\rm
       stir}$ is ``fast'' and much more violent as the disk reverts to the equilibrium
     state. Lines are offset slightly to remove ambiguities. The disks extend from
     5--150\,AU around an A2 star and have parameters: $\eta = 1$, $Q_{\rm D}^\star =
     150$\,J\,kg$^{-1}$, $e = 0.05$, $x_{\rm delay} = 0.01$, and $D_{\rm c} = 60$\,km
     (\emph{left panel}) and $D_{\rm c} = 1$\,km (\emph{right panel}). The peak surface
     density is larger for the fast self-stirred disk because more mass is concentrated
     in smaller size objects.}\label{fig:egs}
  \end{center}
\end{figure*} 

Debris disks around some of the closest stars have been resolved, providing spatial
information that cannot be derived from unresolved photometry. In this section we show
disk profiles derived from our model for comparison. With delayed stirring there are two
possible ways for the disk to evolve when stirring begins. These are set by the
collisional ($t_{\rm c}$) and stirring ($t_{\rm stir}$) times. These two modes of
evolution are shown in Figure \ref{fig:egs}, which shows the $\tau_{\rm eff}$ evolution
vs. radius in an extended disk at several times.

In Figure \ref{fig:egs}, we also compare the evolution with a pre-stirred disk---one that
is stirred when the star is born at $t = 0$ (dotted lines). In this case the surface
density profiles increase with radius to a broad peak. The slope of the inner region,
where mass at all radii is being depleted collisionally, is set by $\tau_{\rm eff,max}
\propto r^{7/3}$ (Eq. \ref{eq:tmax}). Outside the peak the primordial \taueff\ remains
because the collisional time is longer than the disk age there (i.e. $\tau_{\rm eff,max}
\propto r^{-3/2}$). The location of the peak moves outward with time ($r \propto
t^{6/23}$, Eq. \ref{eq:tc}) as more distant regions begin to decay.

The solid curves in Figure \ref{fig:egs} (left panel) show the self-stirred evolution
when the collisional time is longer than the stirring time. Compared to the pre-stirred
case, the difference is the \xdel\ drop outside \rstir\ where the disk has not been
stirred and accretion is ongoing. For this disk little decay happens at \rstir\
immediately, but begins later when $t > t_{\rm c}$. This evolution is therefore similar
to the pre-stirred case, and we term this mode of evolution ``slow'' for a self-stirred
disk. Compared to a pre-stirred disk, the factor \xdel\ leads to a narrower observed
annulus if the disk is too faint to be detected outside \rstir. For comparison, see
Figure 9a of \citet{2004AJ....127..513K}, which shows a qualitatively similar profile.

In the other limit, when the collisional time is shorter than the stirring time
(Fig. \ref{fig:egs}, right panel), for a given age the disk at \rstir\ has much more mass
than it would if it were pre-stirred (the same mass as at $t = 0$). The onset of stirring
is therefore violent, with the rapid shedding of mass just inside \rstir\ as the disk
reverts to its ``equilibrium'' state, that of a pre-stirred disk where the decay is
independent of initial mass. This evolution was hinted at by \citet{2003ApJ...598..626D},
where the excesses of disks with longer delay times decayed faster with time (their
Fig. 3). We call this evolution the ``fast'' mode. These disks appear transient because
the surface density can be significantly above the expected pre-stirred level for their
age (see \S \ref{sec:trans}).

However, we consider fast self-stirred disks with very sharp surface density profiles
unlikely, because disks with short collisional times will start to decay before
Pluto-size objects form. Planetesimal eccentricities increase as the largest objects
grow, and \tc\ decreases accordingly. Thus, a disk that would stir in the fast mode in
our model would in fact begin to decay earlier when $t > t_{\rm c}$ (unless planetesimal
eccentricities increase rapidly, in which case the time difference is small).\footnote{In
  \S \ref{sec:selfvplanet} we show that disks that evolve in the fast mode are possible
  if the stirring mechanism is secular perturbations from an eccentric planet.} Because
the largest objects continue to grow after the disk is stirred, eccentricities continue
to increase. After stirring, \tc\ continues to decrease until the disk begins to
decay. Thus, we also do not expect disks where the stirring time is significantly shorter
than the collision time.

By equating the stirring and collisional timescales (Eqs. \ref{eq:tstir} \& \ref{eq:tc}),
we can derive the conditions required for the boundary between these regimes of evolution
at \rstir\ (assuming $X_{\rm c} \ll 1$). Evolution will be ``slow'' at \rstir\ if
\begin{equation}\label{eq:rapidcond}
  \mathcal{R} \equiv r_{\rm stir}^{5/6} \, D_{\rm c} \, {Q_{\rm D}^\star}^{5/6} \,
  e^{-5/3} \, M_\star^{-5/6} \, \eta^{-1/2} > 2 \times 10^6 \, .
\end{equation}
At 50\,AU on the 10\,Myr curves, the disks illustrated in Figure \ref{fig:egs} have
$\mathcal{R} = 4 \times 10^6$ (left panel) and $6 \times 10^4$ (right panel), 2 times
greater, and 33 times smaller than Condition (\ref{eq:rapidcond}) respectively. Because
the collisional time increases more strongly with distance than the stirring time, inner
disk regions are more prone to fast evolution, as are disks with weak or small
planetesimals. There is some tendency towards fast evolution for more massive disks,
though if the stirring time were simply the Pluto formation time (Eq. \ref{eq:tpluto})
this condition would be independent of surface density. Because disks with $\mathcal{R}
\sim 10^6$ are the most physically plausible within our model, we check the distribution
of $\mathcal{R}$ when generating population models in \S \ref{sec:astar}.

The evolution of slow self-stirred disks makes clear predictions for resolved
observations of debris disks. While the brightest disks will yield broad surface density
profiles, fainter disks only have an annulus of detectable emission because the surface
density drops away interior and exterior to where the peak emission occurs. The observed
width of the disk depends on how fast the disk is decaying relative to its age; however,
we do not expect surface density profiles as sharp as shown in the right panel of Figure
\ref{fig:egs} for self-stirred disks. Of course, the disk extent is also set by where
planetesimals can form and where they are not affected by external effects such as
clearing, accretion, or ejection by planets.

For slow self-stirred disks the peak emission is set by the collisional time (i.e. the
peak surface density in the left panel of Fig. \ref{fig:egs} is slightly interior to
\rstir\ at 100\,Myr)
\begin{equation}\label{eq:rcoll}
  r \propto t^{6/23} \, .
\end{equation}
If the difference between the level of emission at the peak and at \rstir\ is small (as
in the left panel of Fig. \ref{fig:egs} at 1\,Myr), and the disk beyond \rstir\ is too
faint to detect, then the outer edge of slow self-stirred disks will still appear to
increase with time as $r \propto t^{1/3}$ (Eq. \ref{eq:tstir}).

\subsection{Transience}\label{sec:trans}

In the \citet{2007ApJ...658..569W} model, narrow pre-stirred planetesimal belts have a
maximum luminosity. This property arises because the decay time depends on the disk mass,
and all disks were assumed to be stirred at $t = 0$. With an assumption about their width
$dr$, the value \fmax\ can be calculated for unresolved belts with observations in
several bands, which gives an estimate of their radii. If the disk luminosity is
significantly higher than \fmax\ the excess is unlikely to arise from steady-state
evolution.

Ideally, comparisons would be made between resolved disks and the predicted maximum
surface density $\tau_{\rm eff,max}$. As discussed above, we do not expect self-stirred
disks to have \taueff\ significantly greater than $\tau_{\rm eff,max}$; those that do may
be transient. However, because disks may be secularly stirred in the fast mode (\S
\ref{sec:selfvplanet}), they can appear transient without being so in the sense meant by
\citet{2007ApJ...658..569W}. Another issue, whether disks are extended or narrow belts is
important because disks that are truly transient may not appear to be if they are assumed
to be wider than they actually are. Therefore, if disks are narrow belts then \fmax\ is a
reasonable indicator of transience.

\subsection{Radius evolution}\label{sec:rinf}

\begin{figure}
  \begin{center}
    \vspace{-0.2in}
    \hspace{-0.35in} \psfig{figure=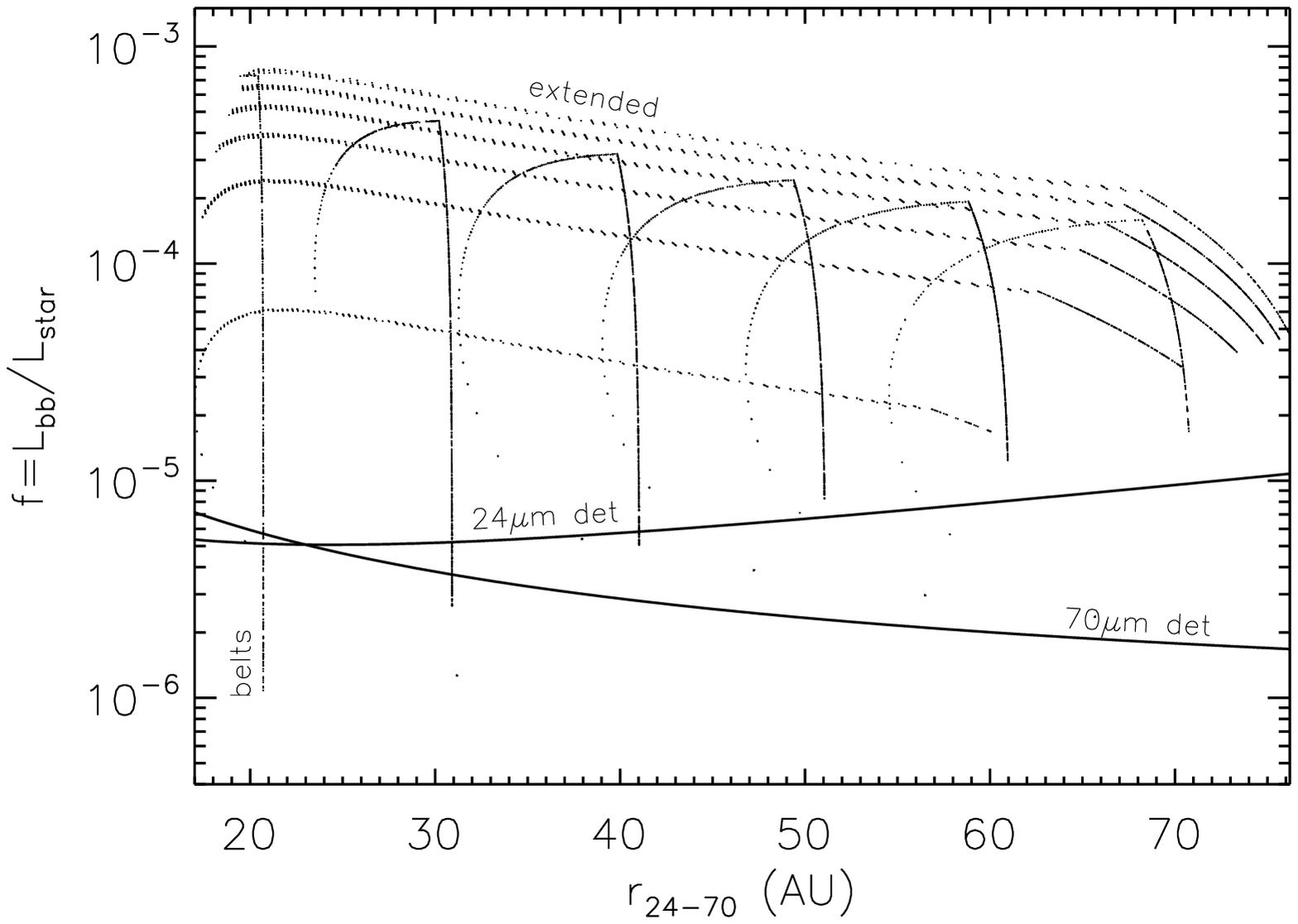,width=0.52\textwidth}
    \caption{Evolution of narrow belts and extended disks (as labelled) in $f$
      vs. $r_{24-70}$ space from 1\,Myr to 1\,Gyr. Disk \rinf\ increases as stirring
      moves outward (so evolution is left to right). Here, $f$ is derived from a
      blackbody at \rinf\ (i.e. what detection at 24 and 70\,\um\ would imply). Solid
      lines show 24 and 70\,\um\ blackbody detection limits from
      \citet{2007ApJ...663..365W}. Extended disks have $r_{\rm in-out} = 15$--120\,AU and
      $\eta = 0.05-1.05$. For belts $r_{\rm mid} = 20$, 30, 40, 50, 60, and 70\,AU, $dr =
      r_{\rm mid}/2$, and $\eta = 1$. Belts increase in radius from left to right, and
      extended disks increase in $f$ with higher $\eta$. Other model parameters are as in
      the left panel of Figure \ref{fig:egs}.}\label{fig:fvsr}
  \end{center}
\end{figure}

There are considerable differences between the evolution of narrow belts and extended
disks. Figure \ref{fig:fvsr} shows the $f$ vs. \rinf\ evolution for disks extending from
15--120\,AU for a range of initial surface densities, and a series of narrow belts with
the same surface density centred at different $r_{\rm mid}$ between 20 and 70\,AU (\rinf\
is the radius inferred from the $24 - 70$\,\um\ colour assuming blackbody grains at a
single temperature, Eq. \ref{eq:tbb}).

For extended disks \rinf\ increases as \rstir\ moves outward. The disk \rinf\ is located
between the inner edge and where the surface density peaks, so moves outward with time.
It is not simply at the peak surface density because the inner disk contribution to the
SED is non-negligible when $\tau_{\rm eff} \propto r^{7/3}$. Thus, for extended disks the
increase in \rinf\ with time is in fact slower than expected by Equation
(\ref{eq:rcoll}). The fractional excess decreases somewhat because the decay of emission
interior to \rstir\ is stronger than the increased emission from newly stirred regions.

In contrast, belts show a small change in radius and then decay at near constant radius
once stirred to the outer disk edge. The narrower disk extent means that \rinf\ is nearer
the peak excess. When stirring reaches the outer edge, all disks break to a faster
decline in $f$ because no new regions can be stirred. Because the peak surface density is
still evolving outward (i.e. these disks are evolving in the slow mode), \rinf\ continues
to increase. When $t \gg t_{\rm c}$ at the outer edge, $f$ decreases at constant
$r_{24-70}$ (because the disk has $\tau_{\rm eff} \propto r^{7/3}$ everywhere and \rinf\
cannot change). Because the collision time scales more strongly with radius than stirring
($t_{\rm c} \propto r^{23/6}$ vs. $t_{\rm stir} \propto r^{3}$), most disks do not reach
evolution at constant \rinf\ until long after the outer disk edge is stirred (and long
after the maximum 1\,Gyr age shown in Figure \ref{fig:fvsr}). That is, slow evolution is
more likely to occur at large radii (Eq. \ref{eq:rapidcond}).

For comparison, pre-stirred disks show similar trends to Figure \ref{fig:fvsr}, but $f$
can only decrease because pre-stirred disks start evolving at $t = 0$ everywhere. Thus,
pre-stirred disks do not show the initial increase in $f$ seen for the self-stirred
evolution. Also, there is no break to a faster decline in $f$, as occurs when
self-stirring reaches the outer disk edge.

\section{Application to A-star statistics}\label{sec:astar}

In this section we apply our model to evolution of 24 and 70\,$\mu$m excesses around
A-stars. We first outline the observations and the limitations of the pre-stirred model.
Our aim is to test whether self-stirring can successfully reproduce the rise in
24\,$\mu$m emission at $\sim$10--30\,Myr (and the 24 and 70\,\um\ statistics at other
times), and see what constraints the statistics set on model parameters.

\subsection{The A-star sample}\label{sec:obs}

The observations are compiled from several sources. \citet{2005ApJ...620.1010R} and
\citet{2006ApJ...653..675S} observed large unbiased samples of A-stars at 24 and
70\,$\mu$m with \emph{Spitzer}. While these samples provided the basis for the
\citet{2007ApJ...663..365W} study of pre-stirred evolution, they comprise stars mostly
older than 10\,Myr.

\begin{figure*}
  \begin{center}
    \vspace{-0.2in}
    \begin{tabular}{cc}
      \hspace{-0.35in} \psfig{figure=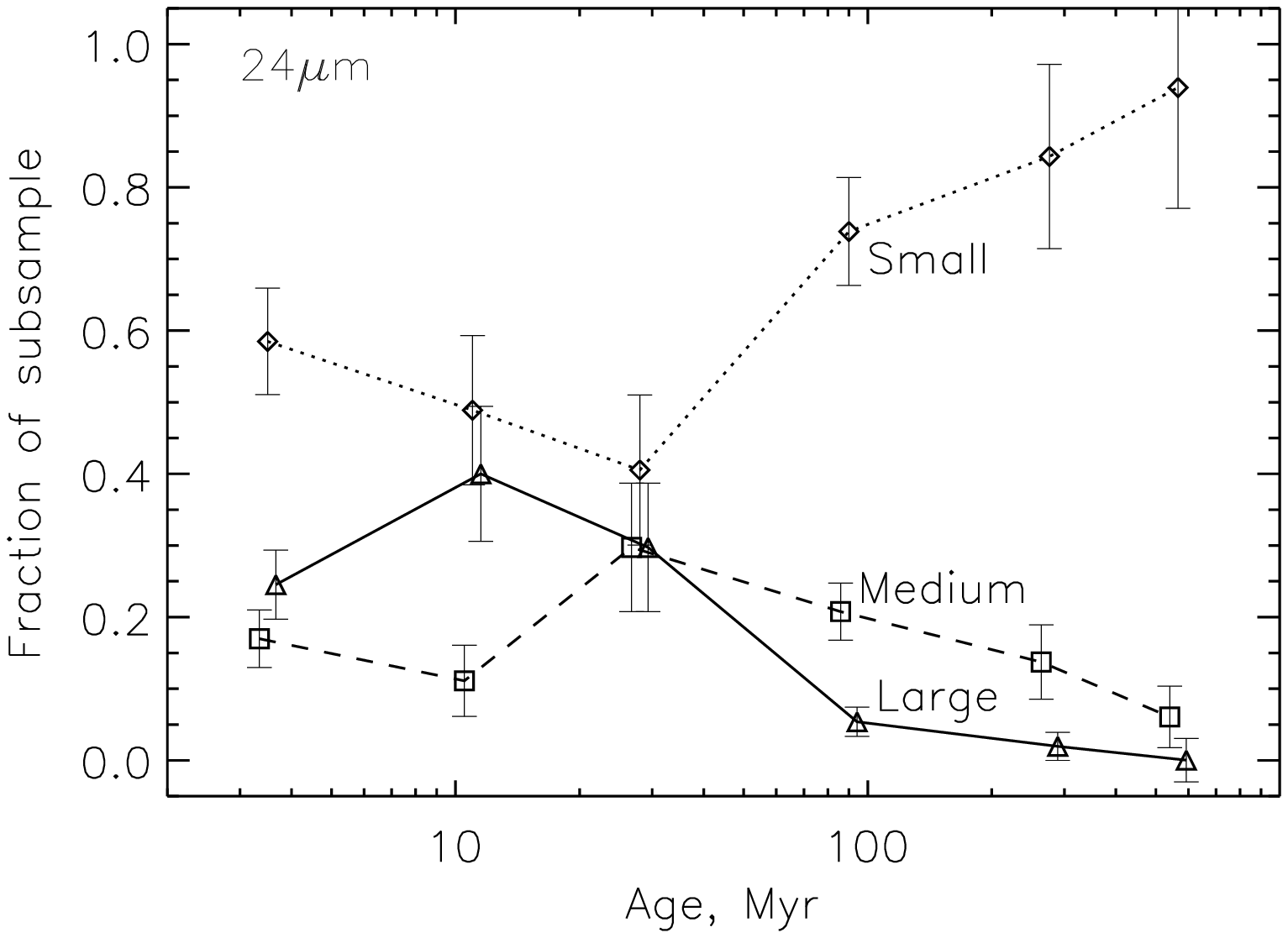,width=0.52\textwidth} &
      \hspace{-0.35in} \psfig{figure=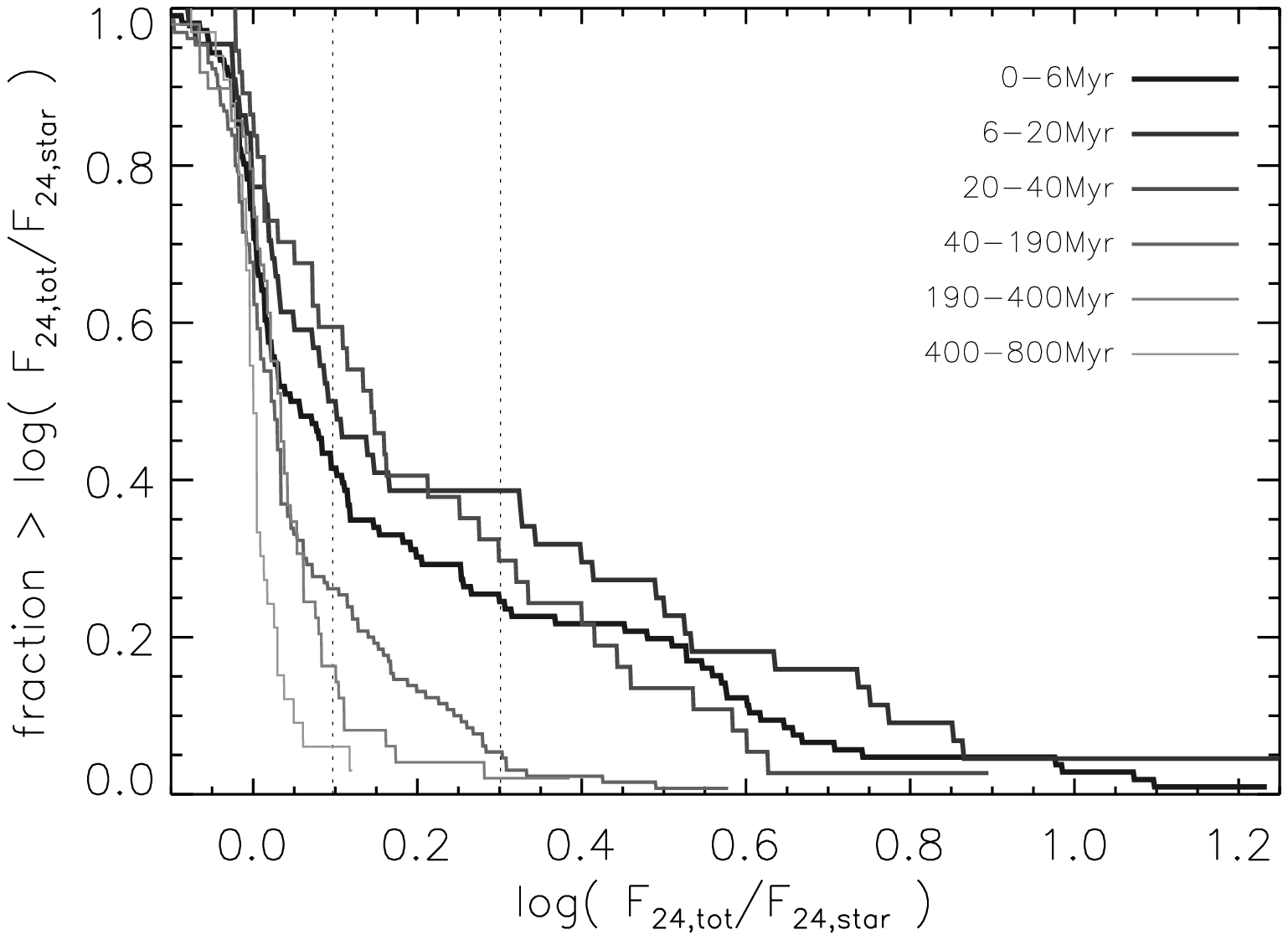,width=0.52\textwidth} \\
    \end{tabular}
    \caption{Evolution of A-star 24\,\um\ excesses. The left panel shows stars binned by
      age and large (triangles), medium (squares) and small (diamonds) excesses (ages
      offset slightly for clarity). Error bars are $\sqrt{N}$ in each bin and for the
      oldest large excess bin we assume 1 disk to calculate the error. Excesses reach an
      overall peak around 30\,Myr. The right panel shows cumulative distributions of
      24\,\um\ excesses for the same age bins, from dark to light grey lines. The dotted
      lines indicate the bin edges for small, medium, and large excesses. In the second
      youngest bin HR4796A and $\beta$ Pic are beyond the right edge of the
      plot.}\label{fig:data24}
  \end{center}
\end{figure*}

\begin{figure*}
  \begin{center}
    \vspace{-0.2in}
    \begin{tabular}{cc}
      \hspace{-0.35in} \psfig{figure=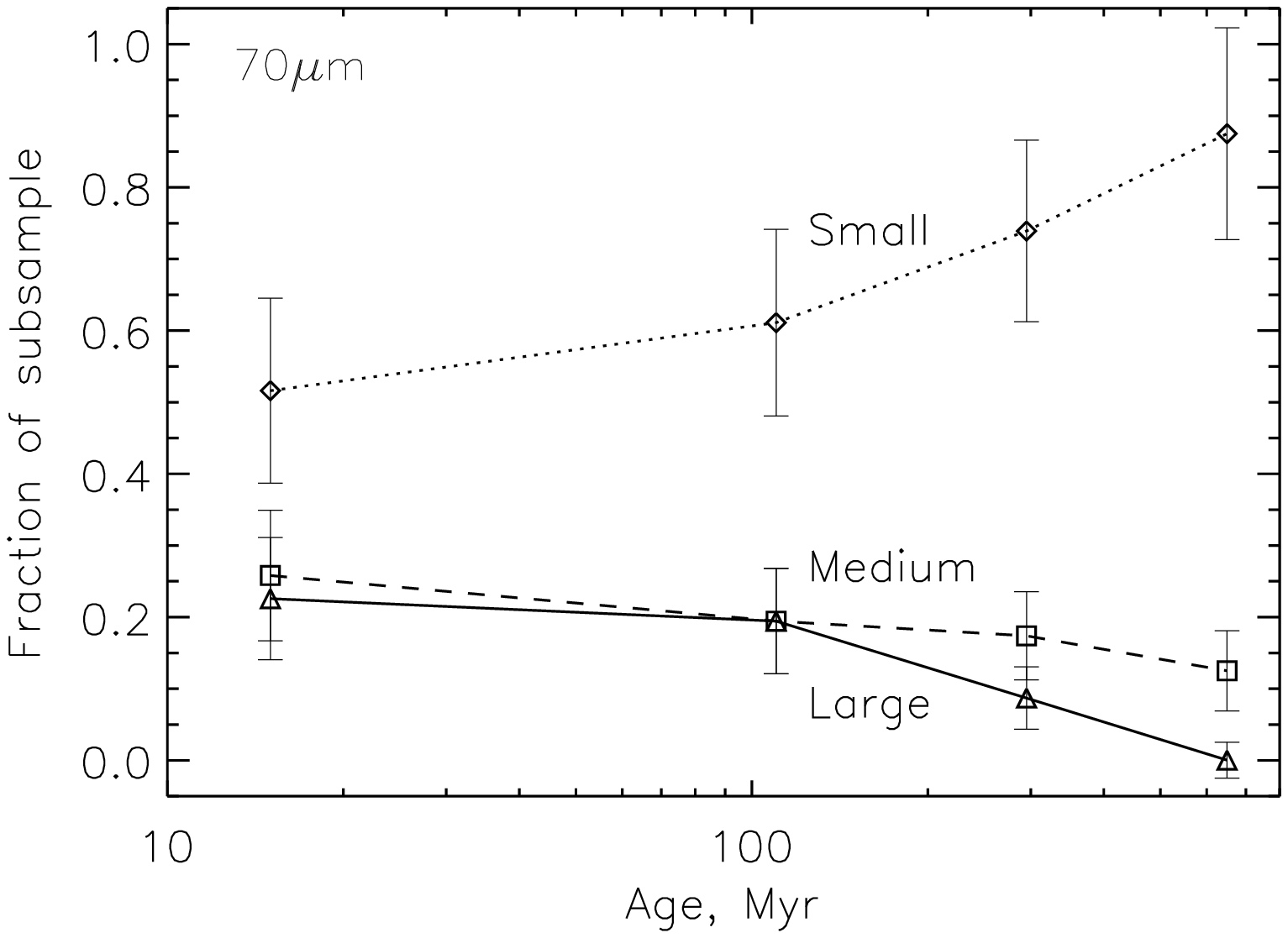,width=0.52\textwidth} &
      \hspace{-0.35in} \psfig{figure=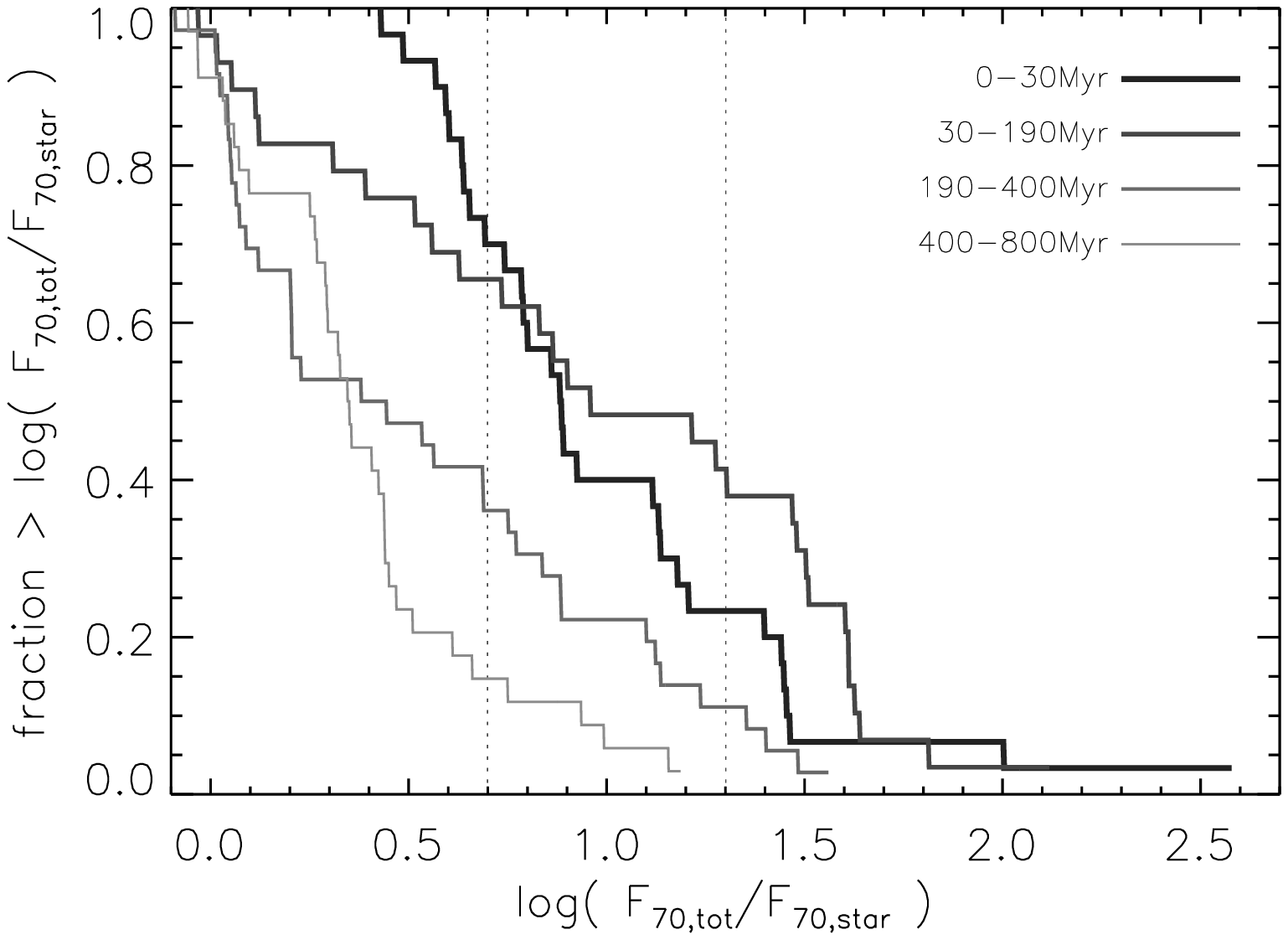,width=0.52\textwidth} \\
    \end{tabular}
    \caption{Same as Figure \ref{fig:data24}, but for 70\,\um\
      excesses.}\label{fig:data70}
  \end{center}
\end{figure*}

To supplement these data, we collect 24\,$\mu$m data for B8--A9 stars in the following
young clusters/associations: $\sigma$ Ori \citep{2007ApJ...662.1067H}, OB1a/b
\citep{2006ApJ...652..472H}, $\lambda$ Ori \citep{2009ApJ...707..705H}, $\gamma$ Velorum
\citep{2008ApJ...686.1195H}, Upper Sco \citep{2009ApJ...705.1646C}, $\beta$ Pic Moving
Group \citep[BPMG][]{2008ApJ...681.1484R},\footnote{$\beta$ Pic is removed from the
  \citet{2005ApJ...620.1010R} sample. We set the excesses of the Be stars HD 21362 (Rieke
  sample), HD 67985 ($\gamma$ Vel), and HIP 78207 (Upper Sco) to $F_{\rm 24,tot}/F_{\rm
    24,\star} = 1$.} NGC 2232 \citep{2008ApJ...688..597C}, and IC 2391
\citep{2007ApJ...654..580S}. We use the K$_{\rm s} -$[24] excess to derive $F_{\rm
  24,disk}/F_{24,\star}$ for these objects and use spectral types from
\citet{2001KFNT...17..409K} where needed. These data make a sample of about 400 A-stars
observed at 24\,$\mu$m with ages between $\sim$3--800\,Myr. All stellar photospheres in
these samples are detectable and there are no upper limits (i.e. disk detections are
calibration limited). We do not include stars in more distant regions, where observations
are sensitivity limited and excess fractions are lower limits.

Because our 24\,\um\ sample includes stars younger than 10\,Myr old, it is possible that
some are protoplanetary disks. However, consistently shorter disk lifetimes around
intermediate mass stars mean that few A-star protoplanetary disks survive beyond a few
Myr \citep{2009ApJ...695.1210K}. Indeed we exclude only three stars; V346 Ori in OB1a, HD
290543 in OB1b, and HD 245185 in $\lambda$ Ori, which have K$_{\rm s} - [24] \approx
6.5$--7.5 and strong IRAC excesses \citep{2006ApJ...652..472H}. Less certain is whether
the youngest stars in our sample contain dust left over from the protoplanetary disk
phase, rather than from that created by fragmentation from self-stirring
\citep[e.g.][]{2008ARA&A..46..339W}.

Figure \ref{fig:data24} (left panel) shows the 24\,$\mu$m excesses as a function of time
binned into ``large'' ($F_{\rm 24,tot}/F_{\rm 24,\star} > 2$), ``medium'' ($1.25 < F_{\rm
  24,tot}/F_{\rm 24,\star} < 2$), and ``small'' ($F_{\rm 24,tot}/F_{\rm 24,\star} <
1.25$) excesses. The age bins are 0--6\,Myr, 6--20\,Myr, 20--40\,Myr, 40--190\,Myr,
190--400\,Myr, and 400--800\,Myr. There are 106, 45, 37, 130, 51, and 33 stars in these
bins respectively. The primary result at 24\,\um\ is a decline in excesses on a
$\sim$150\,Myr timescale \citep{2005ApJ...620.1010R}.\footnote{This decay is more obvious
  when viewed on a linear-time plot
  \citep[e.g.][]{2005ApJ...620.1010R,2006ApJ...653..675S,2007ApJ...663..365W} and is only
  seen in the oldest three bins in Figure \ref{fig:data24}.} The large and medium excess
bins add to give a peak in debris disk fraction at 30\,Myr
\citep[e.g.][]{2008ApJ...688..597C}. By separating the excesses into three bins, there is
perhaps evidence for an additional trend: the large excesses peak around 10\,Myr, and the
medium excesses peak around 30\,Myr. This behaviour suggests a population of large excess
disks that form around 10\,Myr, and then decay relatively rapidly to produce the medium
excess population a few tens of Myr later.

Though these features are tantalising evidence of systematic trends that may be caused by
self-stirring, their existence is motivated more by expectations than by statistical
significance. In addition, \citet{2009ApJ...705.1646C} note that because A-stars take
$\sim$10\,Myr to reach the main-sequence, stars of fixed spectral type decrease in
luminosity from 1--10\,Myr. The decreasing luminosity decreases $D_{\rm bl}$, and can
cause at least some of the rise in 24\,\um\ excesses that has been attributed to
self-stirring. Using the cumulative distributions shown in Figure \ref{fig:data24} (right
panel) we estimate the significance of changes between different age bins with the KS
test. According to this test, each pair of cumulative distributions for the three
youngest bins (three upper curves) could have all come from the same distribution (max KS
probability of difference is 82\%). There is at least a 99\% chance that all other
combinations of distributions are not drawn from the same distribution, with the
exception of the fourth age bin when compared to the fifth age bin (78\%). That is, there
is a robust decay in excesses from early to late times. For the individual clusters
collected above, aside from the 50\,Myr IC 2391, there is no pair different at more than
the 98\% level according to the KS test, though NGC 2232 is different from $\sigma$ Ori,
Orion OB1b, and $\gamma$ Vel at 93--98\%, and Upper Sco is different from the BPMG at
92\%. Therefore, while there are interesting trends at early times, the only formally
secure result is that 24\,\um\ excesses decrease after $\sim$50\,Myr.

While there are far fewer observations of A-stars at 70\,$\mu$m, we use the
\citet{2006ApJ...653..675S} stars, comprising 153 stars (including 19 observed with IRAS,
and excluding Be stars HD 21362 \& HD 58715 and Herbig Ae/Be star HD 58647). These data
include many upper limits, which are largely for stars farther than 100\,pc. For the
model comparison undertaken here, we use the sample as given by Table 4 of
\citet{2006ApJ...653..675S} (shown in Fig. 9 of that paper, or Fig. 2 of
\citet{2007ApJ...663..365W}). As in \citet{2006ApJ...653..675S} and
\citet{2007ApJ...663..365W}, the data are binned into ``large'' ($F_{\rm 70,tot}/F_{\rm
  70,\star} > 20$), ``medium'' ($5 < F_{\rm 70,tot}/F_{\rm 70,\star} < 20$), and
``small'' ($F_{\rm 70,tot}/F_{\rm 70,\star} < 5$) excesses. The data are also binned by
age: 0--30\,Myr, 30--190\,Myr, 190--400\,Myr, and 400--800\,Myr.

The inclusion of upper limits in the 70\,\um\ sample means that the excess fractions are
upper limits. As a check, we compare this sample to an unbiased subset of stars closer
than 100\,pc (and exclude IRAS sources, all but 2/19 of which are upper limits). In this
subset only two stars (HD 27962 \& HD 142703) with upper limits fall above the small excess
bin (with $F_{\rm 70,tot}/F_{\rm 70,\star} < 5.61$ \& $<$ 7.63 respectively). The only
significant difference between these two samples lies in the youngest age bin, where the
$<$100\,pc subset has only one star (of 11) with a small excess, compared to 16 (of 31)
for the full sample. This difference arises because the main sample includes many
Upper-Sco sources ($\sim$150\,pc away) that have upper limits in the small excess
bin. However, the paucity of young star forming regions within 100\,pc also results in
poor statistics in this bin (with only 11 stars). An alternative approach, setting all
stars with upper limits to have small excesses, gives a $\sim$10\% increase in small
excesses for the youngest two age bins, and little difference for the older two. For both
approaches, there is little change in the excess fractions in all but the youngest age
bin (relative to Poisson errors). Based on these comparisons, we conclude that
differences in the youngest age bin simply reflect poor statistics due to a lack of very
close young stars. The small differences between (sub)samples in the older age bins mean
that sample choice does not affect our results. Thus, we use the entire
\citet{2006ApJ...653..675S} 70\,\um\ sample for comparison with our model.

Figure \ref{fig:data70} shows the evolution of excesses at 70\,\um. As with the older age
bins at 24\,\um, these decay monotonically with time, but with a slower timescale of
$\gtrsim$400\,Myr \citep{2006ApJ...653..675S}. A KS test shows that the differences
between adjacent age bins are not particularly strong ($\sim$90-98\% chance of not being
from the same distribution), but it is very unlikely that non-adjacent bins could be
taken from the same distribution (with probabilities $>$99.99\%). Therefore, the decay
from a higher to lower fraction of 70\,\um\ excesses over time is robust.

Finally, we use the \citet{2007ApJ...663..365W} 46 star subsample of stars for which
excess emission has been detected at both 24 and 70\,\um\ by \emph{Spitzer} (or 25 and
60\,\um\ by \emph{IRAS}). With the assumption of blackbody grains at a single
temperature, this sample provides a set of disk radii (\rinf) for comparison with our
model.

\subsection{Method}\label{sec:method}

We now explore various population models and their ability to reproduce the A-star
statistics. For these models, we select 30,000 stars randomly in the spectral type range
B8--A9 with uniform $\log ({\rm age})$, assuming all stars are born with planetesimal disks. As
in \citet{2007ApJ...663..365W} we fix some parameters and allow others to vary. We fix
$\rho = 2700$\,g\,cm$^{-3}$, $q = 11/6$, $I = e$ and do not consider the effects of
varying them. These assumptions leave \qd, \dc, $e$, $\eta$, and the disk radii as
parameters.

In addition to comparison of fractional excesses in different age bins, our modelling
process also involves comparison of model and observed disk radius distributions using
\rinf.

The low significance of the 24\,\um\ trends means that we measure the ability of the
self-stirring model to fit A-star statistics at early times partly with a formal $\chi^2$
measure for the 24 and 70\,\um\ statistics and the \rinf\ distribution, but also the
success of the model at reproducing trends in the data. In some cases we can rule out
models based purely on a poor fit to the statistics for the older A-stars, in which case
our conclusions are independent of the low significance of trends for the younger stars.

\subsection{Pre-stirred disks}\label{sec:pre}

\begin{figure*}
  \begin{center}
    \vspace{-0.2in}
    \hspace{-0.35in} \psfig{figure=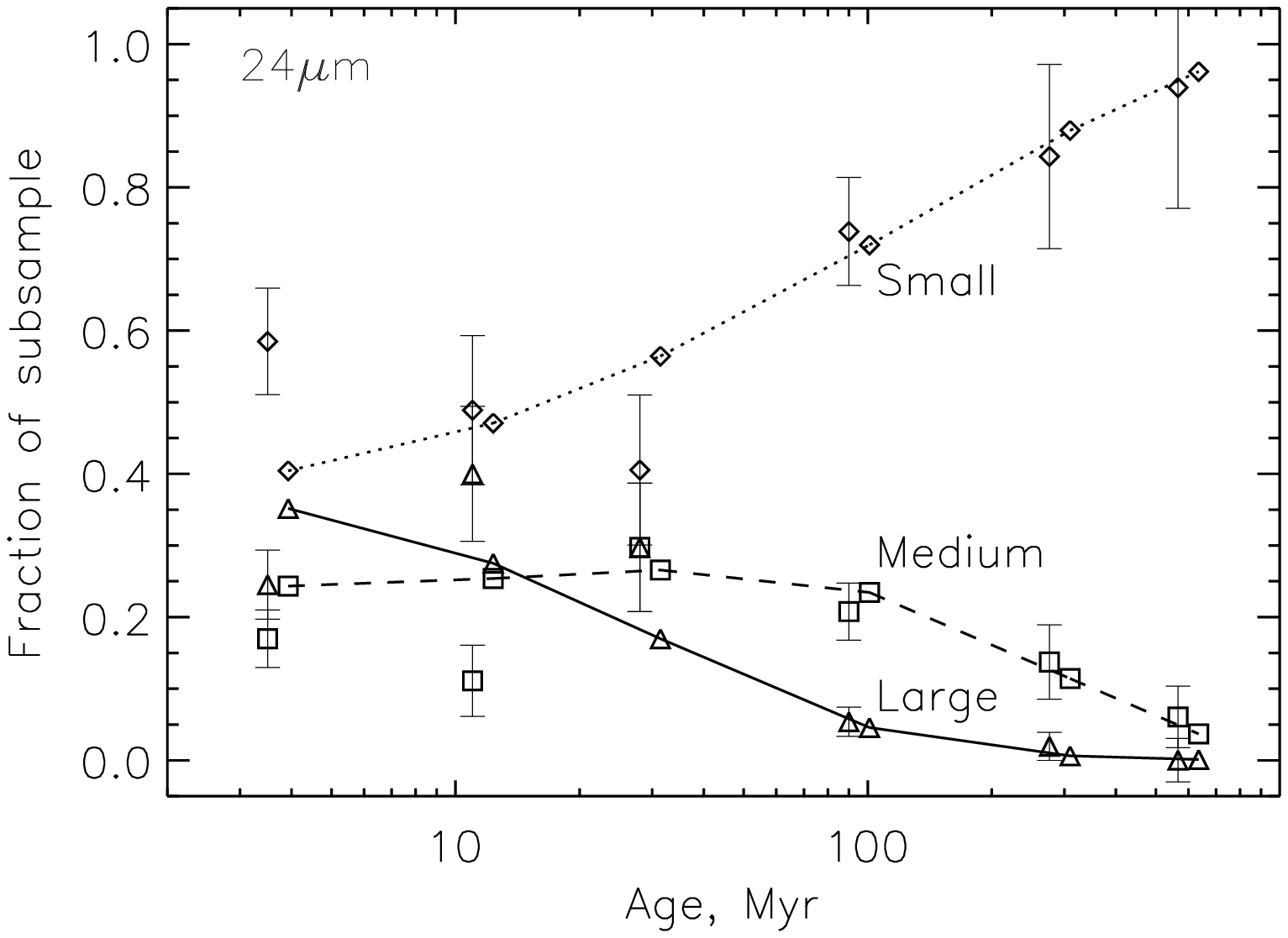,width=0.52\textwidth}
    \hspace{-0.35in} \psfig{figure=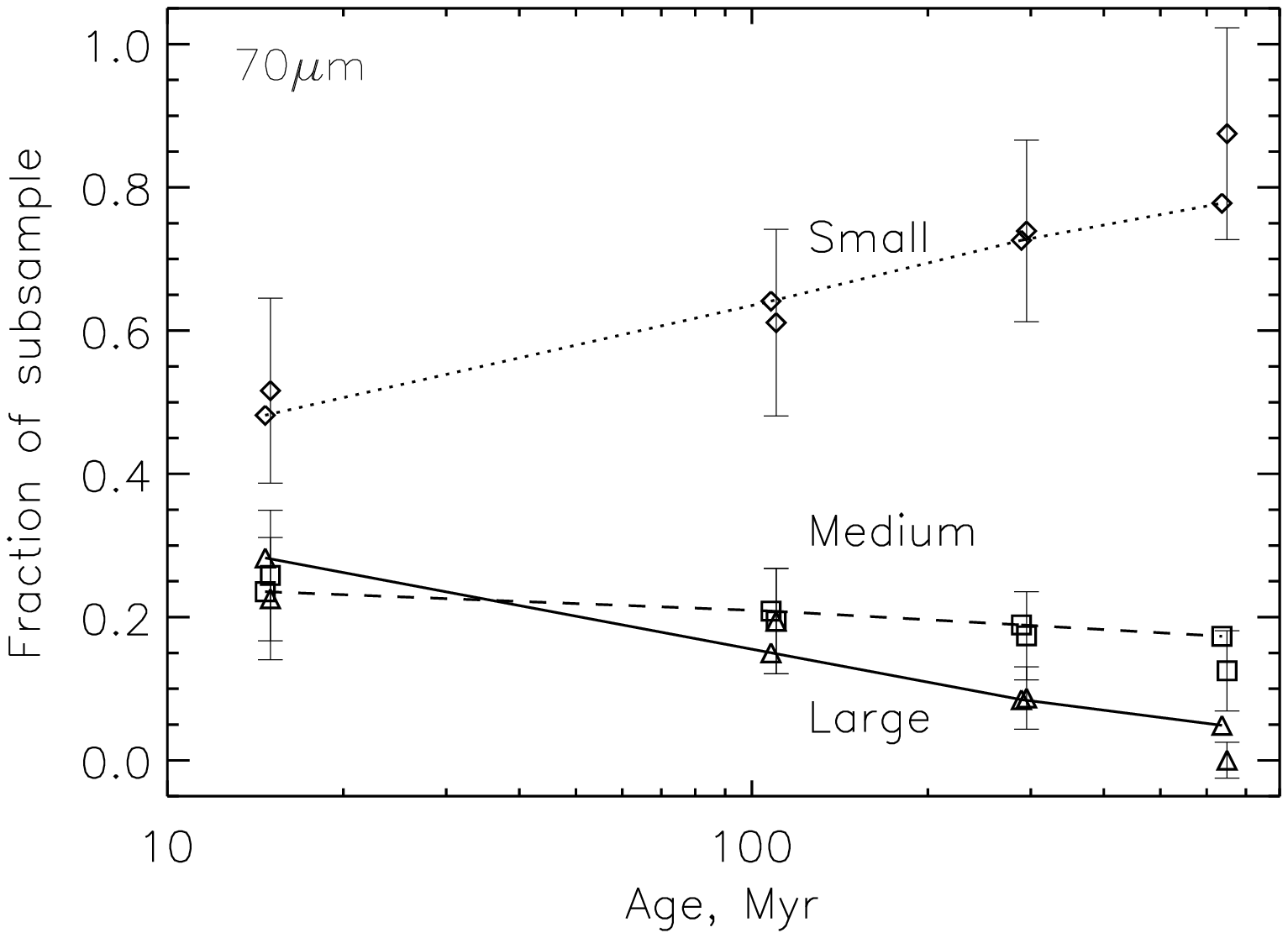,width=0.52\textwidth}
    \caption{Best fit parameters for the \citet{2007ApJ...663..365W} pre-stirred model at
      24 and 70\,\um\ (same symbols as Figure \ref{fig:data24}, joined by solid, dashed,
      and dotted lines for large, medium, and small excesses respectively). Symbols with
      error bars show observed A-star statistics. At 24\,\um\ the model predicts too many
      stars with large excesses at the earliest times, and a monotonic decline in the
      overall fraction of stars with excesses. There are no new 70\,\um\ data so the
      right panel is the same as \citet{2007ApJ...663..365W} Figure 2.}\label{fig:pre}
  \end{center}
\end{figure*}

To illustrate the shortcomings of the pre-stirred model for young stars, the best fit
parameters from \citet{2007ApJ...663..365W} are compared to the A-star observations in
Figure \ref{fig:pre}. This Figure uses the model of \S \ref{sec:model} with $t_{\rm stir}
= 0$ so that disks are pre-stirred. The parameters are: $Q_{\rm D}^\star = 150$\,J
kg$^{-1}$, $D_{\rm c} = 60$\,km, $e = 0.05$, and a planetesimal belt radius distribution
$N(r) \propto r^\gamma$ with $\gamma = -0.8$ and disk width $dr = r_{\rm mid}/2$. We find
that $\eta_{\rm mid} = 0.2$ gives the best results (\citeauthor{2007ApJ...663..365W}
specified the disk mass distribution with $M_{\rm mid} = 10\,M_\oplus$). The Figure shows
that the model still reproduces the 24 and 70\,\um\ A-star statistics for stars 100\,Myr
and older, but fails to reproduce the trend in large and overall excesses for younger
stars at 24\,\um. The model shows only a monotonic decline in the overall fraction of
stars with excesses in contrast to the observations.

As a metric for comparing models, we quote the $\chi^2$ value at 24 and 70\,\um, for the
distribution of radii \rinf, and the sum of all three. These are computed as the sum of
squared differences between the data and model divided by the Poisson error for all 18
data points at 24\,\um\ (six age bins and three excess bins), 12 points at 70\,\um\ (four
age and three excess bins), and 5 points for the \rinf\ distribution. In the pre-stirred
case shown in Figure \ref{fig:pre}, the values are $\chi^2_{24} = 32.4$, $\chi^2_{70} =
5.2$, and $\chi^2_{\rm \chir} = 17.0$, for a total of $\chi^2_{\rm tot} = 54.6$. Though
these numbers give no indication of why a particular model succeeds or fails, or whether
it reproduces the desired trends, they formally show how well the statistics are
reproduced and provide a benchmark for measuring the success of different self-stirring
models.

\subsection{Delayed stirring: the simplest case}\label{sec:delayed}

\begin{figure*}
  \begin{center}
    \vspace{-0.2in}
    \hspace{-0.35in} \psfig{figure=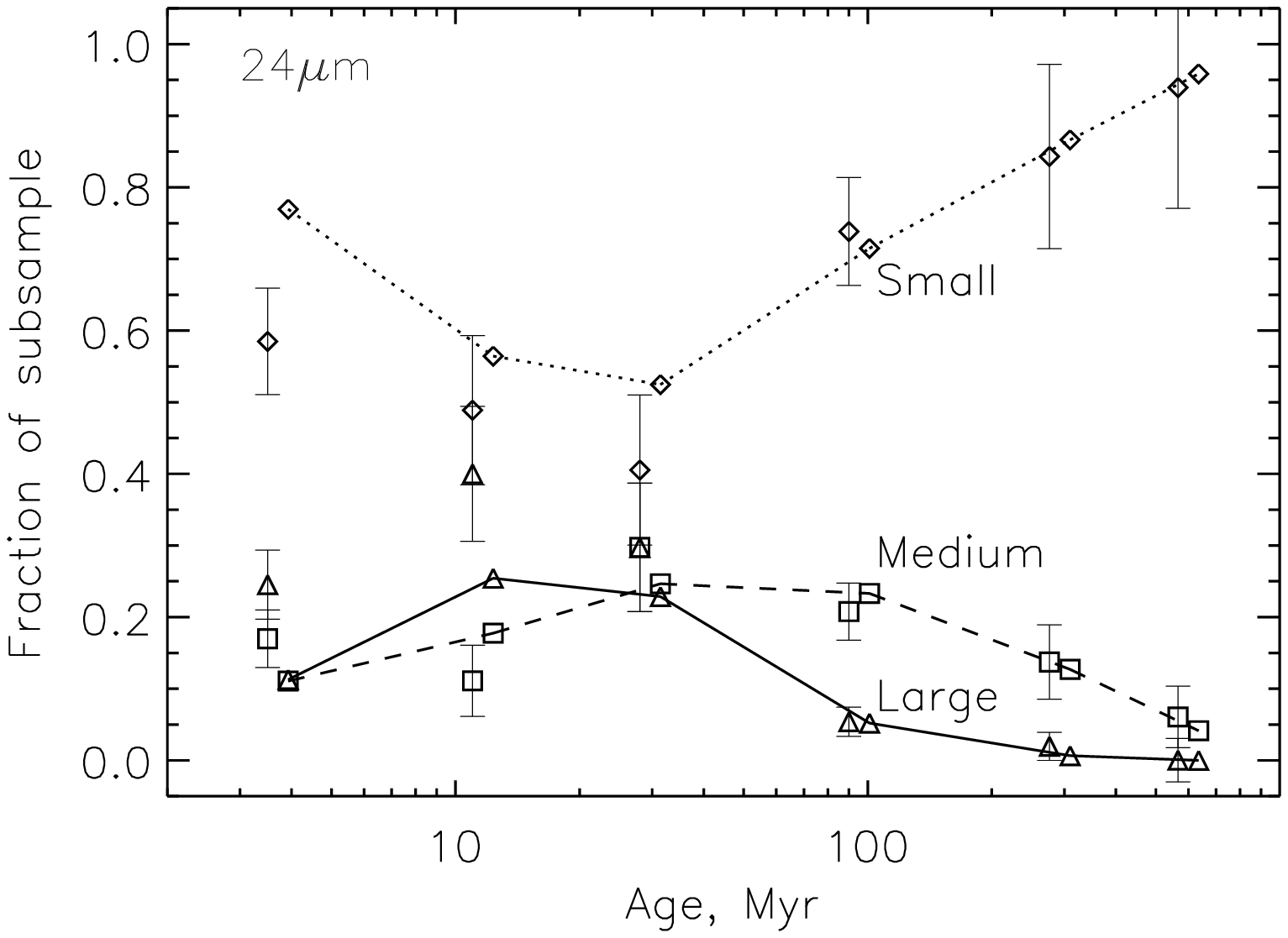,width=0.52\textwidth}
    \hspace{-0.35in} \psfig{figure=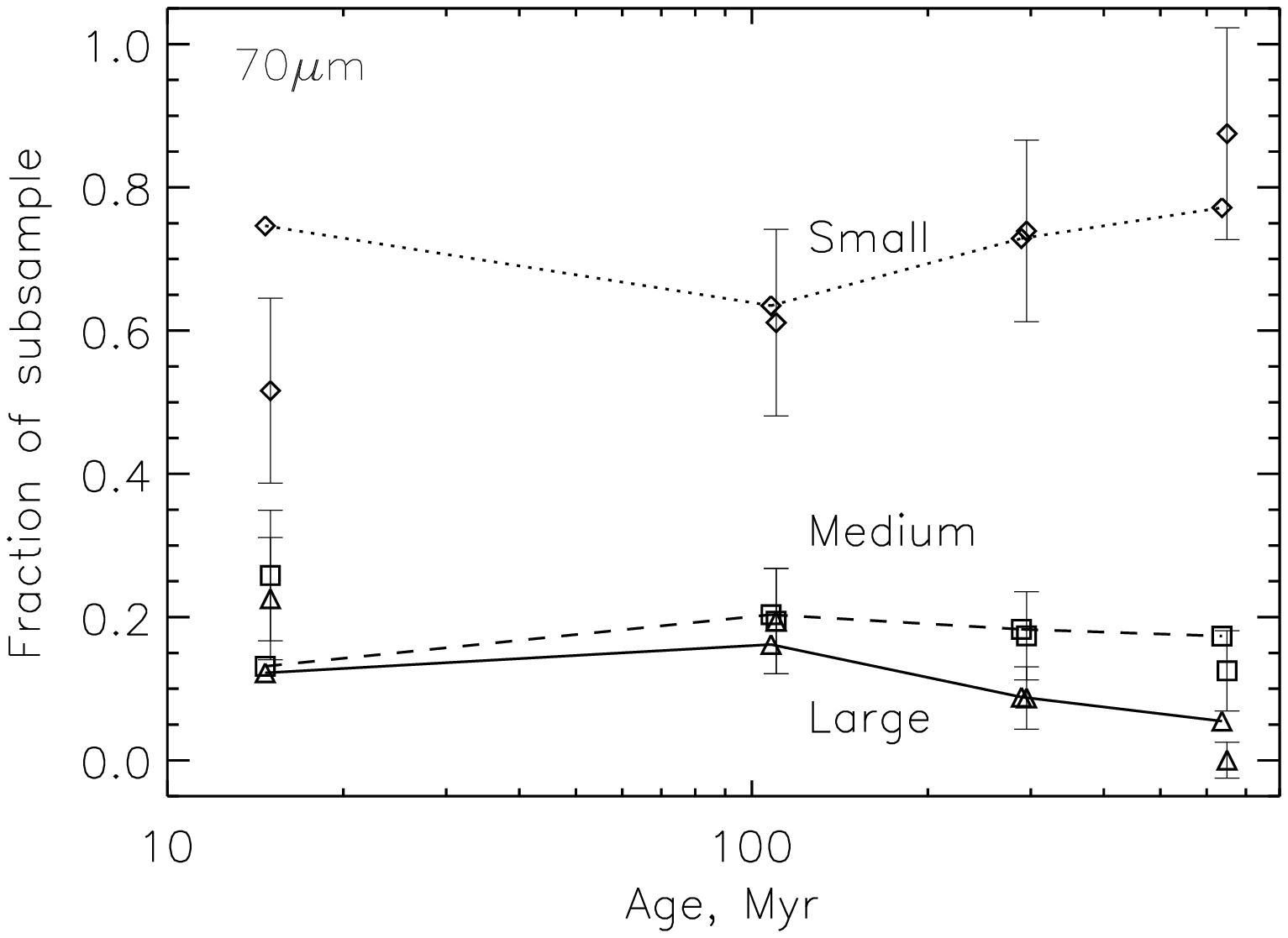,width=0.52\textwidth}
    \caption{Simple delayed stirring model compared with 24 and 70\,$\mu$m
      statistics. Symbols and lines are as in Figure \ref{fig:pre}. All disks begin their
      evolution after $\sim$10\,Myr.}\label{fig:pop_del}
  \end{center}
\end{figure*}

By taking the peak in large excesses at 10\,Myr as an indication of the delay time, we
can add the simplest possible model of delayed stirring to our model. Figure
\ref{fig:pop_del} shows a model with almost the same parameters as the pre-stirred model
with the addition of a 10\,Myr delay (1$\sigma$ width 0.1\,dex) before the onset of
stirring for all disks (here $\eta_{\rm mid} = 0.25$). After this delay all disks begin
evolving, regardless of radius. This simple modification to the model reproduces the
24\,$\mu$m excesses better than the pre-stirred model with some change at 70\,\um,
yielding $\chi^2_{24} = 19.9$, $\chi^2_{70} = 12.5$, and $\chi^2_{\rm \chir} = 21.9$ for
a total of $\chi^2_{\rm tot} = 54.3$. The three 24\,\um\ excess bins now show the right
general trends: the large excesses peak at 10\,Myr, the medium excesses show a broad peak
over 10--100\,Myr, and the small excesses reach a minimum around 10--30\,Myr. The
fraction of small 70\,\um\ excesses at the earliest times increases due to more unstirred
disks. An obvious physical motivation for this prescription is unclear, though there are
some possibilities that may merit further study.

One possibility might seem to be objects emerging from a protoplanetary disk phase
somewhat longer than the ``typical'' 6\,Myr lifetime \citep[e.g.][]{2001ApJ...553L.153H}.
However, this explanation is unlikely for two reasons. Primarily, only a small fraction
of stars have long lived disks, not the $\sim$40\% suggested by Figure \ref{fig:data24}.
Further, primordial disk lifetimes around A-stars are consistently shorter than for less
massive stars \citep{2009ApJ...695.1210K}. Therefore, the delay between primordial disk
dispersal and the creation of large 24\,$\mu$m excesses is too long for a plausible
direct link.

A possible delay mechanism is scattering of protoplanets and embryos to unstirred
locations in the outer disk, where they stir smaller objects to collision velocities. In
this scenario the same fraction as in the 10\,Myr peak, $\sim$40\%, of stars must harbour
planets that undergo scattering events. This figure is higher than the fractions of
intermediate mass stars both known \citep[9\% for
1.3--1.9\,$M_\odot$,][]{2007ApJ...670..833J} and predicted \citep[$\sim$20\% for
2--3\,$M_\odot$,][]{2008ApJ...673..502K} to have gas giants. However, stellar systems may
contain many more undetectable lower mass and/or more distant planets so the fraction of
stars with planetary systems is probably not a major problem for this scenario.

The main issues with a scattering scenario are timing and depletion. For all scattering
events to occur at around 10\,Myr seems to produce a fine tuning problem, when
simulations of scattering around Solar-type stars show that instabilities can happen at a
wide range of epochs. For example, in the case of the ``Nice'' model for the Late Heavy
Bombardment (LHB) in the Solar System, \citet{2005Natur.435..466G} show that the LHB
epoch varies strongly with the primordial Kuiper Belt's inner edge location. In addition,
this type of simulation is very sensitive to the initial conditions and it is unlikely
that all systems undergo instability at any particular time.

The way that planetesimal belts are depleted during scattering events also presents a
problem. Based on an analysis of the Nice model, \citet{2009MNRAS.399..385B} show that
the signature of an LHB like event is a drop in 24 and 70\,\um\ excess ratios of around 4
orders of magnitude with a brief ($\sim$15\,Myr) peak in 24\,\um\ excess, meaning that
excess fractions should drop, not peak when LHBs occur. Therefore, planet-planet
scattering is an unlikely source of the trends seen in the 24\,\um\ A-star statistics.

Another delay mechanism arises if planetesimals are initially large \citep[``born
big'',][]{2009Icar..204..558M}. The lack of an initial population of small grains means
that observable dust is not generated until the large planetesimals collide and begin to
decay, thus causing a rise and subsequent fall in the small dust population. This delay
could plausibly be 10--30\,Myr depending on the planetesimal properties and locations.

\subsection{Self-stirring}\label{sec:ss}

We now turn to the self-stirring model. As discussed in \S \ref{sec:comp}, \dc\ is of
order 10\,km. The eccentricity when 1000\,km objects form in the
\citeauthor{2008ApJS..179..451K} models is roughly constant with radius, and around $e =
0.01$. We therefore use these numbers as a order of magnitude guide and attempt to
reproduce the A-star statistics with similar values. The remaining model parameters are
\qd, $\eta_{\rm mid}$, and the disk radii. For self-stirring, \xdel\ has little impact on
the results (see \S \ref{sec:surf}). Of particular importance is the inner hole specified
by \rin\ (or the minimum \rmid\ for narrow belts), needed to ensure a peak in the overall
fraction of stars with disks.

We consider two types of disks in our population models. These disks are motivated by
observations that show debris disks are commonly narrow belts, in contrast to the
structure of protoplanetary disks that extend from several stellar radii to hundreds of
AU. Though resolved observations may not discern between a truly narrow belt and an
annulus of bright emission in an extended disk, Figure \ref{fig:fvsr} shows that the
excesses for these disks evolve quite differently. Comparing population models to the
A-star statistics may therefore suggest which kind of disk is more common. Because
protoplanetary disks are the precursors of debris disks, knowing whether they have the
same spatial extent will yield information about likely mechanisms and locations for
planetesimal and planet formation.

\subsubsection{Narrow Belts}\label{sec:belts}

\begin{figure*}
  \begin{center}
    \vspace{-0.2in}
    \begin{tabular}{cc}
     \hspace{-0.35in} \psfig{figure=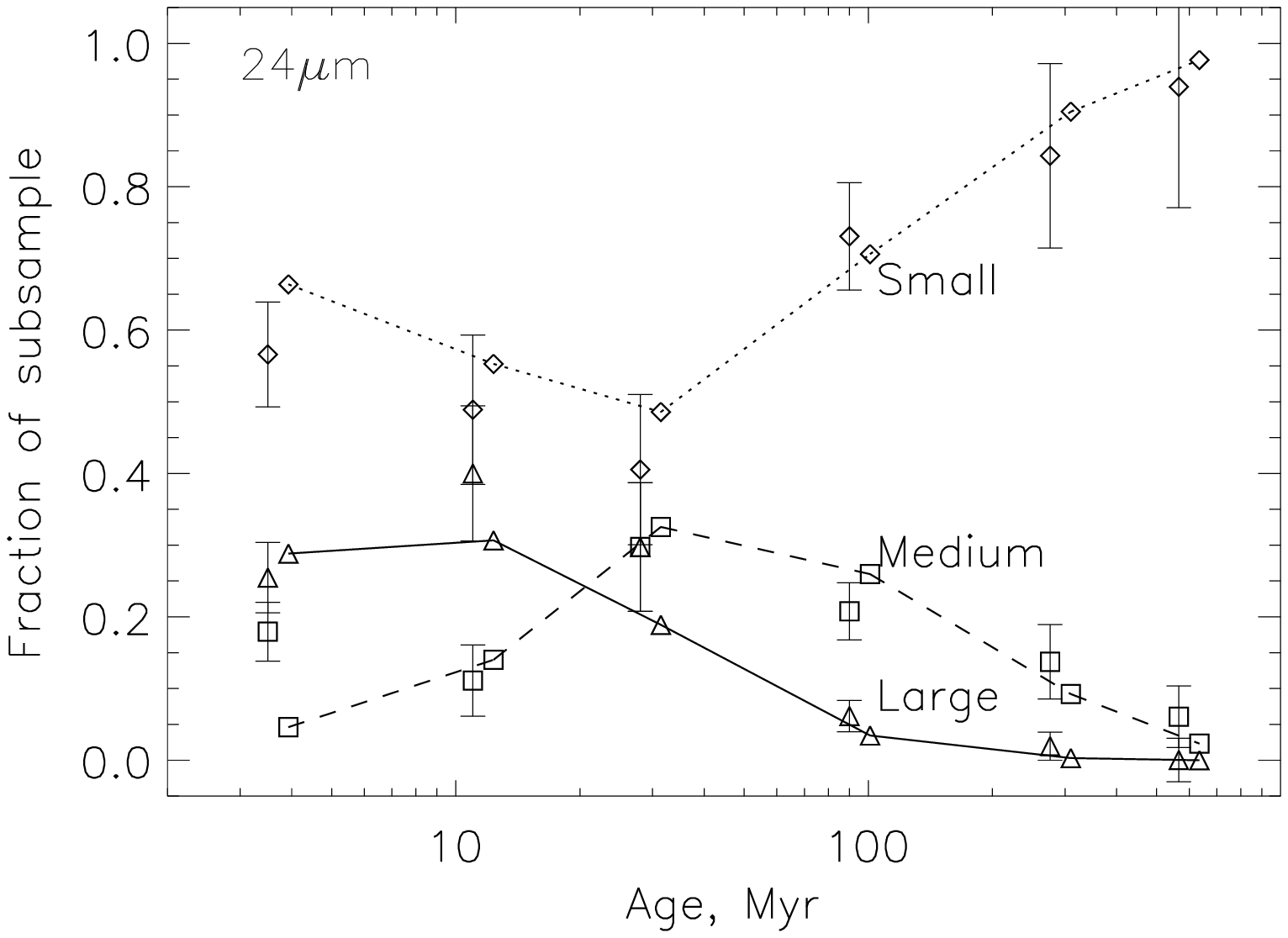,width=0.52\textwidth}&
     \hspace{-0.35in} \psfig{figure=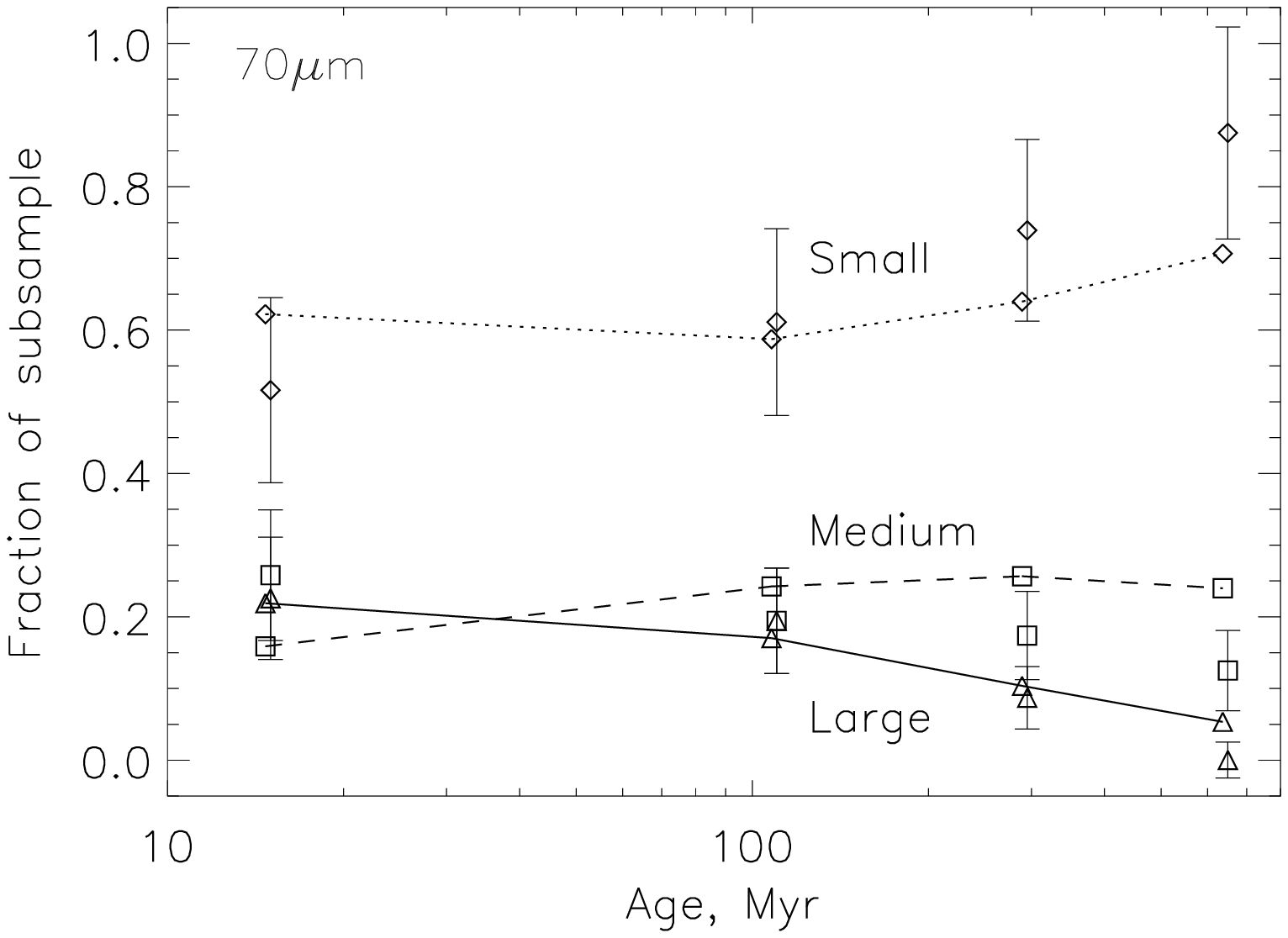,width=0.52\textwidth}\\
     \hspace{-0.35in} \psfig{figure=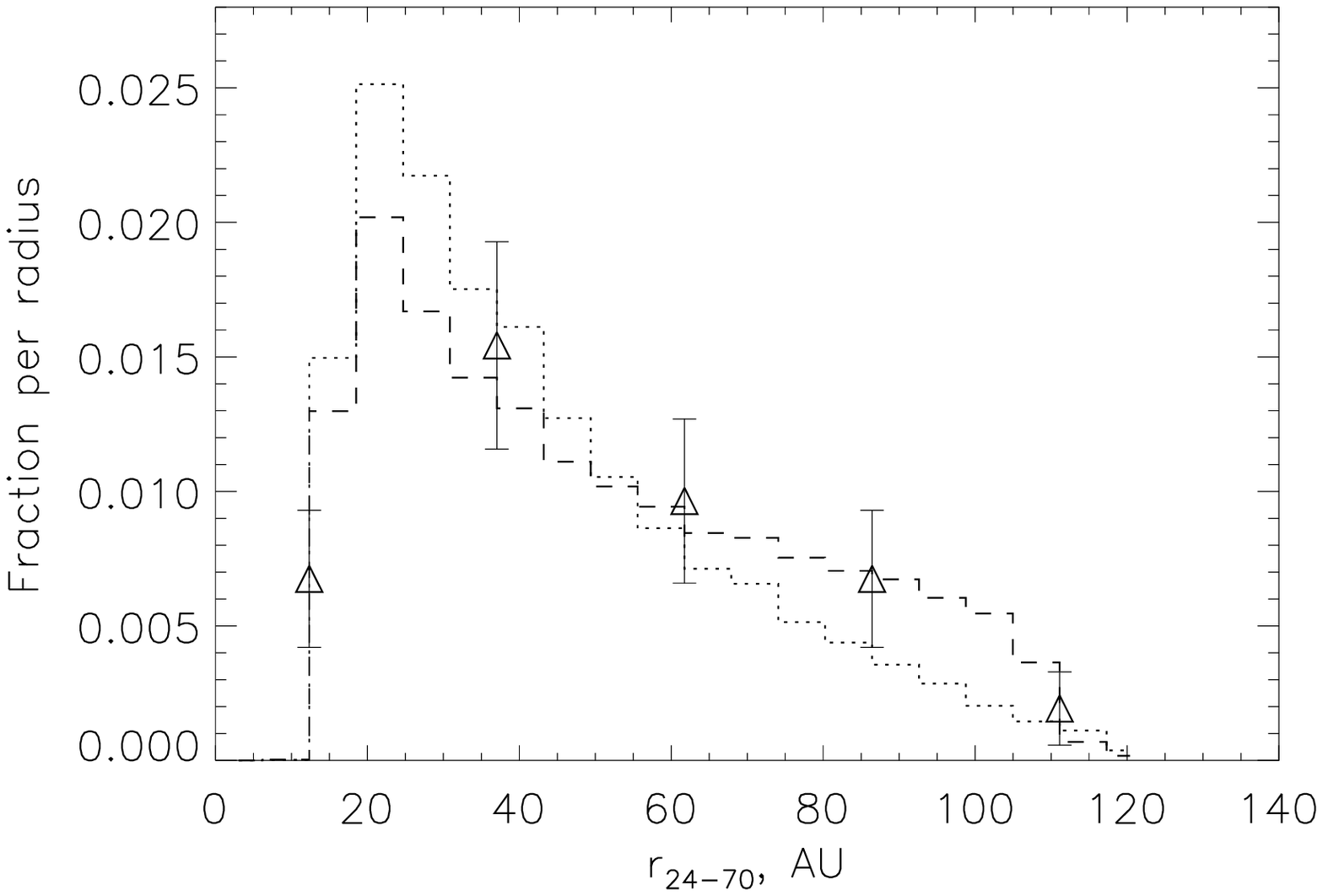,width=0.52\textwidth}&
     \hspace{-0.35in} \psfig{figure=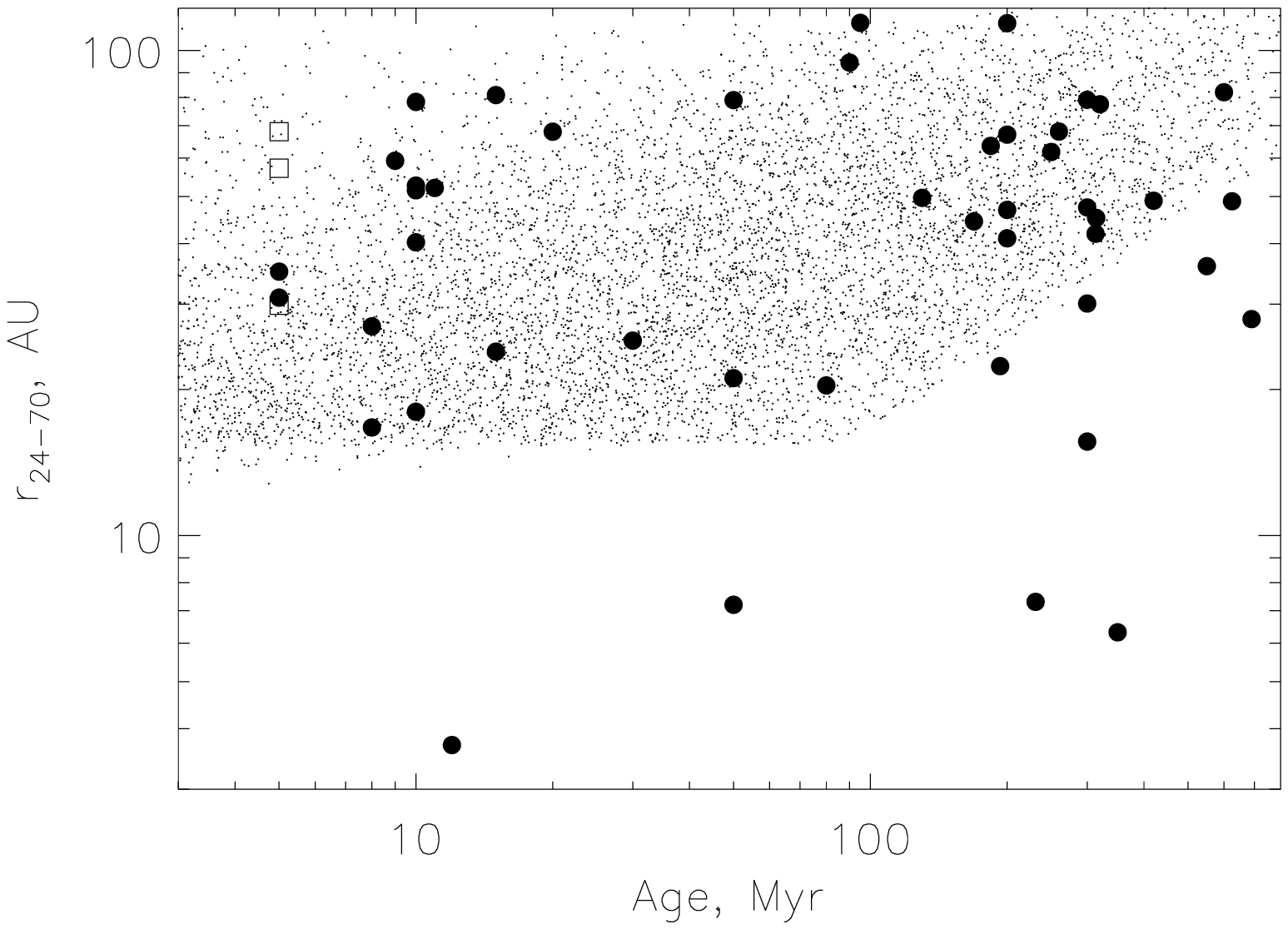,width=0.52\textwidth}\\
   \end{tabular}
   \caption{Best fit of the model population to 24 and 70\,$\mu$m statistics with narrow
     belts. \emph{Top left and right panels}: binned 24\,\um\ and 70\,\um\ evolution,
     \emph{Lower left panel}: overall (\emph{dashed line}) and detectable (\emph{dotted
       line}) model \rinf\, (omitting the four possibly transient sources noted in the
     text), and \emph{Lower right panel}: $r_{24-70}$ vs. $t$ compared to the 46 star
     subsample \citep[open squares show three additional sources detected by][HIP 76310,
     HIP 77911, and HIP 80088]{2009ApJ...705.1646C}.}\label{fig:belts}
  \end{center}
\end{figure*}

Figure \ref{fig:belts} shows the model that best reproduces the A-star trends for narrow
belts. Belt radii are taken from a distribution with $r_{\rm mid} = 15$--120\,AU and
$\gamma = -0.8$. Belts have width $dr = r_{\rm mid}/2$, $Q_{\rm D}^\star =
45$\,J\,kg$^{-1}$, and $\eta_{\rm mid} = 0.15$. We set $D_{\rm c} = 10$\,km and $e =
0.025$, which are similar to the values expected from the \citet{2008ApJS..179..451K}
models. The fit has improved significance over the pre-stirred and simple delay models
with $\chi^2_{24} = 22.4$, $\chi^2_{70} = 15.0$, $\chi^2_{\rm \chir} = 2.1$ and
$\chi^2_{\rm tot} = 39.5$.

Initially we expected that the 10--30\,Myr delay for the rise in excesses would lead
directly to an inner disk radius using Equation (\ref{eq:tstir}) for \tstir. Therefore,
with $\eta \sim 0.15$ we expect $r_{\rm mid,min} \sim 30$\,AU for stirring to occur
around 10\,Myr (Eq. \ref{eq:tstir}). However, setting \rmidmin\ to values larger than
$\approx$15\,AU means that fewer disks are concentrated at small radii, and the
distribution of inferred radii and the 70\,\um\ excesses become inconsistent with the
observations (see \S \ref{sec:best}). Because we cannot simultaneously reproduce the
statistics and have a peak in large excesses at 10\,Myr, it is actually the 70\,\um\
excesses and \rinf\ distribution that sets \rmidmin. The timing of the medium excess peak
is partly set by \qd\ because many of these disks decay from younger large excess disks.
The relative fractions in the excess bins are set by the distribution of surface
densities.

Though a reasonable fit to the 24 and 70\,\um\ statistics and overall distribution of
radii is possible, there is some difference in the distribution of radii with time (Fig.
\ref{fig:belts}, lower right). At 100\,Myr, the lower envelope of model disk radii starts
to increase from 15\,AU to $\sim$60\,AU by 800\,Myr. This change is caused by the decay
of smaller disks below detectable levels. Seven disks lie well away from the region
populated by the model. The most discrepant disks are the same six noted by
\citet{2007ApJ...663..365W} when compared to the pre-stirred population model. All disks
with radii less than 10\,AU have $f/f_{\rm max} > 20$ and may therefore be transient (HDs
38678, 115892, 3003, and 172555). The potential influence of PR drag was offered as an
explanation for the remaining two, HD 2262 and HD 106591, which we maintain here and also
apply to HD 97633.

As discussed in \S \ref{sec:surf}, we expect self-stirred debris disks to have stirring
times similar to their collisional times (i.e. $\mathcal{R} \sim 10^6$). For the narrow
belt model we find that the $\mathcal{R}$ distribution peaks around $2 \times 10^6$ with
nearly all disks within $\pm$1\,dex.

In summary, we find a reasonable fit to the A-star statistics for self-stirred
planetesimal belts with physically plausible parameters. The model does not show a strong
peak in large 24\,\um\ excesses at 10\,Myr because the disk masses and minimum belt radii
required mean that many of the smallest disks are stirred at their inner edges before
they are observed. However, the model does reproduce all A-star statistics and trends at
both 24\,\um\ and 70\,\um.

\subsubsection{Extended disks}\label{sec:extended}

We consider three possible populations of extended disks, loosely motivated by potential
processes that may shape protoplanetary and debris disks: (1) Disks with some fixed inner
hole size of a few tens of AU that extend to some variable outer radius, perhaps due to
their natural size or truncation by stellar encounters or companions, (2) disks with a
fixed outer radius and a variable inner hole size, perhaps cleared by planets that form
at various locations, and (3) disks with fixed inner and outer radii, similar in
structure to young protoplanetary disks. We do not consider disks with both variable
\rin\ and \rout\ because these are effectively a mix of belts and extended disks and
provide little useful information about which type of disk is more likely. We again use
$D_{\rm c} = 10$\,km, and $e = 0.01$ as a starting point. When it is fixed, we set
$r_{\rm out} = 150$\,AU. The variable inner or outer radii are chosen from a power-law
distribution between the maximum and minimum with power law index $\gamma$.

For the most extended disks, those with fixed \rin\ and \rout, a reasonable fit to the
24\,\um\ statistics can be found with \rin\ around 40\,AU ($\chi^2_{24} \sim 35$), but
the 70\,\um\ statistics and radius distributions are significantly different
($\chi^2_{70} \sim 190$ and $\chi^2_{\rm \chir} = 50$). Similarly, for variable \rin\
with $\gamma = -0.8$ and fixed \rout, good fits to the 24\,\um\ statistics ($\chi^2_{24}
\approx 30$) and radius distribution ($\chi^2_{\rm \chir} \approx 5$) can be achieved,
but at the cost of a poor fit to the 70\,\um\ statistics ($\chi^2_{70} \sim 120$).  The
high $\chi^2_{70}$ in both cases is due to $\sim$40--50\% of the model population having
medium 70\,\um\ excesses at late ($>$300\,Myr) times.

This high fraction can be understood by looking at how extended disks evolve, shown in
Figures \ref{fig:24comp} and \ref{fig:fvsr}. While the 24\,\um\ excess declines as
stirring moves further out in the disk, 70\,\um\ excesses for extended disks only show
small decreases while the disk is still being stirred. The $\eta = 1/3$ disk in Figure
\ref{fig:24comp} (middle solid curves) has a 24\,\um\ excess $>$2 between 7--100\,Myr,
and would thus contribute to the large 24\,\um\ excess peak in the population model. The
70\,\um\ excess for the same disk rises above the medium excess ratio of 5 at 6\,Myr, and
remains higher until after 1\,Gyr. Even when stirring reaches the outer disk edge, the
strong radial dependence on the collisional time means that at $\sim$100\,AU the excess
only decreases slowly. Therefore, extended disks with fixed \rout\ that have large
24\,\um\ excesses at early times have medium (or large) 70\,\um\ excesses for long
periods of time. This evolution means that the A-star statistics rule out self-stirred
extended disks with fixed outer radii that evolve like our model. This conclusion is
independent of the significance of the 24\,\um\ trends for the younger A-stars.

Though fixed \rout\ leads to problems with the 70\,\um\ excesses at late times, this
issue can be addressed by allowing \rout\ to vary. With fixed $r_{\rm in} = 15$\,AU and
the distribution of \rout\ set with $\gamma = -0.8$, the 70\,\um\ statistics can be
reasonably reproduced ($\chi^2_{70} \sim 6$). However, this radius distribution has
trouble reproducing the 24\,\um\ statistics ($\chi^2_{24} \approx 100$) and the \rinf\
distribution for the 46 star subsample, predicting too many disks with small radii
($\chi^2_{\rm \chir} \approx 13$). Attempts to remedy this problem by increasing the
number of wider disks (increasing the power law index $\gamma$) results in essentially
fixing \rout, which leads to the previous problem of too many medium excess disks at
70\,\um\ at late times. Thus, extended disks with fixed \rin\ also have trouble
reproducing the A-star statistics.

The conclusion that extended disks cannot reproduce the A-star statistics may depend on
the particular simplifying assumptions needed to make an analytic population model. A
possible key difference is illustrated by Figure \ref{fig:24comp}: the
\citet{2008ApJS..179..451K} models decay at 70\,\um\ more rapidly than ours and the
excesses are $\sim$2--5 times lower after 100\,Myr, probably due to continued accretion
and stirring (see \S \ref{sec:comp}).

Whether continued accretion and stirring will lower the predicted 70\,\um\ excesses for
extended disks at late times without making the radius distribution significantly steeper
(i.e. more disks at large $r$ being undetectable) is not clear, but the comparison in \S
5.2.1 of \citet{2009arXiv0911.4129K} gives some indication. Their model predictions of
70\,\um\ excess fractions for disks with fixed \rin\ and \rout\ are several times higher
than the \citet{2006ApJ...653..675S} observations at the latest times, as we found in \S
\ref{sec:extended}. Therefore, it appears that extended disk models that include
continued accretion and stirring still evolve too slowly at 70\,\um, providing further
evidence for the belt-like nature of debris disks.

\subsubsection{Summary}\label{sec:modelsummary}

Our population model suggests that debris disks are more likely to be narrow belts than
extended disks. The formal $\chi^2$ for narrow belts is an improvement over the
pre-stirred model, and reproduces all A-star statistics, including the rise and fall in
24\,\um\ excesses. The inability of extended disks to reproduce the statistics is due to
how they evolve. Disks with fixed \rin\ predict too many disks at small radii and disks
with fixed \rout\ predict too many 70\,\um\ medium excess disks at late times.

The reason belts work well appears to be because they restrict the evolution of
excesses. The power-law distribution of \rmid\ means that most disks are at relatively
small $r$, and the evolution is truncated when stirring reaches the outer edge (see
Fig. \ref{fig:fvsr}). The \rinf\ distribution is therefore a closer reflection of the
initial power-law distribution, rather than being set by self-stirred disk evolution as
for extended disks.

\subsection{Constraints on the best fitting model}\label{sec:best}

With the self-stirring model we have somewhat more power to constrain parameters than the
\citet{2007ApJ...663..365W} pre-stirred model. This ability arises because we have
estimates of \dc\ and $e$ based on the \citet{2008ApJS..179..451K} results. The remaining
parameters left to fit the statistics are \qd, $\eta_{\rm mid}$, and the disk radii
parameters $r_{mid,min}$ and $\gamma$

For the belt model surface density, we find $\eta_{\rm mid} = 0.15$ gives the best fit.
Because we find that debris disks tend to be belts, our use of a surface density law has
only a minor impact on our model. There is a factor 5 difference in surface density from
inner to outer disk edges, much smaller than the range of $\eta$ we consider.
Consequently, the model does not strongly constrain the initial surface density power-law
index $\delta$. For our best fit model we use $\delta = 1.5$, but find that populations
with $\delta = 1$ produce similar results.

We find $Q_{\rm D}^\star = 45$\,J\,kg$^{-1}$ gives the best fit to the peak in medium
24\,\um\ excesses, similar to that used for the comparison in \S \ref{sec:comp}. This
value is best considered an effective value for the evolution, since a range of \qd\ are
expected for different size planetesimals and at different radii. Our \qd\ is reasonable
for weak rock and ice for $D \sim 10$\,km for the $\sim$100-1000\,m\,s$^{-1}$ range of
collision velocities \citep{2009Icar..199..542L}. The value of 45\,J\,kg$^{-1}$ is a third
of that found for the best fit model in \citep{2007ApJ...663..365W}, but as in that paper
\qd\ and $e$ are degenerate. Thus, our results do not change if ${Q_{\rm D}^\star}^{5/6}
\, e^{-5/3} \approx 11100$. \citet{2008ApJ...673.1123L} show that the assumption of a
single planetesimal strength is within an order of magnitude of a more complex model that
includes a size-dependent planetesimal strength (their Fig. 11), with the largest
differences occurring at late times $\gtrsim$1\,Gyr.

Constraining the range of disk radii within the narrow belt model is also possible to
some degree. The model has trouble producing a stronger peak in large excesses at 10\,Myr
while retaining a reasonable fit to the statistics. The model shows a slightly stronger
peak in large excesses at 24\,\um\ when the minimum \rmid\ is increased to 20\,AU, but
the fit to the 70\,\um\ excess fractions becomes worse because the relative fraction of
wider disks increases (for fixed $\gamma$). Therefore, at late times there are more disks
with medium 70\,\um\ excesses, essentially the same problem faced for extended disks with
fixed \rout. We find it difficult to fit the A-star statistics with $r_{\rm mid,min}
\gtrsim 20$\,AU (where $\chi^2_{24} = 32.5$, $\chi^2_{70} = 24.8$, $\chi^2_{\rm \chir} =
5.2$ and total $\chi^2_{\rm tot} = 62.5$).

On the other hand, decreasing the inner hole size actually betters the formal
significance of the model fit to the A-stars. With $r_{\rm mid} = 3$--120\,AU and $\eta =
0.45$ we find $\chi^2_{24} = 14.0$, $\chi^2_{70} = 6.6$, $\chi^2_{\rm \chir} = 10.6$ and
$\chi^2_{\rm tot} = 31.2$. This model shows a monotonic decline in excesses as for the
pre-stirred model. However, in contrast to our favoured belt model, this model predicts a
population of disks younger than 100\,Myr with \rinf\ less than 15\,AU. This region is
empty in the lower right panel of Figure \ref{fig:belts} (though we ignore disks we deem
to be transient). Thus, while the typical minimum belt radius could lie between
3--15\,AU, we favour the model with minimum belt radii $\approx$15\,AU.

The power-law distribution of disk radii is fairly well constrained to $\gamma \approx
-0.8$. This constraint arises due to the belt-like nature of disks, which means that the
model $\gamma$ must be similar to the observed distribution.

The chosen age distribution has a small effect on the large excess peak at 10\,Myr. If we
set the minimum age to 1\,Myr (instead of 3\,Myr based on $\sigma$ Ori stars) then the
peak becomes slightly stronger. The difference arises because including younger stars
results in more unstirred disks (with small excesses) in the youngest age bin. Though
this effect is minor, it shows that uncertainty in stellar ages can affect the results of
the population model.

\section{Discussion}\label{sec:discussion}

In the previous sections we have shown how self-stirred debris disks evolve. The model
makes predictions, some of which can be compared to photometric observations, such as how
disk radii inferred from blackbody models are distributed and evolve.

The most basic prediction of this kind for pre- and self-stirred models is that the radial
location of peak emission should increase with time (Fig. \ref{fig:egs}). However, most
surveys have failed to find any evidence for this trend
\citep[e.g.][]{2006ApJ...653..675S,2005ApJ...635..625N}. \citet{2007ApJ...660.1556R} find
an apparent increase in disk radii inferred from IRAS colours. Unfortunately, all sources
with disk radii greater than 100\,AU---old stars largely responsible for the trend---have
yet to be confirmed with new observations (i.e. with \emph{Spitzer}). This picture is
also complicated by the expected increase in the lower envelope of radii at late times as
close-in disks drop below detection limits \citep[Fig. \ref{fig:belts} lower
right;][]{2007ApJ...663..365W}. Therefore, our model shows that the expected increase in
radii may not be as obvious as predicted by Equation (\ref{eq:rcoll}), and that other
predictions of self-stirring could be more useful. For example, another prediction of
self-stirring is that disks with large radii for their age should have higher than
average disk surface density, because these disks stir to large radii the fastest.

However, it is important that these predicted trends are not just compared with
photometric observations, because disk models based solely on SEDs can be degenerate
and/or uncertain. Resolved imaging is necessary to confirm or correct SED-derived
estimates. Imaging is also needed to test predictions such as the surface density
profiles shown in \S \ref{sec:surf}. Below we take a detailed look at a subset of
resolved debris disks with the aim of comparing observed disk characteristics with those
predicted by models.

Also, we have not addressed an alternative possibility to self-stirring, that disks are
stirred by secular perturbations from planets not co-located with the disk
\citep{2009MNRAS.399.1403M}. In this section we show that a model with \tstir\ set by
planet-stirring can reproduce the A-star statistics and suggest that these planets could
cause debris disks to be narrow belts. Because both mechanisms can fit the A-star
statistics, high resolution imaging is the best way to differentiate between
self-stirring and planet-stirring.

\subsection{Self-stirring vs. planet stirring}\label{sec:selfvplanet}

The fact that the A-star observations are well reproduced with a self-stirring model
shows that this mechanism may be important for debris disks. However, self-stirring is
not needed as an explanation for systems such as $\beta$ Pic where a planet is the likely
stirrer. It is therefore important to predict features that allow the stirring mechanism
to be identified. In the case of individual systems influenced by planets these features
are well known, and stem largely from the same influence that causes planetesimal random
velocities to increase and collisions to be destructive. Secular perturbations both warp
the disk (e.g. $\beta$ Pic) and cause it to be offset from the star
(e.g. Fomalhaut). These features may be imaged directly, or an offset inferred from
peri/apocenter glow \citep[e.g.][]{1999ApJ...527..918W}. On shorter timescales, objects
on unstable orbits too close to planets are ejected, which can result in sharp disk edges
(e.g. Fomalhaut). The remaining way of identifying whether a planet may be the stirrer is
to detect it directly (e.g. Fomalhaut, and perhaps $\beta$ Pic).

Though these features provide a way of inferring a stirring mechanism (and discovering
planets), they can only be applied to individual systems. Distinguishing the dominant
stirring mechanism at a population level is more difficult, because planet-stirring
introduces yet more parameters to the model. To briefly look at whether a planet-stirred
population model can reproduce the A-star statistics, we use the stirring time assuming
internal perturbers \citep{2009MNRAS.399.1403M}
\begin{equation}\label{eq:tpl}
  t_{\rm stir} = 5 \times 10^{-5} \, \frac{ \left( 1 - e_{\rm pl}^2 \right)^{3/2} }{e_{\rm pl}}
  \, \frac{ r^{9/2} \, \sqrt{M_\star}}{M_{\rm pl,Jup} \, r_{\rm pl}^3}
\end{equation}
in Myr, where the `pl' subscripts indicate planet properties and $r$ is the disk
location where the stirring time applies.

The increased number of parameters allows more flexibility in reproducing the observed
A-star statistics. For example, the A-star statistics can be reproduced as well as in
Figure \ref{fig:belts} for narrow belts if we set the stirring time with planet
properties: $M_{\rm pl} = 0.5\,M_{\rm Jup}$, $e_{\rm pl} = 0.1$, $r_{\rm pl} = r_{\rm
  mid} / 3$. That is, each belt is assumed to have a 0.5\,$M_{\rm Jup}$ planet with
eccentricity 0.1 located at one third of its average radius.\footnote{Setting Equations
  (\ref{eq:tstir}) and (\ref{eq:tpl}) equal and solving for $r_{\rm pl}$ suggests $r_{\rm
    pl} \propto \sqrt{r_{\rm mid}}$. However, the stronger scaling may be needed to
  account for the stirring time being independent of disk mass.} We assume that planets
form early---during the protoplanetary disk phase---so the formation time can be
ignored. For comparison with the previous models, this planet stirred model has
$\chi^2_{24} = 19.7$, $\chi^2_{70} = 7.3$, $\chi^2_{\rm \chir} = 4.4$ and $\chi^2_{\rm
  tot} = 31.3$. This example is unlikely to be the only type of planet distribution that
reproduces the observations, and shows that distinguishing between self-stirring and
planet-stirring is not yet possible by this method. Future studies of this type can use
distributions of known exoplanet properties as input, though these are only complete to
$\sim$5\,AU, and disks may be perturbed by planets at much larger radii (e.g. Fomalhaut).

Some other interesting points can be made if disks are stirred by planets. The decreasing
upper envelope of 24\,\um\ excesses for A-stars suggests that the stirrers are not too
far interior (or exterior) to the disk, because the strong radial dependence on the
stirring time means that disks far from their planets will be unstirred early and then
luminous at late times. This conclusion explains the success of the above example, where
the planet location scales with the disk radius.

Because the stirring time is set by planet properties and not the disk mass, disks can
stir in the fast mode and no longer have a maximum surface density $\tau_{\rm{eff,max}}$
(see \S \ref{sec:surf}). Neighbouring planetesimal orbits can begin to cross at non-zero
eccentricity as they precess, so the collision velocity steps from zero to the forced
eccentricity times the Keplerian velocity when the disk is stirred
\citep{2009MNRAS.399.1403M}. We can also derive a condition similar to
(\ref{eq:rapidcond}), but now using Equation (\ref{eq:tpl}) for the stirring time. For an
interior planet-stirred disk to evolve in the slow mode:
\begin{eqnarray}\label{eq:rapidcondpl}
  \mathcal{R_{\rm pl}} & \equiv &
  \frac {e_{\rm pl}}{ \left( 1 - e_{\rm pl}^2 \right)^{3/2} }
  \, r_{\rm pl}^3 \, M_{\rm pl,Jup} \, r_{\rm stir}^{-2/3} \, M_\star^{-17/6} \nonumber \\
  & & \times \, D_{\rm c} \, {Q_{\rm D}^\star}^{5/6} \, e^{-5/3} \, \eta^{-1}
     > 2.3 \times 10^5 \, .
\end{eqnarray}
This relation is qualitatively different to Condition (\ref{eq:rapidcond}), because the
planet-stirring time increases more strongly with radius than the decay time
\citep[see Fig. 6 of][]{2009MNRAS.399.1403M}. Thus, disks are more likely to evolve in the
fast mode at large radii, because the disk is stirred so late that it would have decayed
earlier if it were pre-stirred. Parameters that shorten the planet-stirring time, such as
higher $e_{\rm pl}$ or $M_{\rm pl}$, or larger $r_{\rm pl}$ (bringing the planet closer
to the disk because this example is for interior planets) make the disk more likely to
stir in the slow mode.

In summary, a population model of disks stirred by secular perturbations from planets can
reproduce the A-star statistics if the planets are located near the disk. This model is
unlikely to be unique, as different distributions of planet properties can probably give
similar results. Therefore, it is not possible to distinguish between self-stirring and
planet-stirring for debris disk populations by this method yet. The features shown by
high resolution imaging of individual objects, such as warps and offsets, remain the best
marker of debris disks influenced by planets. Our planet-stirred example also motivates
planetary system architecture as a possible reason for debris disks to be narrow belts.

\subsection{The origin of narrow belts}\label{sec:beltorigin}

Returning to the idea that resolved disks can be roughly split into extended disks and
belts \citep{2006ApJ...637L..57K}, our results suggest that debris disks tend to be
narrow belts that have minimum radii of $\sim$15\,AU. A similar conclusion was reached by
\citet{2006ApJS..166..351C}, who found that most of their objects' IRS spectra were best
fit by single temperature blackbodies colder than 130\,K (thus also suggesting that disks
have inner holes). In contrast, nearly all young stars have evidence for protoplanetary
dust and gas disks that extend from very near the star
\citep[e.g.][]{2001ApJ...553L.153H} to hundreds of AU
\citep[e.g.][]{1996AJ....111.1977M,2007prpl.conf..523W}. The implication is that not only
the primordial disk extent sets where debris disks reside, but other influences such as
photoevaporation, disk fragmentation, and truncation and clearing by planetary and
stellar companions. Within the context of the previous section, planets that stir the
disk may also be responsible for clearing it at other locations. As is likely the case
with $\beta$ Pic \citep{2001A&A...370..447A}, apparently extended disks may result from
the blowout of small grains created in a relatively narrow planetesimal belt.

To produce debris disks that are narrow belts, these mechanisms need to plausibly
reproduce two qualitative trends: (1) disk inner and outer radii are positively
correlated because we find that they are narrow belts, and (2) most disks have relatively
small radii, to reproduce the observed power-law distribution of disk radii.

One possibility is that the belts are locations where planetesimal formation was possible
or favoured. For example, one process that both clears dust from inner regions and
enhances more distant regions is the influence of photoevaporative disk clearing.
\citet{2007MNRAS.375..500A} show that after the inner gas disk has cleared, small grains
($\lesssim$10--100\,cm) are dragged outward as the inner edge of the gas disk moves
outward. At some point, the gas disk either becomes too tenuous to keep moving the
grains, or the dust density becomes comparable to the gas density. Either way, a
concentrated mass of grains is left behind by the gas (though objects larger than
$\sim$1\,m are less affected by this process). Formation of Pluto-size objects will be
enhanced here, either simply due to the faster growth time, or a more rapid instability
\citep[e.g.][]{2002ApJ...580..494Y}. Another mechanism that may result in an inner hole
and planetesimals at a particular location is the direct or rapid formation of
planetesimals in the spiral arms of a self-gravitating disk
\citep{2006MNRAS.372L...9R,2009MNRAS.398L...6C}. This process necessarily occurs beyond
tens of AU where the disk is marginally stable and dust may be concentrated in spiral
arms on a timescale shorter than the orbital period. This process may result in narrow
belts, because at $\gtrsim$100\,AU distances the disk is unstable to fragmentation and
may form companions that truncate the disk. However, given that only $\sim$10\% of the
\citet{2006ApJ...653..675S} sample have known companions, binary truncation seems an
unlikely process for setting disk outer radii.

The alternative is that the belts are locations where systems are able to retain
planetesimals. For example, one possibility is truncation by exterior stellar companions,
or within a cluster environment. However, stellar flybys are unlikely to be what sets
disk outer radii because the cross section for a close encounter suggests that this
mechanism should more often result in large disks, rather than small ones.

Planet formation provides a possible explanation of debris disk radii, and is naturally
consistent with both self-stirring and planet-stirring models. It is reasonable to think
that planetesimals form out to radii some fraction farther than where planets can
form. Dynamical clearing by planets can then set disk inner radii, analogous to how Solar
System planets dictate the Asteroid and Kuiper belt locations. In this picture debris
disk systems therefore consist of an inner planetary system with some radial extent and a
narrow planetesimal belt that extends somewhat further. This picture is essentially that
of the planet-stirred example that reproduces the A-star statistics above in \S
\ref{sec:selfvplanet}.

The range and distribution of disk radii may be linked to the initial protoplanetary disk
mass. Higher surface density disks are expected to form more giant planets over a wider
range of radii \citep[e.g.][]{2008ApJ...673..502K} and these systems are likely more
susceptible to scattering, resulting in more extended dynamical clearing and debris disks
with larger inner radii (and provides a simplistic explanation for why the BPMG and TW
Hydrae A-star disks with the largest excesses have the largest radii as discussed
below). If most protoplanetary disks are relatively low mass
\citep[e.g.][]{2005ApJ...631.1134A}, this scenario would also typically result in debris
disks at relatively small radii in agreement with the \rinf\ from our A-star sample.

The degree to which planets influence debris disk structure probably varies. In some
cases planets may simply dynamically clear inner regions while the rest of the disk is
self-stirred, whereas in other cases the disk structure may be entirely set by migration
and shepherding, scattering, and secular stirring. We now turn to a small sample of
resolved A-stars that allow us to study these possibilities for individual systems.

\subsection{Comparison with resolved imaging}\label{sec:res}

Resolved debris disks that show structures such as warps and offsets reveal planets that
may remain otherwise invisible. In these cases, planet-stirring is probably more
important than self-stirring. Therefore, resolved imaging allows estimation of the
stirring mechanism in individual cases.

Resolved imaging also allows other comparisons between observations and models to be
made, such as with the surface density profiles shown in Figure \ref{fig:egs}. Though the
distribution of radii in our model is set by comparing model and observed \rinf, this
measure tends to underestimate disk radii and cannot account for disks with several dust
components, providing further motivation for imaging.

As a sample of resolved debris disks, we use stars in the $\sim$12\,Myr old $\beta$ Pic
Moving Group \citep[BPMG,][]{2004ARA&A..42..685Z} and the similarly aged $\sim$8\,Myr old
TW Hydrae Association. This sample is therefore roughly coeval and about the age of the
24\,\um\ excess peak.

\begin{table}
  \caption{BPMG and TW Hydrae A-stars with disks. Inferred disk radii \rinf\ and $f = L_{\rm
      disk}/L_\star$ from \citet{2007ApJ...663..365W}. Real disk radii $r_{\rm real}$ are
    derived from imaging and modelling \citep[$\beta$ Pic and HR
    4796A,][]{2001A&A...370..447A,2005Natur.433..133T,1999ApJ...513L.127S,2009AJ....137...53S}
    or detailed SED modelling \citep{2006ApJS..166..351C,2009A&A...493..299S}. The
    fractional luminosities $R_{24} = F_{\rm 24,tot}/F_{24,\star}$ are from
    \citet{2008ApJ...681.1484R} and \citet{2005ApJ...620.1010R}.}\label{tab:bpmg}
  \begin{center}
    \begin{tabular}{lcccr}
      \hline
      Name & \rinf\ & $r_{\rm real}$ & $L_{\rm disk}/L_\star$ & $R_{\rm 24}$ \\
      & (AU) & (AU) & & \\
      \hline
      HR 4796A    &           27 & 70      & 330e-5            & 97\phantom{.1} \\
      $\beta$ Pic &           24 & 70      & 140e-5            & 26\phantom{.1} \\
      HR 7012     & \phantom{2}4 & 0.9, 6  & \phantom{1}50e-5  & 5.9 \\
      $\eta$ Tel  &           25 & 3.9, 24 & \phantom{1}20e-5  & 5.5 \\
      \hline
    \end{tabular}
  \end{center}
\end{table}

The A-stars in this sample are HR 4796A (HD 109573), $\beta$ Pic (HD 39060), $\eta$ Tel
(HD 181296), HR 7012 (HD 172555), HR 6070 (HD 146624), and HR 6749/HR 6750 (HD 165189/HD
165190). Of these, HR 6070 and HR 6749/6750 have no excesses at 24 or 70\,\um\
\citep{2008ApJ...681.1484R}. Characteristics of the remaining four are shown in Table
\ref{tab:bpmg}. The actual radii ($r_{\rm real}$) of HR 4796A and $\beta$ Pic are several
times larger than \rinf, and in reasonable agreement for HR 7012 and the outer dust
component of $\eta$ Tel. The difference in the case of $\beta$ Pic and HR 4796A is
explained by the presence of small grains, which emit inefficiently at long wavelengths,
and are thus hotter than a blackbody grain at the same stellocentric distance.

Compared to the entire A-star sample, the 24\,\um\ excesses in this 10\,Myr sample are
among the largest (see Fig. \ref{fig:data24}). Comparing the fraction of stars with disks
with $R_{24} \ge 5.5$ (i.e. the same or brighter than $\eta$ Tel), the $\sim$10\,Myr
sample has $67 \pm 33$\% (4/6) while the fraction for the remaining stars in the
6--20\,Myr age bin in the overall A-star sample is $8 \pm 5$\% (3/38). Whether this
relatively high excess fraction is the result of evolution within a low density
association or simply due to a small sample size is unclear.

Starting with $\beta$ Pic, this star harbours the second brightest known debris disk,
which appears extended in both scattered light and at IR and longer wavelengths
\citep[e.g.][]{1984Sci...226.1421S,1991A&A...252..220C,1998Natur.392..788H,2005Natur.433..133T}. The
disk appears over a wide range of radii, at least partly due to blowout of small grains
created in collisions \citep{2001A&A...370..447A}. Modelling of the scattered light and
mid-IR emission suggests that the parent disk where these grains originate is centred at
around 70--80\,AU \citep{2001A&A...370..447A,2005Natur.433..133T}. In addition, a
$\sim$10\,$M_{\rm Jup}$ planet at $\sim$10\,AU has been inferred as the cause of a warp
in the disk \citep{1997MNRAS.292..896M} that reaches a peak near the same location as the
parent belt \citep{2000ApJ...539..435H}. This proposed companion has recently been imaged
\citep{2009A&A...493L..21L}, though confirmation with second epoch observations is needed
\citep{2009A&A...497..557L,2009A&A...506..927L}. Finally, \citet{2005Natur.433..133T}
find an apparent clump of small grains in the South-West wing of the disk at
$\sim$50\,AU.

\begin{figure}
  \begin{center}
    \vspace{-0.2in}
    \hspace{-0.35in} \psfig{figure=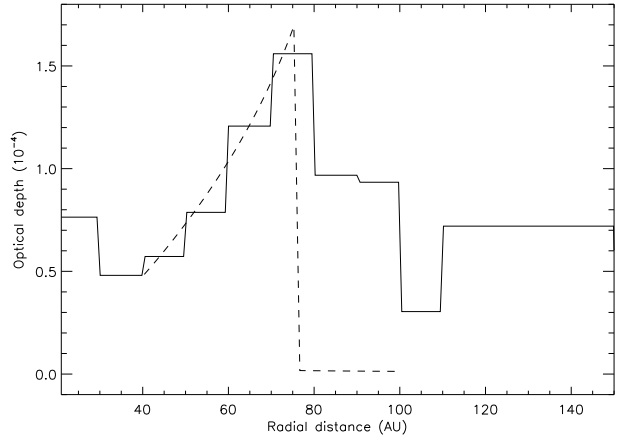,width=0.52\textwidth}
    \caption{Optical depth in the $\beta$ Pictoris debris disk derived from Mid-IR
      imaging \citep{2005Natur.433..133T}. Overplotted is the $\tau_{\rm eff}$ profile
      for a secular-stirred planetesimal belt (\emph{dashed line}) between 40--100\,AU
      (see text for details).}\label{fig:betapic}
  \end{center}
\end{figure}

The solid line in Figure \ref{fig:betapic} shows the optical depth profile of $\beta$ Pic
derived from mid-IR imaging of $\beta$ Pic \citep[North-East
wing,][]{2005Natur.433..133T}. The optical depth peaks at around 75\,AU, with a decrease
to larger and smaller radii. This profile suggests that the disk has been stirred to
75\,AU. Comparison with Figure \ref{fig:egs} suggests that the profile is more akin to a
fast self-stirred disk because the profile drops immediately interior to 75\,AU. That is,
the collisional time is shorter than the stirring time at \rstir\ so Condition
(\ref{eq:rapidcond}) is not satisfied. As discussed in \S \ref{sec:surf} this profile is
unlikely to be the result of self-stirring (though can be fit with a self-stirred model).

There are, of course, other possible explanations for the origin of the $\beta$ Pic
optical depth profile. The surface density of small grains interior to 75\,AU may drop
more rapidly due to continued accretion and stirring by Pluto-size objects for
example. Alternatively, if we still assume that the profile is due to collisional
evolution, planet-stirring is a possible scenario. As noted is \S \ref{sec:selfvplanet},
this evolution can produce disks that stir in the fast mode. Indeed, a planet has already
been proposed as the cause for a warp in the $\beta$ Pic disk, and the maximum extent of
the warp suggests that stirring due to the planet has reached $\sim$75\,AU, the location
of the peak optical depth.

The dashed line in Figure \ref{fig:betapic} shows the optical depth profile of a
planet-stirred model with good agreement between 40--80\,AU. The model has a planet with
mass $M_{\rm pl}/M_\odot = 16 \times 10^{-3}$ at $a_{\rm pl} = 10$\,AU with $e_{\rm pl} =
0.1$, and the disk is therefore stirred to $\sim$75\,AU in 12\,Myr (though
Eq. \ref{eq:tpl} shows that $M_{\rm pl}$, $a_{\rm pl}$, \& $e_{\rm pl}$ are degenerate in
setting \tstir). The value $e = 0.025$ is roughly the forced eccentricity set by the
planet which in turn sets the collision velocities as $\sim$150\,m\,s$^{-1}$.  For the
disk to be depleted interior to 75\,AU we need fairly small planetesimals, with $D_{\rm
  c} = 0.1$\,km, $Q_{\rm D}^\star = 40$\,J\,kg$^{-1}$, and $\eta = 0.01$. To match the
overall level of optical depth requires decreasing $q$ to 1.8 (from 1.83). The collision
velocities may be higher due to increased planetesimal inclinations, expected if the
planet is inclined relative to the disk, which is the interpretation of the observed warp
\citep{1997MNRAS.292..896M,2001A&A...370..447A}. In this case, a model with larger
planetesimals can reproduce the observed optical depth profile.

To match the optical depth exterior to 80\,AU requires very large $x_{\rm delay} \sim
0.5$. It is more likely that the emission outside 80\,AU is due to small grains created
in collisions at $<$80\,AU being blown out of the system, thus making the disk appear
more extended than it really is \citep{2001A&A...370..447A}.  Because we infer small \dc,
an alternative explanation for the clump of small grains in the South-West wing at about
50\,AU is needed, because the clump has about as much mass as a 100\,km planetesimal
\citep{2005Natur.433..133T}. Collective phenomena such as resonance trapping
\citep{2003ApJ...598.1321W} or dust avalanches \citep{2007A&A...461..537G} would be
required to explain the clump.

The disk around HR 4796A has a similar radius to $\beta$ Pic. Detailed modelling suggests
a $\sim$15\,AU wide parent belt at 70\,AU with a wider distribution of smaller blowout
grains \citep{2005ApJ...618..385W}. As with $\beta$ Pic, this disk is unlikely to be
self-stirred. The relatively sharp inner disk edge and a brightness asymmetry and
possible offset
\citep[e.g.][]{2000ApJ...530..329T,2005ApJ...618..385W,2009AJ....137...53S}, make HR
4796A reminiscent of the Fomalhaut disk, which is known to harbour an interior planet
that may affect the disk \citep{2009ApJ...693..734C}. If the inner edge of the HR 4796A
disk is truncated by a planet then it is probably stirred by that planet.

The two remaining disks, those around HR 7012 and $\eta$ Tel are fainter than the
previous two, but still have large excesses relative to the overall A-star sample. HR
7012 has a very small disk, with detailed models of IRS spectra suggesting grain
temperatures corresponding to 0.9--6\,AU \citep{2006ApJS..166..351C}. These models
require sub-\um-sized grains, with a composition indicative of dust produced in a recent
collision \citep{2006ApJS..166..351C,2009ApJ...701.2019L}. The mass of grains inferred is
of order $10^{21}$\,g, the mass contained in a planetesimal a few tens of km in diameter.
This disk has an unusually small radius for its age \citep{2007ApJ...663..365W} and a
reasonably high $f/f_{\rm max} \sim 100$, also suggesting that the dust is likely
transient and not due to self-stirring. However, this conclusion does not mean that the
disk was never self-stirred. To have reached what may be an analogous stage to the
giant-impact period that formed the Solar-System's terrestrial planets, objects orbiting
HR 7012 almost certainly went through the stages of growth where self-stirring is
expected. This late stage of chaotic growth could be considered a second phase of
self-stirring, where the stirring this time arises because the surface density of smaller
objects is insufficient to damp the largest objects and big objects stir each other.

Finally, $\eta$ Tel has distinct planetesimal belts at $\sim$4 and 24\,AU, each
contributing about equally to the excess at 24\,\um.
\citep{2006ApJS..166..351C,2009A&A...493..299S}. \citet{2009A&A...493..299S} find that
the outer disk can be explained by a self-stirred model. If stirring has reached 24\,AU,
the hotter dust may be the result of a collision as planets continue to grow in inner
regions. In contrast to HR 7012 however, there is no evidence for sub-\um\ grains in the
$\eta$ Tel disk from which it might be inferred that a recent collision is required to
explain short lived dust \citep{2006ApJS..166..351C}.

In summary, when confronted with detailed observations, the self-stirring model appears
to face competition from the continued growth of planets through stochastic collisions
and their dynamical effects in trying to provide explanations of disk structure. The
$\beta$ Pic disk is likely stirred by the proposed planet at $\sim$10\,AU. The
non-azimuthal symmetry of the HR 4796A disk also suggests a planetary influence. The
$\eta$ Tel disk is consistent with a self-stirring model and HR 7012 is probably a
transient disk resulting from a recent collision.

\section{Conclusions}\label{sec:conclusions}

Recent observations of young A-stars show evidence for an increase in the level and
frequency of 24\,\um\ excesses from $\sim$3 to 10--30\,Myr
\citep{2006ApJ...652..472H,2008ApJ...672..558C,2008ApJ...688..597C}. Excesses then
decline on an timescale of $\sim$150\,yr \citep{2005ApJ...620.1010R}. The rise in debris
disk emission at early times has been interpreted as evidence for self-stirring, where a
collisional cascade begins when Pluto-size objects form and stir planetesimals. Because
the time taken to form Plutos increases with radial distance from the central star,
\citet{2008ARA&A..46..339W} noted that the 10--30\,Myr delay also implied that A-star
disks must have inner holes of order 10\,AU if they are self-stirred. Though there is in
fact little evidence that the fraction of stars with 24\,\um\ excesses changes in the
first 50\,Myr or so (Fig. \ref{fig:data24}), the overall trend shown by the A-star
statistics provides tentative observational evidence of self-stirring. However, a
promising alternative to self-stirring should also be considered, that debris disks are
instead stirred by secular perturbations from an eccentric planet
\citep{2009MNRAS.399.1403M}.

In this paper, we use the analytic model described in \S \ref{sec:model} to study the
evolution of self-stirred disks. Our model is essentially the steady-state evolution
model of \citet{2007ApJ...663..365W}, modified to include self-stirring. We first compare
our model to the detailed \citet{2008ApJS..179..451K} results, using an empirical delay
for self-stirring to ensure we reproduce their excess trends over a range of disk surface
densities and stellar masses (\S \ref{sec:single}). The only difference in evolution is
after the peak excess, with the \citeauthor{2008ApJS..179..451K} models decaying more
rapidly at 70\,\um, probably due to continued accretion.

We illustrate the implications of collisional evolution for resolved disks (\S
\ref{sec:surf}). Because disks process their mass from the inside out, the surface
density profile of any collisionally evolved disk region increases as $r^{7/3}$ and disks
appear to increase in radius over time (pre- and planet-stirred disks show the same
behaviour). The primordial surface density profile remains where the disk has not been
stirred (and protoplanetary growth is ongoing).

Disks with delayed stirring can evolve in two different ways. If the collisional time is
short compared to the stirring time (i.e. has more mass than it would if it were
pre-stirred) then the disk rapidly loses mass as it reverts to its equilibrium
state. This evolution results in a bright narrow ring of emission where stirring is
occurring (Fig. \ref{fig:egs}). While we suggest that this evolution is unlikely for
self-stirred disks, it can occur for planet-stirred disks. Disks that stir before the
collisional time evolve in the same way as pre-stirred disks, with the difference that
there is less emission in exterior regions where the collisional cascade has not
started. This is the typical evolution we expect for self-stirred disks.

In \S \ref{sec:astar}, we turn to the observations and show why the pre-stirred debris
disk model fails to produce the trends in the A-star statistics; with no mechanism to
delay the onset of stirring, 24\,\um\ excesses in the model are highest at the earliest
times. The overall fraction of stars with disks declines monotonically with time in
contrast with the observations, which peak around 30\,Myr. Using the same power-law
planetesimal belt radius distribution as \citet{2007ApJ...663..365W}, and planetesimal
sizes and eccentricities consistent with the \citet{2008ApJS..179..451K} models, we show
that the A-star trends and statistics can be reasonably reproduced by a self-stirring
model with $Q_{\rm D}^\star = 45$\,J\,kg$^{-1}$ and the average disk mass 0.15 times an
MMSN disk. Disks are ``narrow belts'' with width $dr = r_{\rm mid}/2$. The smallest
planetesimal belt has $r_{\rm mid} = 15$\,AU, and the largest 120\,AU.

We have less success fitting the A-star observations with extended disks---disks with
fixed inner and/or outer radii. Although the 24\,\um\ emission can be reasonably
reproduced with disks with fixed $\sim$150\,AU outer radii and fixed or variable inner
radii, these models result in too many 70\,\um\ excess disks at late times. This problem
arises because extended disks evolve at near constant 70\,\um\ fractional luminosity
until their outer edges are stirred (Figs. \ref{fig:24comp} \& \ref{fig:fvsr}). Disks
with fixed inner radii of $\sim$15\,AU and variable outer radii also fail to fit the
observed statistics, because the models over-predict the number of disks with small
radii. Thus, our conclusion that debris disks are narrow belts and not extended is
independent of the A-star trends for ages $\lesssim$50\,Myr.

Progress can be made in several directions to further understand the effects of
self-stirring on model populations. Our model only removes dust from the small end of the
size distribution, whereas the \citet{2008ApJS..179..451K} models show that mass is also
lost as Pluto-size objects continue to accrete fragments, and that mass loss is
accelerated as the largest objects continue to grow. A model including continued
accretion and stirring will lower 70\,\um\ excesses for the oldest disks, perhaps
allowing extended disks to reproduce the A-star statistics. However, the
\citet{2009arXiv0911.4129K} model comparison with A-star data suggests there will still
be difficulties, with extended disk models that include continued accretion and stirring
also evolving too slowly at 70\,\um.

Planets probably stir and set the structure of some disks. In \S \ref{sec:selfvplanet} we
show the A-star statistics can be fit with a population of narrow belts stirred by
secular perturbations from an eccentric planet. The planet-stirred model produces
essentially the same results as the self-stirred one, with the key to reproducing the
A-star observations apparently being the planet location. If planets are located too far
from the disk then stirring occurs too late and the characteristic $\sim$150\,Myr
timescale decay of 24\,\um\ excesses does not occur. Thus, the successful model has
0.5\,$M_{\rm Jup}$ planets with $e=0.1$ that are located at one third the disk radius.

Therefore, population models cannot yet distinguish whether self-stirring or
planet-stirring is more important. Population models are also unlikely to rule out one
stirring mechanism due to the many model parameters. The poor statistics for A-stars
younger than $\sim$50\,Myr also hinder progress. The fact that the rise in 24\,\um\
excesses for young A-stars has marginal statistical significance is unlikely to change in
the near future, as most nearby regions have been studied with \emph{Spitzer} and a
significant increase in numbers awaits the launch of JWST.

In \S \ref{sec:beltorigin} we consider the origin of narrow planetesimal belts, and
suggest that planet formation provides a natural explanation, if planetesimals form to
radii somewhat larger than planets. The debris disk inner holes are then regions cleared
by planets, and the outer extent set by where planetesimals can form. Depending on
planetary system architecture, these planets may also stir the disk as suggested by our
planet-stirred example in \S \ref{sec:selfvplanet}.

In \S \ref{sec:res}, we look more closely at the sample of $\sim$10\,Myr old resolved
disks around A-stars from the BPMG and the TW Hydrae Association, and find that only
$\eta$ Tel allows a reasonable explanation with a self-stirring model. The disks around
$\beta$ Pic and HR 4796A seem more likely to be affected by planets. It is possible that
the $\beta$ Pic debris disk is stirred through secular perturbations from the planet
proposed to orbit at $\sim$10\,AU. The disk around HR 7012 appears transient, though
probably went through a phase of self-stirring when it was younger. These observations
suggest that the degree to which debris disks are influenced by planets varies, and that
the answer to the question of debris disk stirring lies with high resolution imaging.

\section*{Acknowledgements}

It is a pleasure to thank Ken Rice for helpful discussions, Scott Kenyon for discussions
and kindly sharing output from his self-stirring models, and Alexander Krivov for a
thorough review that improved the content and presentation of this contribution. This
study was completed during the Isaac Newton Institute's Dynamics of Disks and Planets
programme in Cambridge, UK.

\bibliography{ref,extras} \bibliographystyle{astroads}

\begin{thebibliography}{77}
\expandafter\ifx\csname natexlab\endcsname\relax\def\natexlab#1{#1}\fi
\expandafter\ifx\csname href\endcsname\relax
  \def\href#1#2{}\fi
\expandafter\ifx\csname urllinklabel\endcsname\relax
  \def\urllinklabel{[LINK]}\fi
\expandafter\ifx\csname adsurllinklabel\endcsname\relax
  \def\adsurllinklabel{[ADS]}\fi

\bibitem[{{Alexander} \& {Armitage}(2007)}]{2007MNRAS.375..500A}
{Alexander}, R.~D. \& {Armitage}, P.~J. 2007, \mnras, 375, 500
 \href{http://adsabs.harvard.edu/abs/2007MNRAS.375..500A}{\adsurllinklabel}

\bibitem[{{Andrews} \& {Williams}(2005)}]{2005ApJ...631.1134A}
{Andrews}, S.~M. \& {Williams}, J.~P. 2005, \apj, 631, 1134
 \href{http://adsabs.harvard.edu/abs/2005ApJ...631.1134A}{\adsurllinklabel}

\bibitem[{{Andrews} \& {Williams}(2007)}]{2007ApJ...671.1800A}
---. 2007, \apj, 671, 1800
 \href{http://adsabs.harvard.edu/abs/2007ApJ...671.1800A}{\adsurllinklabel}

\bibitem[{{Augereau} {et~al.}(2001){Augereau}, {Nelson}, {Lagrange},
  {Papaloizou}, \& {Mouillet}}]{2001A&A...370..447A}
{Augereau}, J.~C., {Nelson}, R.~P., {Lagrange}, A.~M., {Papaloizou}, J.~C.~B.,
  \& {Mouillet}, D. 2001, \aap, 370, 447
 \href{http://adsabs.harvard.edu/abs/2001A&A...370..447A}{\adsurllinklabel}

\bibitem[{{Blum} \& {Wurm}(2008)}]{2008ARA&A..46...21B}
{Blum}, J. \& {Wurm}, G. 2008, \araa, 46, 21
 \href{http://adsabs.harvard.edu/abs/2008ARA&A..46...21B}{\adsurllinklabel}

\bibitem[{{Booth} {et~al.}(2009){Booth}, {Wyatt}, {Morbidelli},
  {Moro-Mart{\'{\i}}n}, \& {Levison}}]{2009MNRAS.399..385B}
{Booth}, M., {Wyatt}, M.~C., {Morbidelli}, A., {Moro-Mart{\'{\i}}n}, A., \&
  {Levison}, H.~F. 2009, \mnras, 399, 385
 \href{http://adsabs.harvard.edu/abs/2009MNRAS.399..385B}{\adsurllinklabel}

\bibitem[{{Burns} {et~al.}(1979){Burns}, {Lamy}, \&
  {Soter}}]{1979Icar...40....1B}
{Burns}, J.~A., {Lamy}, P.~L., \& {Soter}, S. 1979, Icarus, 40, 1
 \href{http://adsabs.harvard.edu/abs/1979Icar...40....1B}{\adsurllinklabel}

\bibitem[{{Carpenter} {et~al.}(2009){Carpenter}, {Mamajek}, {Hillenbrand}, \&
  {Meyer}}]{2009ApJ...705.1646C}
{Carpenter}, J.~M., {Mamajek}, E.~E., {Hillenbrand}, L.~A., \& {Meyer}, M.~R.
  2009, \apj, 705, 1646
 \href{http://adsabs.harvard.edu/abs/2009ApJ...705.1646C}{\adsurllinklabel}

\bibitem[{{Chen} {et~al.}(2006){Chen}, {Sargent}, {Bohac}, {Kim},
  {Leibensperger}, {Jura}, {Najita}, {Forrest}, {Watson}, {Sloan}, \&
  {Keller}}]{2006ApJS..166..351C}
{Chen}, C.~H., {Sargent}, B.~A., {Bohac}, C., {Kim}, K.~H., {Leibensperger},
  E., {Jura}, M., {Najita}, J., {Forrest}, W.~J., {Watson}, D.~M., {Sloan},
  G.~C., \& {Keller}, L.~D. 2006, \apjs, 166, 351
 \href{http://adsabs.harvard.edu/abs/2006ApJS..166..351C}{\adsurllinklabel}

\bibitem[{{Chiang} {et~al.}(2009){Chiang}, {Kite}, {Kalas}, {Graham}, \&
  {Clampin}}]{2009ApJ...693..734C}
{Chiang}, E., {Kite}, E., {Kalas}, P., {Graham}, J.~R., \& {Clampin}, M. 2009,
  \apj, 693, 734
 \href{http://adsabs.harvard.edu/abs/2009ApJ...693..734C}{\adsurllinklabel}

\bibitem[{{Chini} {et~al.}(1991){Chini}, {Kruegel}, {Kreysa}, {Shustov}, \&
  {Tutukov}}]{1991A&A...252..220C}
{Chini}, R., {Kruegel}, E., {Kreysa}, E., {Shustov}, B., \& {Tutukov}, A. 1991,
  \aap, 252, 220
 \href{http://adsabs.harvard.edu/abs/1991A&A...252..220C}{\adsurllinklabel}

\bibitem[{{Clarke} \& {Lodato}(2009)}]{2009MNRAS.398L...6C}
{Clarke}, C.~J. \& {Lodato}, G. 2009, \mnras, 398, L6
 \href{http://adsabs.harvard.edu/abs/2009MNRAS.398L...6C}{\adsurllinklabel}

\bibitem[{{Currie} {et~al.}(2008{\natexlab{a}}){Currie}, {Kenyon}, {Balog},
  {Rieke}, {Bragg}, \& {Bromley}}]{2008ApJ...672..558C}
{Currie}, T., {Kenyon}, S.~J., {Balog}, Z., {Rieke}, G., {Bragg}, A., \&
  {Bromley}, B. 2008{\natexlab{a}}, \apj, 672, 558
 \href{http://adsabs.harvard.edu/abs/2008ApJ...672..558C}{\adsurllinklabel}

\bibitem[{{Currie} {et~al.}(2008{\natexlab{b}}){Currie}, {Plavchan}, \&
  {Kenyon}}]{2008ApJ...688..597C}
{Currie}, T., {Plavchan}, P., \& {Kenyon}, S.~J. 2008{\natexlab{b}}, \apj, 688,
  597
 \href{http://adsabs.harvard.edu/abs/2008ApJ...688..597C}{\adsurllinklabel}

\bibitem[{{Dermott} {et~al.}(2001){Dermott}, {Grogan}, {Durda}, {Jayaraman},
  {Kehoe}, {Kortenkamp}, \& {Wyatt}}]{dermott01}
{Dermott}, S.~F., {Grogan}, K., {Durda}, D.~D., {Jayaraman}, S., {Kehoe},
  T.~J.~J., {Kortenkamp}, S.~J., \& {Wyatt}, M.~C. E.~{Grun}, B.~A.~S.
  {Gustafson}S.~F. {Dermott} \& H.~{Fechtig} (Berlin:Springer-Verlag), 569--639


\bibitem[{{Dohnanyi}(1969)}]{1969JGR....74.2531D}
{Dohnanyi}, J.~S. 1969, \jgr, 74, 2531
 \href{http://adsabs.harvard.edu/abs/1969JGR....74.2531D}{\adsurllinklabel}

\bibitem[{{Dominik} \& {Decin}(2003)}]{2003ApJ...598..626D}
{Dominik}, C. \& {Decin}, G. 2003, \apj, 598, 626
 \href{http://adsabs.harvard.edu/abs/2003ApJ...598..626D}{\adsurllinklabel}

\bibitem[{{Gomes} {et~al.}(2005){Gomes}, {Levison}, {Tsiganis}, \&
  {Morbidelli}}]{2005Natur.435..466G}
{Gomes}, R., {Levison}, H.~F., {Tsiganis}, K., \& {Morbidelli}, A. 2005, \nat,
  435, 466
 \href{http://adsabs.harvard.edu/abs/2005Natur.435..466G}{\adsurllinklabel}

\bibitem[{{Grigorieva} {et~al.}(2007){Grigorieva}, {Artymowicz}, \&
  {Th{\'e}bault}}]{2007A&A...461..537G}
{Grigorieva}, A., {Artymowicz}, P., \& {Th{\'e}bault}, P. 2007, \aap, 461, 537
 \href{http://adsabs.harvard.edu/abs/2007A&A...461..537G}{\adsurllinklabel}

\bibitem[{{Haisch} {et~al.}(2001){Haisch}, {Lada}, \&
  {Lada}}]{2001ApJ...553L.153H}
{Haisch}, Jr., K.~E., {Lada}, E.~A., \& {Lada}, C.~J. 2001, \apjl, 553, L153
 \href{http://adsabs.harvard.edu/abs/2001ApJ...553L.153H}{\adsurllinklabel}

\bibitem[{{Heap} {et~al.}(2000){Heap}, {Lindler}, {Lanz}, {Cornett}, {Hubeny},
  {Maran}, \& {Woodgate}}]{2000ApJ...539..435H}
{Heap}, S.~R., {Lindler}, D.~J., {Lanz}, T.~M., {Cornett}, R.~H., {Hubeny}, I.,
  {Maran}, S.~P., \& {Woodgate}, B. 2000, \apj, 539, 435
 \href{http://adsabs.harvard.edu/abs/2000ApJ...539..435H}{\adsurllinklabel}

\bibitem[{{Hern{\'a}ndez} {et~al.}(2006){Hern{\'a}ndez}, {Brice{\~n}o},
  {Calvet}, {Hartmann}, {Muzerolle}, \& {Quintero}}]{2006ApJ...652..472H}
{Hern{\'a}ndez}, J., {Brice{\~n}o}, C., {Calvet}, N., {Hartmann}, L.,
  {Muzerolle}, J., \& {Quintero}, A. 2006, \apj, 652, 472
 \href{http://adsabs.harvard.edu/abs/2006ApJ...652..472H}{\adsurllinklabel}

\bibitem[{{Hern{\'a}ndez} {et~al.}(2009){Hern{\'a}ndez}, {Calvet}, {Hartmann},
  {Muzerolle}, {Gutermuth}, \& {Stauffer}}]{2009ApJ...707..705H}
{Hern{\'a}ndez}, J., {Calvet}, N., {Hartmann}, L., {Muzerolle}, J.,
  {Gutermuth}, R., \& {Stauffer}, J. 2009, \apj, 707, 705
 \href{http://adsabs.harvard.edu/abs/2009ApJ...707..705H}{\adsurllinklabel}

\bibitem[{{Hern{\'a}ndez} {et~al.}(2008){Hern{\'a}ndez}, {Hartmann}, {Calvet},
  {Jeffries}, {Gutermuth}, {Muzerolle}, \& {Stauffer}}]{2008ApJ...686.1195H}
{Hern{\'a}ndez}, J., {Hartmann}, L., {Calvet}, N., {Jeffries}, R.~D.,
  {Gutermuth}, R., {Muzerolle}, J., \& {Stauffer}, J. 2008, \apj, 686, 1195
 \href{http://adsabs.harvard.edu/abs/2008ApJ...686.1195H}{\adsurllinklabel}

\bibitem[{{Hern{\'a}ndez} {et~al.}(2007){Hern{\'a}ndez}, {Hartmann}, {Megeath},
  {Gutermuth}, {Muzerolle}, {Calvet}, {Vivas}, {Brice{\~n}o}, {Allen},
  {Stauffer}, {Young}, \& {Fazio}}]{2007ApJ...662.1067H}
{Hern{\'a}ndez}, J., {Hartmann}, L., {Megeath}, T., {Gutermuth}, R.,
  {Muzerolle}, J., {Calvet}, N., {Vivas}, A.~K., {Brice{\~n}o}, C., {Allen},
  L., {Stauffer}, J., {Young}, E., \& {Fazio}, G. 2007, \apj, 662, 1067
 \href{http://adsabs.harvard.edu/abs/2007ApJ...662.1067H}{\adsurllinklabel}

\bibitem[{{Holland} {et~al.}(1998){Holland}, {Greaves}, {Zuckerman}, {Webb},
  {McCarthy}, {Coulson}, {Walther}, {Dent}, {Gear}, \&
  {Robson}}]{1998Natur.392..788H}
{Holland}, W.~S., {Greaves}, J.~S., {Zuckerman}, B., {Webb}, R.~A., {McCarthy},
  C., {Coulson}, I.~M., {Walther}, D.~M., {Dent}, W.~R.~F., {Gear}, W.~K., \&
  {Robson}, I. 1998, \nat, 392, 788
 \href{http://adsabs.harvard.edu/abs/1998Natur.392..788H}{\adsurllinklabel}

\bibitem[{{Johnson} {et~al.}(2007){Johnson}, {Butler}, {Marcy}, {Fischer},
  {Vogt}, {Wright}, \& {Peek}}]{2007ApJ...670..833J}
{Johnson}, J.~A., {Butler}, R.~P., {Marcy}, G.~W., {Fischer}, D.~A., {Vogt},
  S.~S., {Wright}, J.~T., \& {Peek}, K.~M.~G. 2007, \apj, 670, 833
 \href{http://adsabs.harvard.edu/abs/2007ApJ...670..833J}{\adsurllinklabel}

\bibitem[{{Kalas} {et~al.}(2008){Kalas}, {Graham}, {Chiang}, {Fitzgerald},
  {Clampin}, {Kite}, {Stapelfeldt}, {Marois}, \& {Krist}}]{2008Sci...322.1345K}
{Kalas}, P., {Graham}, J.~R., {Chiang}, E., {Fitzgerald}, M.~P., {Clampin}, M.,
  {Kite}, E.~S., {Stapelfeldt}, K., {Marois}, C., \& {Krist}, J. 2008, Science,
  322, 1345
 \href{http://adsabs.harvard.edu/abs/2008Sci...322.1345K}{\adsurllinklabel}

\bibitem[{{Kalas} {et~al.}(2006){Kalas}, {Graham}, {Clampin}, \&
  {Fitzgerald}}]{2006ApJ...637L..57K}
{Kalas}, P., {Graham}, J.~R., {Clampin}, M.~C., \& {Fitzgerald}, M.~P. 2006,
  \apjl, 637, L57
 \href{http://adsabs.harvard.edu/abs/2006ApJ...637L..57K}{\adsurllinklabel}

\bibitem[{{Kennedy} \& {Kenyon}(2008)}]{2008ApJ...673..502K}
{Kennedy}, G.~M. \& {Kenyon}, S.~J. 2008, \apj, 673, 502
 \href{http://adsabs.harvard.edu/abs/2008ApJ...673..502K}{\adsurllinklabel}

\bibitem[{{Kennedy} \& {Kenyon}(2009)}]{2009ApJ...695.1210K}
---. 2009, \apj, 695, 1210
 \href{http://adsabs.harvard.edu/abs/2009ApJ...695.1210K}{\adsurllinklabel}

\bibitem[{{Kenyon} \& {Bromley}(2004)}]{2004AJ....127..513K}
{Kenyon}, S.~J. \& {Bromley}, B.~C. 2004, \aj, 127, 513
 \href{http://adsabs.harvard.edu/abs/2004AJ....127..513K}{\adsurllinklabel}

\bibitem[{{Kenyon} \& {Bromley}(2008)}]{2008ApJS..179..451K}
---. 2008, \apjs, 179, 451
 \href{http://adsabs.harvard.edu/abs/2008ApJS..179..451K}{\adsurllinklabel}

\bibitem[{{Kenyon} \& {Bromley}(2009)}]{2009arXiv0911.4129K}
---. 2009, ArXiv e-prints, (0911.4129)
 \href{http://adsabs.harvard.edu/abs/2009arXiv0911.4129K}{\adsurllinklabel}

\bibitem[{{Kharchenko}(2001)}]{2001KFNT...17..409K}
{Kharchenko}, N.~V. 2001, Kinematika i Fizika Nebesnykh Tel, 17, 409
 \href{http://adsabs.harvard.edu/abs/2001KFNT...17..409K}{\adsurllinklabel}

\bibitem[{{Lagrange} {et~al.}(2009{\natexlab{a}}){Lagrange}, {Gratadour},
  {Chauvin}, {Fusco}, {Ehrenreich}, {Mouillet}, {Rousset}, {Rouan}, {Allard},
  {Gendron}, {Charton}, {Mugnier}, {Rabou}, {Montri}, \&
  {Lacombe}}]{2009A&A...493L..21L}
{Lagrange}, A., {Gratadour}, D., {Chauvin}, G., {Fusco}, T., {Ehrenreich}, D.,
  {Mouillet}, D., {Rousset}, G., {Rouan}, D., {Allard}, F., {Gendron}, {\'E}.,
  {Charton}, J., {Mugnier}, L., {Rabou}, P., {Montri}, J., \& {Lacombe}, F.
  2009{\natexlab{a}}, \aap, 493, L21
 \href{http://adsabs.harvard.edu/abs/2009A&A...493L..21L}{\adsurllinklabel}

\bibitem[{{Lagrange} {et~al.}(2009{\natexlab{b}}){Lagrange}, {Kasper},
  {Boccaletti}, {Chauvin}, {Gratadour}, {Fusco}, {Ehrenreich}, {Apai},
  {Mouillet}, \& {Rouan}}]{2009A&A...506..927L}
{Lagrange}, A., {Kasper}, M., {Boccaletti}, A., {Chauvin}, G., {Gratadour}, D.,
  {Fusco}, T., {Ehrenreich}, D., {Apai}, D., {Mouillet}, D., \& {Rouan}, D.
  2009{\natexlab{b}}, \aap, 506, 927
 \href{http://adsabs.harvard.edu/abs/2009A&A...506..927L}{\adsurllinklabel}

\bibitem[{{Lecavelier Des Etangs} \&
  {Vidal-Madjar}(2009)}]{2009A&A...497..557L}
{Lecavelier Des Etangs}, A. \& {Vidal-Madjar}, A. 2009, \aap, 497, 557
 \href{http://adsabs.harvard.edu/abs/2009A&A...497..557L}{\adsurllinklabel}

\bibitem[{{Leinhardt} \& {Stewart}(2009)}]{2009Icar..199..542L}
{Leinhardt}, Z.~M. \& {Stewart}, S.~T. 2009, Icarus, 199, 542
 \href{http://adsabs.harvard.edu/abs/2009Icar..199..542L}{\adsurllinklabel}

\bibitem[{{Lissauer}(1987)}]{1987Icar...69..249L}
{Lissauer}, J.~J. 1987, Icarus, 69, 249
 \href{http://adsabs.harvard.edu/abs/1987Icar...69..249L}{\adsurllinklabel}

\bibitem[{{Lisse} {et~al.}(2009){Lisse}, {Chen}, {Wyatt}, {Morlok}, {Song},
  {Bryden}, \& {Sheehan}}]{2009ApJ...701.2019L}
{Lisse}, C.~M., {Chen}, C.~H., {Wyatt}, M.~C., {Morlok}, A., {Song}, I.,
  {Bryden}, G., \& {Sheehan}, P. 2009, \apj, 701, 2019
 \href{http://adsabs.harvard.edu/abs/2009ApJ...701.2019L}{\adsurllinklabel}

\bibitem[{{L{\"o}hne} {et~al.}(2008){L{\"o}hne}, {Krivov}, \&
  {Rodmann}}]{2008ApJ...673.1123L}
{L{\"o}hne}, T., {Krivov}, A.~V., \& {Rodmann}, J. 2008, \apj, 673, 1123
 \href{http://adsabs.harvard.edu/abs/2008ApJ...673.1123L}{\adsurllinklabel}

\bibitem[{{McCaughrean} \& {O'dell}(1996)}]{1996AJ....111.1977M}
{McCaughrean}, M.~J. \& {O'dell}, C.~R. 1996, \aj, 111, 1977
 \href{http://adsabs.harvard.edu/abs/1996AJ....111.1977M}{\adsurllinklabel}

\bibitem[{{Morbidelli} {et~al.}(2009){Morbidelli}, {Bottke}, {Nesvorn{\'y}}, \&
  {Levison}}]{2009Icar..204..558M}
{Morbidelli}, A., {Bottke}, W.~F., {Nesvorn{\'y}}, D., \& {Levison}, H.~F.
  2009, Icarus, 204, 558
 \href{http://adsabs.harvard.edu/abs/2009Icar..204..558M}{\adsurllinklabel}

\bibitem[{{Moro-Mart{\'{\i}}n} \& {Malhotra}(2003)}]{2003AJ....125.2255M}
{Moro-Mart{\'{\i}}n}, A. \& {Malhotra}, R. 2003, \aj, 125, 2255
 \href{http://adsabs.harvard.edu/abs/2003AJ....125.2255M}{\adsurllinklabel}

\bibitem[{{Mouillet} {et~al.}(1997){Mouillet}, {Larwood}, {Papaloizou}, \&
  {Lagrange}}]{1997MNRAS.292..896M}
{Mouillet}, D., {Larwood}, J.~D., {Papaloizou}, J.~C.~B., \& {Lagrange}, A.~M.
  1997, \mnras, 292, 896
 \href{http://adsabs.harvard.edu/abs/1997MNRAS.292..896M}{\adsurllinklabel}

\bibitem[{{Mustill} \& {Wyatt}(2009)}]{2009MNRAS.399.1403M}
{Mustill}, A.~J. \& {Wyatt}, M.~C. 2009, \mnras, 399, 1403
 \href{http://adsabs.harvard.edu/abs/2009MNRAS.399.1403M}{\adsurllinklabel}

\bibitem[{{Najita} \& {Williams}(2005)}]{2005ApJ...635..625N}
{Najita}, J. \& {Williams}, J.~P. 2005, \apj, 635, 625
 \href{http://adsabs.harvard.edu/abs/2005ApJ...635..625N}{\adsurllinklabel}

\bibitem[{{Natta} {et~al.}(2000){Natta}, {Grinin}, \&
  {Mannings}}]{2000prpl.conf..559N}
{Natta}, A., {Grinin}, V., \& {Mannings}, V. 2000, Protostars and Planets IV,
  559
 \href{http://adsabs.harvard.edu/abs/2000prpl.conf..559N}{\adsurllinklabel}

\bibitem[{{Quillen}(2006)}]{2006MNRAS.372L..14Q}
{Quillen}, A.~C. 2006, \mnras, 372, L14
 \href{http://adsabs.harvard.edu/abs/2006MNRAS.372L..14Q}{\adsurllinklabel}

\bibitem[{{Quillen} {et~al.}(2007){Quillen}, {Morbidelli}, \&
  {Moore}}]{2007MNRAS.380.1642Q}
{Quillen}, A.~C., {Morbidelli}, A., \& {Moore}, A. 2007, \mnras, 380, 1642
 \href{http://adsabs.harvard.edu/abs/2007MNRAS.380.1642Q}{\adsurllinklabel}

\bibitem[{{Rebull} {et~al.}(2008){Rebull}, {Stapelfeldt}, {Werner}, {Mannings},
  {Chen}, {Stauffer}, {Smith}, {Song}, {Hines}, \& {Low}}]{2008ApJ...681.1484R}
{Rebull}, L.~M., {Stapelfeldt}, K.~R., {Werner}, M.~W., {Mannings}, V.~G.,
  {Chen}, C., {Stauffer}, J.~R., {Smith}, P.~S., {Song}, I., {Hines}, D., \&
  {Low}, F.~J. 2008, \apj, 681, 1484
 \href{http://adsabs.harvard.edu/abs/2008ApJ...681.1484R}{\adsurllinklabel}

\bibitem[{{Rhee} {et~al.}(2007){Rhee}, {Song}, {Zuckerman}, \&
  {McElwain}}]{2007ApJ...660.1556R}
{Rhee}, J.~H., {Song}, I., {Zuckerman}, B., \& {McElwain}, M. 2007, \apj, 660,
  1556
 \href{http://adsabs.harvard.edu/abs/2007ApJ...660.1556R}{\adsurllinklabel}

\bibitem[{{Rice} {et~al.}(2006){Rice}, {Lodato}, {Pringle}, {Armitage}, \&
  {Bonnell}}]{2006MNRAS.372L...9R}
{Rice}, W.~K.~M., {Lodato}, G., {Pringle}, J.~E., {Armitage}, P.~J., \&
  {Bonnell}, I.~A. 2006, \mnras, 372, L9
 \href{http://adsabs.harvard.edu/abs/2006MNRAS.372L...9R}{\adsurllinklabel}

\bibitem[{{Rieke} {et~al.}(2005){Rieke}, {Su}, {Stansberry}, {Trilling},
  {Bryden}, {Muzerolle}, {White}, {Gorlova}, {Young}, {Beichman},
  {Stapelfeldt}, \& {Hines}}]{2005ApJ...620.1010R}
{Rieke}, G.~H., {Su}, K.~Y.~L., {Stansberry}, J.~A., {Trilling}, D., {Bryden},
  G., {Muzerolle}, J., {White}, B., {Gorlova}, N., {Young}, E.~T., {Beichman},
  C.~A., {Stapelfeldt}, K.~R., \& {Hines}, D.~C. 2005, \apj, 620, 1010
 \href{http://adsabs.harvard.edu/abs/2005ApJ...620.1010R}{\adsurllinklabel}

\bibitem[{{Schmidt-Kaler}(1982)}]{1982lbor.book.....ASK}
{Schmidt-Kaler}, T. in , Landolt-Bornstein: Numerical Data and Functional
  Relationships in Science and Technology, Group VI, ed. K.~{Schaifers}H.~H.
  {Voigt}, Vol.~2 (Berlin:Springer-Verlag)
 \href{http://adsabs.harvard.edu/abs/1982lbor.book.....A}{\urllinklabel}

\bibitem[{{Schneider} {et~al.}(1999){Schneider}, {Smith}, {Becklin}, {Koerner},
  {Meier}, {Hines}, {Lowrance}, {Terrile}, {Thompson}, \&
  {Rieke}}]{1999ApJ...513L.127S}
{Schneider}, G., {Smith}, B.~A., {Becklin}, E.~E., {Koerner}, D.~W., {Meier},
  R., {Hines}, D.~C., {Lowrance}, P.~J., {Terrile}, R.~J., {Thompson}, R.~I.,
  \& {Rieke}, M. 1999, \apjl, 513, L127
 \href{http://adsabs.harvard.edu/abs/1999ApJ...513L.127S}{\adsurllinklabel}

\bibitem[{{Schneider} {et~al.}(2009){Schneider}, {Weinberger}, {Becklin},
  {Debes}, \& {Smith}}]{2009AJ....137...53S}
{Schneider}, G., {Weinberger}, A.~J., {Becklin}, E.~E., {Debes}, J.~H., \&
  {Smith}, B.~A. 2009, \aj, 137, 53
 \href{http://adsabs.harvard.edu/abs/2009AJ....137...53S}{\adsurllinklabel}

\bibitem[{{Siegler} {et~al.}(2007){Siegler}, {Muzerolle}, {Young}, {Rieke},
  {Mamajek}, {Trilling}, {Gorlova}, \& {Su}}]{2007ApJ...654..580S}
{Siegler}, N., {Muzerolle}, J., {Young}, E.~T., {Rieke}, G.~H., {Mamajek},
  E.~E., {Trilling}, D.~E., {Gorlova}, N., \& {Su}, K.~Y.~L. 2007, \apj, 654,
  580
 \href{http://adsabs.harvard.edu/abs/2007ApJ...654..580S}{\adsurllinklabel}

\bibitem[{{Smith} \& {Terrile}(1984)}]{1984Sci...226.1421S}
{Smith}, B.~A. \& {Terrile}, R.~J. 1984, Science, 226, 1421
 \href{http://adsabs.harvard.edu/abs/1984Sci...226.1421S}{\adsurllinklabel}

\bibitem[{{Smith} {et~al.}(2009){Smith}, {Churcher}, {Wyatt}, {Moerchen}, \&
  {Telesco}}]{2009A&A...493..299S}
{Smith}, R., {Churcher}, L.~J., {Wyatt}, M.~C., {Moerchen}, M.~M., \&
  {Telesco}, C.~M. 2009, \aap, 493, 299
 \href{http://adsabs.harvard.edu/abs/2009A&A...493..299S}{\adsurllinklabel}

\bibitem[{{Su} {et~al.}(2006){Su}, {Rieke}, {Stansberry}, {Bryden},
  {Stapelfeldt}, {Trilling}, {Muzerolle}, {Beichman}, {Moro-Martin}, {Hines},
  \& {Werner}}]{2006ApJ...653..675S}
{Su}, K.~Y.~L., {Rieke}, G.~H., {Stansberry}, J.~A., {Bryden}, G.,
  {Stapelfeldt}, K.~R., {Trilling}, D.~E., {Muzerolle}, J., {Beichman}, C.~A.,
  {Moro-Martin}, A., {Hines}, D.~C., \& {Werner}, M.~W. 2006, \apj, 653, 675
 \href{http://adsabs.harvard.edu/abs/2006ApJ...653..675S}{\adsurllinklabel}

\bibitem[{{Telesco} {et~al.}(2000){Telesco}, {Fisher}, {Pi{\~n}a}, {Knacke},
  {Dermott}, {Wyatt}, {Grogan}, {Holmes}, {Ghez}, {Prato}, {Hartmann}, \&
  {Jayawardhana}}]{2000ApJ...530..329T}
{Telesco}, C.~M., {Fisher}, R.~S., {Pi{\~n}a}, R.~K., {Knacke}, R.~F.,
  {Dermott}, S.~F., {Wyatt}, M.~C., {Grogan}, K., {Holmes}, E.~K., {Ghez},
  A.~M., {Prato}, L., {Hartmann}, L.~W., \& {Jayawardhana}, R. 2000, \apj, 530,
  329
 \href{http://adsabs.harvard.edu/abs/2000ApJ...530..329T}{\adsurllinklabel}

\bibitem[{{Telesco} {et~al.}(2005){Telesco}, {Fisher}, {Wyatt}, {Dermott},
  {Kehoe}, {Novotny}, {Mari{\~n}as}, {Radomski}, {Packham}, {De Buizer}, \&
  {Hayward}}]{2005Natur.433..133T}
{Telesco}, C.~M., {Fisher}, R.~S., {Wyatt}, M.~C., {Dermott}, S.~F., {Kehoe},
  T.~J.~J., {Novotny}, S., {Mari{\~n}as}, N., {Radomski}, J.~T., {Packham}, C.,
  {De Buizer}, J., \& {Hayward}, T.~L. 2005, \nat, 433, 133
 \href{http://adsabs.harvard.edu/abs/2005Natur.433..133T}{\adsurllinklabel}

\bibitem[{{Wahhaj} {et~al.}(2005){Wahhaj}, {Koerner}, {Backman}, {Werner},
  {Serabyn}, {Ressler}, \& {Lis}}]{2005ApJ...618..385W}
{Wahhaj}, Z., {Koerner}, D.~W., {Backman}, D.~E., {Werner}, M.~W., {Serabyn},
  E., {Ressler}, M.~E., \& {Lis}, D.~C. 2005, \apj, 618, 385
 \href{http://adsabs.harvard.edu/abs/2005ApJ...618..385W}{\adsurllinklabel}

\bibitem[{{Watson} {et~al.}(2007){Watson}, {Stapelfeldt}, {Wood}, \&
  {M{\'e}nard}}]{2007prpl.conf..523W}
{Watson}, A.~M., {Stapelfeldt}, K.~R., {Wood}, K., \& {M{\'e}nard}, F. 2007,
  Protostars and Planets V, 523
 \href{http://adsabs.harvard.edu/abs/2007prpl.conf..523W}{\adsurllinklabel}

\bibitem[{{Weidenschilling}(1977)}]{1977Ap&SS..51..153W}
{Weidenschilling}, S.~J. 1977, \apss, 51, 153
 \href{http://adsabs.harvard.edu/abs/1977Ap&SS..51..153W}{\adsurllinklabel}

\bibitem[{{Wyatt}(2003)}]{2003ApJ...598.1321W}
{Wyatt}, M.~C. 2003, \apj, 598, 1321
 \href{http://adsabs.harvard.edu/abs/2003ApJ...598.1321W}{\adsurllinklabel}

\bibitem[{{Wyatt}(2005)}]{2005A&A...433.1007W}
---. 2005, \aap, 433, 1007
 \href{http://adsabs.harvard.edu/abs/2005A&A...433.1007W}{\adsurllinklabel}

\bibitem[{{Wyatt}(2008)}]{2008ARA&A..46..339W}
---. 2008, \araa, 46, 339
 \href{http://adsabs.harvard.edu/abs/2008ARA&A..46..339W}{\adsurllinklabel}

\bibitem[{{Wyatt} \& {Dent}(2002)}]{2002MNRAS.334..589W}
{Wyatt}, M.~C. \& {Dent}, W.~R.~F. 2002, \mnras, 334, 589
 \href{http://adsabs.harvard.edu/abs/2002MNRAS.334..589W}{\adsurllinklabel}

\bibitem[{{Wyatt} {et~al.}(1999){Wyatt}, {Dermott}, {Telesco}, {Fisher},
  {Grogan}, {Holmes}, \& {Pi{\~n}a}}]{1999ApJ...527..918W}
{Wyatt}, M.~C., {Dermott}, S.~F., {Telesco}, C.~M., {Fisher}, R.~S., {Grogan},
  K., {Holmes}, E.~K., \& {Pi{\~n}a}, R.~K. 1999, \apj, 527, 918
 \href{http://adsabs.harvard.edu/abs/1999ApJ...527..918W}{\adsurllinklabel}

\bibitem[{{Wyatt} {et~al.}(2007{\natexlab{a}}){Wyatt}, {Smith}, {Greaves},
  {Beichman}, {Bryden}, \& {Lisse}}]{2007ApJ...658..569W}
{Wyatt}, M.~C., {Smith}, R., {Greaves}, J.~S., {Beichman}, C.~A., {Bryden}, G.,
  \& {Lisse}, C.~M. 2007{\natexlab{a}}, \apj, 658, 569
 \href{http://adsabs.harvard.edu/abs/2007ApJ...658..569W}{\adsurllinklabel}

\bibitem[{{Wyatt} {et~al.}(2007{\natexlab{b}}){Wyatt}, {Smith}, {Su}, {Rieke},
  {Greaves}, {Beichman}, \& {Bryden}}]{2007ApJ...663..365W}
{Wyatt}, M.~C., {Smith}, R., {Su}, K.~Y.~L., {Rieke}, G.~H., {Greaves}, J.~S.,
  {Beichman}, C.~A., \& {Bryden}, G. 2007{\natexlab{b}}, \apj, 663, 365
 \href{http://adsabs.harvard.edu/abs/2007ApJ...663..365W}{\adsurllinklabel}

\bibitem[{{Yi} {et~al.}(2001){Yi}, {Demarque}, {Kim}, {Lee}, {Ree}, {Lejeune},
  \& {Barnes}}]{2001ApJS..136..417Y}
{Yi}, S., {Demarque}, P., {Kim}, Y., {Lee}, Y., {Ree}, C.~H., {Lejeune}, T., \&
  {Barnes}, S. 2001, \apjs, 136, 417
 \href{http://adsabs.harvard.edu/abs/2001ApJS..136..417Y}{\adsurllinklabel}

\bibitem[{{Youdin} \& {Shu}(2002)}]{2002ApJ...580..494Y}
{Youdin}, A.~N. \& {Shu}, F.~H. 2002, \apj, 580, 494
 \href{http://adsabs.harvard.edu/abs/2002ApJ...580..494Y}{\adsurllinklabel}

\bibitem[{{Zuckerman} \& {Song}(2004)}]{2004ARA&A..42..685Z}
{Zuckerman}, B. \& {Song}, I. 2004, \araa, 42, 685
 \href{http://adsabs.harvard.edu/abs/2004ARA&A..42..685Z}{\adsurllinklabel}

\end{thebibliography}

\end{document}